\documentclass[11pt]{article}
\pdfoutput=1
\usepackage{jheppub}
\usepackage{amsmath,amssymb}

\usepackage[utf8]{inputenc}
\usepackage[english]{babel}

\usepackage{amsthm}
\usepackage{mathtools}
\usepackage{physics}
\usepackage{xcolor}
\usepackage{graphicx}
\usepackage{adjustbox}
\usepackage{placeins}
\usepackage[T1]{fontenc}
\usepackage{lipsum}
\usepackage{csquotes}
\usepackage{algorithm}
\usepackage{algpseudocode}
\usepackage{MnSymbol} 
\usepackage[normalem]{ulem}
\usepackage{tcolorbox}

\newcommand{\widesim}[2][1.5]{
  \mathrel{\overset{#2}{\scalebox{#1}[1]{$\sim$}}}
}

\usepackage{subfiles} 

\title{Evolution of Parton Distribution Functions in the \\Short-Distance Factorization Scheme}

\newcommand*{\WM}{Physics Department, William \& Mary, Williamsburg, VA 23187, USA}
\newcommand*{\JLAB}{Thomas Jefferson National Accelerator Facility, Newport News, VA 23606, USA}
\newcommand*{\CNRS}{Aix Marseille Univ, Universit\'e de Toulon, CNRS, CPT, Marseille, France.}

\author[a]{Herv\'e Dutrieux,}
\author[b]{ Joseph Karpie,}
\author[a]{ Christopher Monahan,}
\author[a,b]{ Kostas Orginos,}
\author[c]{ Savvas Zafeiropoulos}
\author{\\(on behalf of the HadStruc Collaboration)}
\affiliation[a]{\WM}
\affiliation[b]{\JLAB}
\affiliation[c]{\CNRS}

\emailAdd{hldutrieux@wm.edu}
\emailAdd{jkarpie@jlab.org}
\emailAdd{cmonahan@wm.edu}
\emailAdd{kostas@wm.edu}
\emailAdd{savvas.zafeiropoulos@cpt.univ-mrs.fr}

\date{\today}
\preprint{JLAB-THY-23-3951}

\abstract{
Lattice QCD offers the possibility of computing parton distributions from first principles, although not in the usual $\overline{MS}$ factorization scheme. We study in this paper the evolution of non-singlet parton distribution functions (PDFs) in the short-distance factorization scheme which notably arises in lattice calculations in the pseudo-distribution approach. We provide an assessment of non-perturbative evolution of PDFs from already published lattice matrix elements, and show how this evolution can be used to reduce the fluctuation of the lattice data. We compare our result with expectations obtained thanks to a perturbative matching to $\overline{MS}$. By highlighting the limitations of the current computations, we advocate for a new strategy using lattice calculations in small volume.}

\begin{document}
\maketitle

\section{Introduction}
\label{sec:Intro}
The introduction of the Large-Momentum Effective Theory (LaMET) by X. Ji \cite{PhysRevLett.110.262002,Ji:2014gla} opened a new window for calculations of light-cone correlation functions using lattice quantum chromodynamics (QCD). Numerous works have implemented both LaMET and many variations, such as the pseudo-distribution approach at the center of this paper~\cite{Radyushkin:2017cyf,Radyushkin:2018cvn,Orginos:2017kos} and the equal time current-current correlator approach \cite{Chambers:2017dov, Ma:2017pxb,Bali:2018nde} (Good Lattice Cross Sections). We refer the reader to the recent reviews \cite{Cichy:2018mum, Radyushkin:2019mye,Constantinou:2020hdm} for a panorama of this field of research. The fundamental principle underlying these approaches is that equal-time non-local matrix elements in Euclidean space -- with space-like separation of the constituent fields -- contain the same soft physics as light-cone distributions, although they differ by their regularization of collinear divergences \cite{Braun:2007wv, Ma:2017pxb, Izubuchi:2018srq}. 
Evolution in $z^2$ of pseudo-distributions in this so-called ``short-distance factorization'' scheme (SDF) offers the remarkable possibility of accessing non-perturbative evolution from first principles. 

Perturbation theory provides significant inputs for evolution in SDF: the matching kernels that relate SDF to $\overline{MS}$ for parton distribution functions (PDFs) are available to order ${\cal O}(\alpha_s^2)$ \cite{Chen:2020ody,Li:2020xml}, and PDF evolution equations in $\overline{MS}$ are known exactly to order ${\cal O}(\alpha_s^3)$ \cite{Moch:2004pa, Vogt:2004mw}. However, due to the perturbative divergence of the strong coupling at $\Lambda_{QCD}$, matching derived from perturbation theory can induce a sizeable theoretical uncertainty even at quite small values of $z^2$. We demonstrate in this paper that enforcing the existence of evolution operators directly in the SDF scheme -- which can be tied to the validity of the operator product expansion (OPE) at small values of $z^2$ -- allows us to place a non-perturbative requirement on the $z^2$ dependence of the pseudo-distributions, and therefore improve their extraction with a much lesser degree of model dependence compared to imposing $\overline{MS}$ perturbative results. Furthermore, one can then assess the degree of compatibility of the two approaches.

Before going further, it is useful to retrace some of the difficulties that lattice computations of non-local matrix elements face. As for any lattice calculation with a position-space cut-off $a$ (\textit{i.e.}~a space-time lattice) and a finite volume $L$, the physical box size must be much larger than the inverse of the hadronic scale $\Lambda_{QCD}^{-1}$ to avoid finite volume effects, and the lattice spacing must be much smaller than $\Lambda_{QCD}^{-1}$ to avoid discretization errors. Thus, we have the scale hierarchy:
\begin{align}
    a \ll \Lambda_{QCD}^{-1} \ll L\,.
\end{align}
Similarly, the hadron momentum $P$ and the separation $z$ between space-like separated fields must obey:
\begin{align}
    a \ll P^{-1}\,,\ z \ll L\,. \label{eq:rveiuowihjpo}
\end{align}
On the other hand, probing the parton distributions in SDF with a fine resolution in $x$ requires large Ioffe times, $\nu = z \cdot P$, which can be achieved by either large momentum, $P$, or large separations, $z$, or both, challenging the hierarchy expressed in Eq.~\eqref{eq:rveiuowihjpo}. For large $z$, the increasing presence of higher twist effects and the breakdown of perturbation theory prevent a clean use of these data in our current state of understanding of the matrix element \cite{Ji:2022ezo}, although the ratio renormalization in SDF reduces some higher twist effects \cite{Orginos:2017kos, Karpie:2018zaz}. The major source of improvement in terms of kinematic range on the lattice -- which is desirable both for the pseudo- and quasi-distribution formalism -- resides therefore in an access to larger momenta $P$. However, two obvious issues plague the quest for larger momenta: the requirement of finer lattices, and the issue of the exponential decrease of the signal-to-noise ratio.

A similar problem arises for heavy quark physics in lattice QCD, where the lattice spacing must be much smaller than the inverse of the heavy quark mass \cite{deDivitiis:2003iy}. A solution arises through the relaxation of the requirement that $L \gg \Lambda_{QCD}^{-1}$. Working in volumes smaller than the hadronic scale allows a considerable alleviation of the computational burden, but requires a careful connection to the infinite volume limit, obtained thanks to the step-scaling function technique \cite{Guagnelli:2002jd}. Step-scaling is also used to determine the running of the strong coupling constant in lattice QCD over a range of scales spanning several orders of magnitude \cite{Luscher:1991wu,Luscher:1993gh}. We will explore at the end of this paper the possibility to compute parton distributions in small volumes with larger hadon boost and therefore wider ranges of Ioffe time. Inspired by our work on the non-perturbative evolution operator in SDF, we will propose a step-scaling technique to recover the hadronic parton distribution in infinite volume.

In order to keep the discussion simple and highlight the main points of our work, in this paper we concentrate on the flavor non-singlet unpolarized PDF case. The outline of the paper is as follows. First, we briefly introduce the SDF formalism in section II. Then we derive non-singlet PDF evolution operators in SDF from $\overline{MS}$ perturbative results at all orders in subsection III.A. Finding an appropriate truncation and evaluating some of the systematic effects it generates is discussed in subsection III.B. We perform a numerical application with the perturbative ingredients in subsection III.C. After that, we extract in section IV a non-perturbative SDF evolution operator from actual lattice data for the first time by several methods, and discuss the results in view of the previous perturbative calculations. Finally, in section V, we describe a program to access larger momenta thanks to small volume calculations of non-local matrix elements.

\section{PDFs in the short-distance factorization approach}

PDFs can be probed through the computation of 
objects in which  quark fields are displaced at  space-like 
separations rather than  by  light-cone  intervals. In this work, we will focus on a non-singlet quark PDFs.
The basic building block in this paradigm is
the matrix element (in the case of unpolarized quark PDFs of the nucleon):
 \begin{equation}
 {\cal M}^\alpha  (z,P) \equiv \langle  P |  \bar \psi (0) \frac{\lambda_3}2 \,
 \gamma^\alpha \,  { \hat E} (0,z; A) \psi (z) | P \rangle\,, 
\label{Malpha}
\end{equation}
where $\lambda_3$ is a non-singlet flavor projection and $
{ \hat E}(0,z; A)$ is  the  $0\to z$ straight Wilson line gauge link formed by the gauge field $A_\mu$
 in the fundamental representation of SU(3). The external hadronic states $ | P  \rangle$ carry momentum $P$. A Lorentz covariant decomposition of this matrix element yields:
 \begin{equation}
{\cal M}^\alpha  (z,P) = P^\alpha M(\nu, z^2) + z^\alpha N(\nu, z^2)\,,
 \end{equation}
 where we have introduced the quantity $\nu = z\cdot P$ known as Ioffe time. 
 In the pseudo-distribution approach, one uses  $z=(0,0,0,z_3)$, \mbox{$\alpha=0$} and the  momentum  $P=(P_0,0,0,P_3)$ such that ${\cal M}^0  (z,P) = 
P^0 M(\nu, z^2)$, and forms the Lorentz invariant ratio:
 \begin{equation}
{\mathfrak M} (\nu, z^2) \equiv \frac{ {\cal M}^0 (z, P)}{{\cal M}^0 (z, 0)} \frac{ {\cal M}^0 (0, 0)}{{\cal M}^0 (0, P)} \,,
 \label{redm}
\end{equation}
which is finite in the continuum limit, requires no renormalization, and is directly related to the PDF. 
The fact that the soft physics contained in the $z^2$-dependent matrix element is similar to that of ordinary PDFs up to higher-twist contributions allows one to interpret these matrix elements as PDFs expressed within another factorization scheme, the SDF. The approach is formalized through a non-local OPE, in which short distance contributions are computed perturbatively, as discussed, \textit{e.g.},~in~\cite{Braun:2007wv, Ma:2017pxb,Radyushkin:2018cvn,Izubuchi:2018srq}.

As a consequence, the matrix elements computed on the lattice can be related to the Fourier transform of $\overline{MS}$ PDFs by:
\begin{equation} 
{\mathfrak M}(\nu,z^2)= \int_0^1 d\alpha\, {\cal C}(\alpha,z^2\mu^2,\alpha_s(\mu^2)) {\cal Q}(\alpha\nu,\mu^2) \ + z^2 {\cal B}(\nu, z^2)\,,
 \label{eq:factorization}
\end{equation}
where $\alpha_s(\mu^2)$ is the strong coupling. ${\cal C}(\alpha,z^2\mu^2,\alpha_s(\mu^2))$ is called the matching kernel, computed in perturbation theory, and ${\cal Q}(\nu,\mu^2)$ is the  (normalized) Ioffe-time distribution (ITD) \cite{Braun:1994jq}, defined by:
\begin{equation}
     {\cal Q}(\nu,\mu^2) \equiv \int_{-1}^1 dx\, e^{i\nu x} q(x, \mu^2) \bigg/ \int_{-1}^1 dx\,q(x, \mu^2)\,, \label{eq:ahonlvz}
\end{equation}
where $q(x, \mu^2)$ is the $\overline{MS}$ PDF. Notice that the ITD is a complex quantity, whose real part probes the $x$-even part of $q(x, \mu^2)$ and the imaginary part probes the $x$-odd part of $q(x, \mu^2)$. The term $z^2 {\cal B}(\nu, z^2)$ in Eq.~\eqref{eq:factorization} captures additional corrections to the leading order expression, which vanish in the limit of $z^2=0$.  In particular, because the reduced function by construction satisfies  ${\mathfrak M}(0,z^2)=1$, one  can see that $ {\cal B}(0, z^2) = 0$ and
\begin{equation} 
 \int_0^1 d\alpha\, {\cal C}(\alpha,z^2\mu^2,\alpha_s(\mu^2)) = 1 \,.
 \label{eq:kernel_norm}
\end{equation}
The expansion to fixed order ${\cal O}(\alpha_s)$ of the matching kernel ${\cal C}(\alpha,z^2\mu^2,\alpha_s(\mu^2))$ yields \cite{Radyushkin:2018cvn,Zhang:2018ggy,Izubuchi:2018srq}:
\begin{equation}
{\cal C}(\alpha,z^2\mu^2,\alpha_s(\mu^2)) = \delta(1-\alpha) -  \frac{\alpha_s(\mu^2)}{2\pi} \left[ \ln\left(-z^2\mu^2\frac{e^{2\gamma_E+1}}{4}\right)B_1(\alpha) + D(\alpha)\right] + {\cal O}(\alpha_s^2)\,,
\label{eq:bvjeknks}\end{equation}
where $\gamma_E$ is the Euler-Mascheroni constant, and $B_1(\alpha)$ and $D(\alpha)$ are given by:
\begin{equation}\label{eq:APdef}
B_1(\alpha) = C_F\left[\frac{1+\alpha^2}{1-\alpha}\right]_+ \;\;,\;\; D(\alpha) = C_F\left[ 4 \frac{\ln(1-\alpha)}{1-\alpha} - 2 (1-\alpha)\right]_+\,.
\end{equation}
Here we follow the two prescriptions,
\begin{equation}
f(x) = \int_0^1 d\alpha\, f(\alpha x) \delta(1-\alpha)\,,
\end{equation}
and
\begin{equation}
 G(\alpha)_+ \equiv G(\alpha) - \delta(\alpha) \int_0^1 d\alpha'\,G(\alpha') \,. \label{eq:bhevwnjka}
\end{equation}
The expansion of the matching kernel up to order ${\cal O}(\alpha_s^2)$ can be found, \textit{e.g.}, in \cite{Chen:2020ody,Li:2020xml}.

\section{Evolution operators in perturbation theory}

In this section, we will derive evolution equations in the SDF scheme based on the perturbative matching kernels. For convenience, we will only consider a non-singlet PDF so as to not have to deal with the complication of gluon mixing. 

\subsection{All-order expression}

The DGLAP differential evolution equation of unpolarized non-singlet PDFs reads \cite{Dokshitzer:1977sg,Gribov:1972ri,Altarelli:1977zs}:
\begin{align}
    \frac{d}{d \ln \mu^2} \,  
q (x, \mu^2)    &= 
\int_x^1  \frac{d\alpha}{\alpha}\,   B (\alpha, \alpha_s(\mu^2))\,   q \left(\frac{x}{\alpha}, \mu^2\right) \,, \label{eq:grvbdsisjqs}
 \end{align}
 where $B(\alpha, \alpha_s(\mu^2))$ is known as the DGLAP splitting function and admits a perturbative expansion as:
 \begin{equation}
  B(\alpha, \alpha_s(\mu^2))  = \sum_{n=1}^\infty \left( \frac{\alpha_s(\mu^2)}{2\pi}\right)^n B_n(\alpha)\,.\label{eq:bhirefodaj}
 \end{equation}
 The expression of $B_1(\alpha)$ has already been given in Eq.~\eqref{eq:APdef}, obtained by expanding the matching kernel relating $z^2$-dependent matrix elements to $\mu^2$-dependent PDFs. We clarify the reason the matching kernel contains the expansion of the DGLAP splitting function later on. 
 
Performing the integration of the differential evolution equation \eqref{eq:grvbdsisjqs}, we introduce the DGLAP integrated evolution operator from $\mu_0^2$ to $\mu_1^2$, denoted ${\cal E}(\alpha; \mu_0^2, \mu_1^2)$, such that:
 \begin{equation}
     q (x, \mu_1^2)    = 
\int_x^1  \frac{d\alpha}{\alpha}\,   {\cal E} (\alpha; \mu_0^2, \mu_1^2)\,   q \left(\frac{x}{\alpha}, \mu_0^2\right) \,.
 \end{equation}
Anticipating the concepts presented at the end of this paper, we will refer to ${\cal E}(\alpha; \mu_0^2, \mu_1^2)$ also as the ``step-scaling'' function in the $\overline{MS}$ scheme, as it characterizes scale dependence in a ``step'' from $\mu_0^2$ to $\mu_1^2$. For brevity, we will from now on use the symbolical notation for the DGLAP convolution:
 \begin{equation}
     [f(\textrm{param}_f) \otimes g(\textrm{param}_g)](x) \equiv \int_x^1  \frac{d\alpha}{\alpha}\,   f(\alpha, \textrm{param}_f )\,  g \left(\frac{x}{\alpha}, \textrm{param}_g\right)\,.\label{eq:vhicbejks}
 \end{equation}
 It is trivial to verify that $f \otimes g = g \otimes f$, and that $1_\otimes :\alpha \mapsto \delta(1-\alpha)$ is the identity element ($f \otimes 1_\otimes = f$). We will also use the notation $f^{\otimes n}$ to denote the repeated convolution of $f$ with itself $n$ times, and $f^{\otimes -1}$ the inverse for the convolution defined by:
 \begin{equation}
     f^{\otimes -1} \otimes f = 1_\otimes\,.
 \end{equation}
 There exist some non-trivial considerations on the existence and properties of this inverse, which we will address in the next section. To exemplify the use of the notation, the DGLAP integrated evolution equation \eqref{eq:vhicbejks} now reads:
 \begin{equation}
     q (\mu_1^2)    = 
{\cal E} (\mu_0^2, \mu_1^2)\otimes   q \left( \mu_0^2\right)\,, \label{eq:vhcdkbjnkm}
 \end{equation}
 and ${\cal E} (\mu_0^2, \mu_1^2)^{\otimes-1}$ represents the DGLAP backward evolution operator, that is ${\cal E} (\mu_1^2, \mu_0^2)$.

The Fourier transform of the DGLAP differential evolution equation \eqref{eq:grvbdsisjqs} gives the differential evolution equation of the ITD  \cite{Braun:1994jq}:
 \begin{align}
    \frac{d}{d \ln \mu^2} \,{\cal Q} (\nu, \mu^2)    &= 
\int_0^1  d\alpha \,   B(\alpha, \alpha_s(\mu^2))   \,{\cal Q} (\alpha \nu, \mu^2)  \,.
\label{EEg}
 \end{align}
Applying the Fourier transform has changed the nature of the convolution compared to the one we have dubbed by $\otimes$, although the splitting function $B$ is unchanged. Among the obvious differences, the integral runs from 0 to 1 instead of $x$ to 1. Consequently, the evolution of ITDs probes the entire range of the evolution operator, regardless of the value of the Ioffe time $\nu$, whereas evolution of momentum-dependent PDFs only probes the evolution operators for $\alpha > x$. Therefore, extracting ITDs provides, in principle, a better setting for the study of evolution than working in $x$ space -- although obtaining a good resolution requires as always to access large values of Ioffe time.

\begin{figure}[h]
    \centering
    \includegraphics[scale=.35]{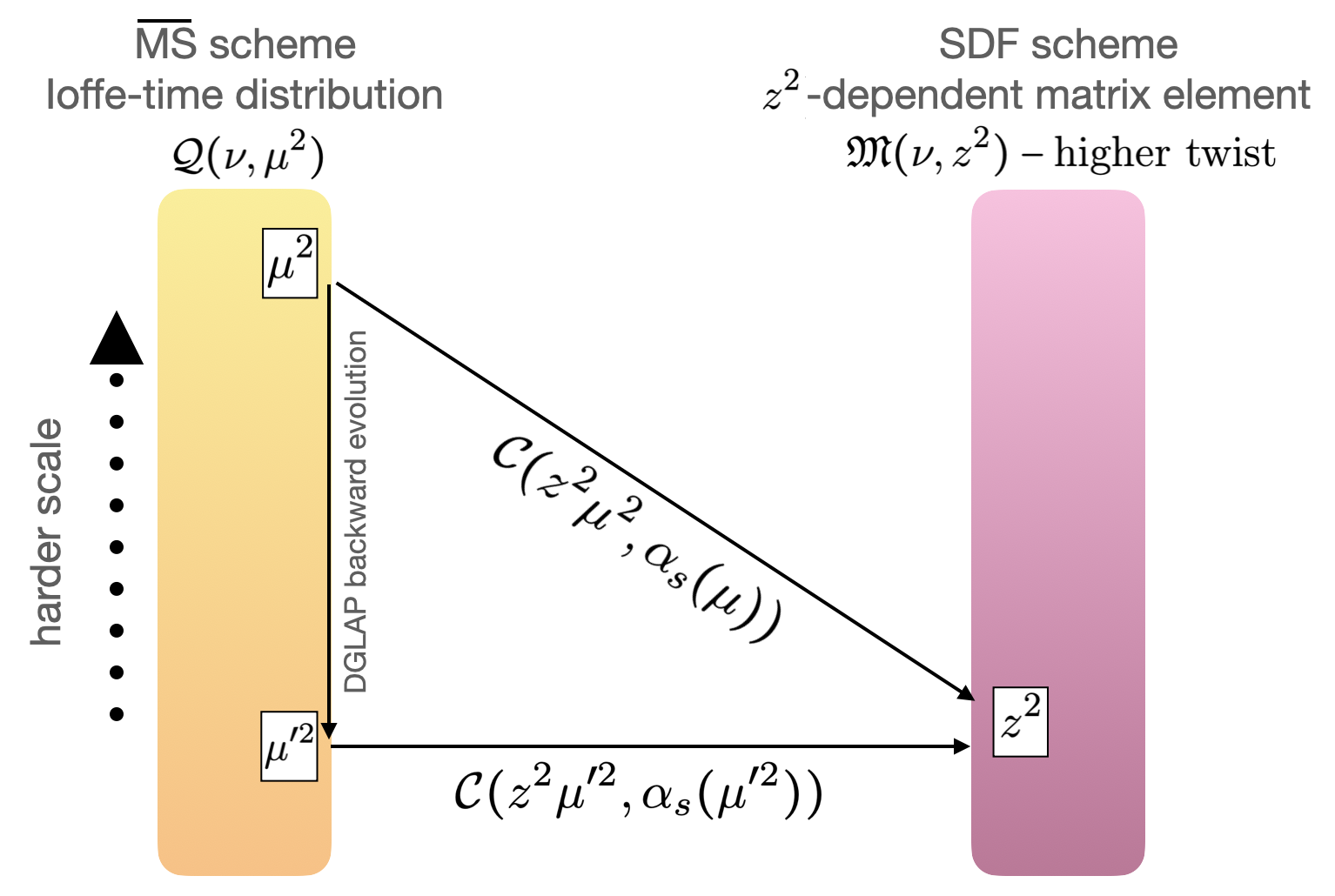}
    \caption{The matching to any scale in $\overline{MS}$ can be performed by dividing the operation into matching to an intermediate scale and evolution to the desired final scale as long as all orders are considered.}
    \label{fig1}
\end{figure}

Observing that the $z^2$-dependent matrix element, whose relation to the ITD is stated in Eq.~\eqref{eq:factorization}, has no $\mu^2$ dependence, we write, assuming $z^2$ is sufficiently small that power corrections are suppressed:
\begin{align}
    0 = \frac{d}{d\ln \mu^2}{\mathfrak M}(\nu,z^2)  &= \int_0^1 d\alpha\, \frac{d}{d\ln \mu^2}{\cal C}(\alpha,z^2\mu^2,\alpha_s(\mu^2)) \,{\cal Q}(\alpha\nu,\mu^2) \nonumber \\
    &\hspace{-30pt}+ \int_0^1 d\beta\, {\cal C}(\beta,z^2\mu^2,\alpha_s(\mu^2)) \int_0^1  d\alpha \,   B (\alpha, \alpha_s(\mu^2))   \,{\cal Q} (\alpha \beta\nu, \mu^2)+ {\mathcal O}(z^2)\,,\label{eq:vrfebjdnsk}
\end{align}
with the leading power term giving:
\begin{align}
    &\int_0^1 d\beta\, {\cal C}(\beta,z^2\mu^2,\alpha_s(\mu^2)) \int_0^\beta  \frac{d\alpha}{\beta} \,   B \left(\frac{\alpha}{\beta} , \alpha_s(\mu^2) \right)  \,{\cal Q} (\alpha \nu, \mu^2) = \nonumber \\
    &\hspace{30pt}\int_0^1 d\alpha\, {\cal Q} (\alpha \nu, \mu^2) \int_\alpha^1 \frac{d\beta}{\beta}\,{\cal C}(\beta,z^2\mu^2,\alpha_s(\mu^2)) B \left(\frac{\alpha}{\beta} , \alpha_s(\mu^2) \right)\,.\label{eq:tvuoajs}
\end{align}
The latter term contains the DGLAP convolution, and the combination of Eqs.~\eqref{eq:vrfebjdnsk} and \eqref{eq:tvuoajs} gives therefore that the renormalization group equation of the matching kernel is exactly the opposite of the ordinary DGLAP equation:
\begin{equation}
    \label{eq:vebcdjlkej}
    \frac{d}{d\ln \mu^2}{\cal C}(z^2\mu^2,\alpha_s(\mu^2)) = -B (\alpha_s(\mu^2)) \otimes {\cal C}\left(z^2\mu^2,\alpha_s(\mu^2)\right)\,.
\end{equation}
This relation already expressed in various forms in the literature (see \textit{e.g.} \cite{Monahan:2016bvm,Gao:2021hxl, Su:2022fiu}) explains the observation made earlier that the matching kernel contains the terms of the expansion of the DGLAP splitting function. Due to the sign flip compared to the ordinary differential equation, the $\mu^2$ dependence of the matching kernel is dictated by the DGLAP backward evolution operator (compare to Eq.~\eqref{eq:vhcdkbjnkm}): 
\begin{equation}
    {\cal C}( z^2\mu^2,\alpha_s(\mu^2)) = {\cal E}(\mu^2, \mu'^2) \otimes {\cal C}( z^2\mu'^2,\alpha_s(\mu'^2))\,. \label{eq:efvdvnjke}
    \end{equation}
In short, the matching can be performed at an intermediate scale $\mu'^2$ and evolved later without any consequences, because of a trade-off between the matching and the DGLAP evolution that occurs if all orders are considered. The physical content of Eq.~\eqref{eq:efvdvnjke} is very intuitive, and summarized in Figure \ref{fig1}. Of course, if one uses a finite truncation, then the scale at which the matching is done actually matters. 

We have reviewed the evolution equations of the PDF \eqref{eq:vhcdkbjnkm}, Ioffe-time distribution \eqref{EEg} and matching kernel \eqref{eq:efvdvnjke}. We are now ready to derive the final result of this subsection, namely the evolution equation with respect to $z^2$ of the matrix element.

Expressing the matrix element (with the subtraction of the higher-twist corrections) as a function of itself at a different value of $z^2$, some algebra gives: 
\begin{align}
    {\mathfrak M}(\nu,z_1^2) &= \int_0^1 d\alpha\, {\cal C}(\alpha,z_1^2\mu^2,\alpha_s(\mu^2)) {\cal Q}(\alpha\nu,\mu^2) \,, \\
    &= \int_0^1 d\alpha{\mathfrak M}(\alpha\nu,z_0^2) \int_{\alpha}^1 \frac{d\beta}{\beta} {\cal C}\left(\frac{\alpha}{\beta},z_1^2\mu^2,\alpha_s(\mu^2)\right) {\cal C}^{\otimes-1}(\beta,z_0^2\mu^2,\alpha_s(\mu^2)) \,.
\end{align}

\begin{figure}[h]
    \centering
    \includegraphics[scale=.35]{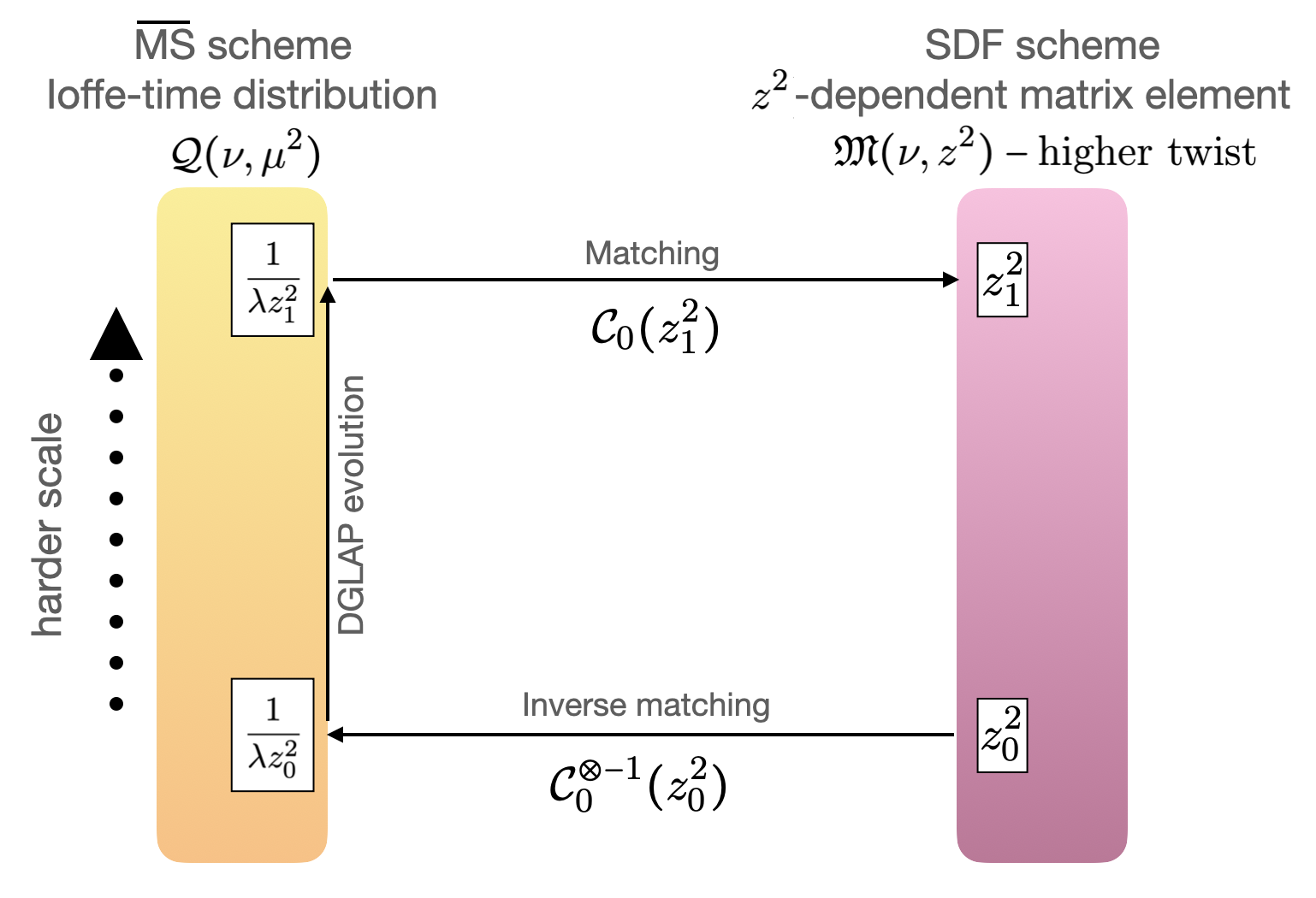}
    \caption{The evolution in SDF depending on $z^2$ is derived from the $\overline{MS}$ evolution by a back-and-forth matching procedure.}
    \label{fig2}
\end{figure}

The SDF step-scaling function is therefore given by:
\begin{equation} 
\Sigma(z_0^2,z_1^2) \equiv {\cal C}(z_1^2\mu^2,\alpha_s(\mu^2)) \otimes  {\cal C}^{\otimes-1}(z^2_0\mu^2,\alpha_s(\mu^2))\,.
\end{equation}
The SDF step-scaling function is independent of $\mu^2$ if all orders are considered. To see this, we notice that introducing an intermediate scale $(\lambda z^2)^{-1}$ where $\lambda \equiv -e^{2\gamma_E+1}/4$ gives a simplified form of the matching kernel without logarithmic dependence in Eq.~\eqref{eq:bvjeknks}:
\begin{equation}
    {\cal C}_0(z^2) \equiv {\cal C}\left( \frac{1}{\lambda},\alpha_s((\lambda z^2)^{-1})\right) = 1_\otimes - \frac{\alpha_s((\lambda z^2)^{-1})}{2\pi} D + {\cal O}(\alpha_s^2)\,.
    \label{eq:bvekjcnsds}
\end{equation}
We explore in more depth the perturbative expansion of the SDF matching kernel, how it relates to the DGLAP splitting function at higher orders, and the question of the scale setting subtended by this coefficient $\lambda$ in Appendix \ref{canonical_scale}.
Then Eq.~\eqref{eq:efvdvnjke} gives:
\begin{align} 
\Sigma(z_0^2,z_1^2) &= {\cal E}\left(\mu^2, \frac{1}{\lambda z_1^2}\right) \otimes {\cal C}_0(z_1^2) \otimes {\cal E}\left(\frac{1}{\lambda z_0^2}, \mu^2\right) \otimes {\cal C}_0^{\otimes-1}(z_0^2)\,, \\
&= {\cal E}\left(\frac{1}{\lambda z_0^2}, \frac{1}{\lambda z_1^2}\right) \otimes {\cal C}_0(z_1^2) \otimes {\cal C}_0^{\otimes-1}(z_0^2)\,.
\label{eq:vihcdsjak}\end{align}
The latter expression shows how the SDF step-scaling function can be computed perturbatively to any order from the $\overline{MS}$ one. The content of Eq.~\eqref{eq:vihcdsjak} is again intuitive, and summarized in Figure \ref{fig2}.

By differentiating Eq.~\eqref{eq:vihcdsjak} with respect to $\ln z_1^2$, one can derive SDF splitting functions:
\begin{align}
    \frac{d}{d\ln z_1^2}\Sigma(z_0^2,z_1^2) &= \frac{d}{d\ln z_1^2}{\cal E}\left(\frac{1}{\lambda z_0^2}, \frac{1}{\lambda z_1^2}\right) \otimes {\cal C}_0(z_1^2) \otimes {\cal C}_0^{\otimes-1}(z_0^2) \nonumber \\
    & \qquad +{\cal E}\left(\frac{1}{\lambda z_0^2}, \frac{1}{\lambda z_1^2}\right) \otimes \frac{d}{d\ln z_1^2}{\cal C}_0(z_1^2) \otimes {\cal C}_0^{\otimes-1}(z_0^2)\,.
\end{align}
The first term produces the opposite of the ordinary DGLAP splitting function:
\begin{equation}
    \frac{d}{d\ln z_1^2}{\cal E}\left(\frac{1}{\lambda z_0^2}, \frac{1}{\lambda z_1^2}\right) = - B(\alpha_s((\lambda z_1^2)^{-1})) \otimes {\cal E}\left(\frac{1}{\lambda z_0^2}, \frac{1}{\lambda z_1^2}\right)\,,
\end{equation}
whereas the second term can be reshaped by noticing that:
\begin{equation}
    {\cal E}\left(\frac{1}{\lambda z_0^2}, \frac{1}{\lambda z_1^2}\right) \otimes {\cal C}_0^{\otimes-1}(z_0^2) = \Sigma(z_0^2, z_1^2) \otimes {\cal C}_0^{\otimes-1}(z_1^2)\,.\label{eq:fghjbckjaxn}
\end{equation}
Therefore, we obtain:
\begin{equation}
    \frac{d}{d\ln z_1^2}\Sigma(z_0^2,z_1^2) = \bigg\{- B(\alpha_s((\lambda z_1^2)^{-1})) + \frac{d}{d \ln z_1^2}{\cal C}_0(z_1^2)\otimes {\cal C}_0^{\otimes-1}(z_1^2)\bigg\} \otimes \Sigma(z_0^2,z_1^2)\,.
\end{equation}
Alternatively, using the running of the strong coupling defined by:
\begin{equation}
    \frac{d}{d\ln\mu^2}\alpha_s(\mu^2) = \beta(\alpha_s(\mu^2))\,,
\end{equation}
we find:
\begin{equation}
    \frac{d}{d\ln z_1^2}\Sigma(z_0^2,z_1^2) = \bigg\{-B(\alpha_s((\lambda z_1^2)^{-1})) - \beta(\alpha_s((\lambda z_1^2)^{-1}))\frac{d}{d\alpha_s}{\cal C}_0(z_1^2)\otimes {\cal C}_0^{\otimes-1}(z_1^2)\bigg\} \otimes \Sigma(z_0^2,z_1^2)\,. \label{eq:gheiufwjdlwkn}
\end{equation}
The bracketed result of Eq.~\eqref{eq:gheiufwjdlwkn} is the SDF splitting function. This result closely resembles the ``physical anomalous dimensions'' derived in \cite{Catani:1996sc,Blumlein:2000wh,Blumlein:2021lmf} for the $Q^2$ dependence of the structure functions of deep inelastic scattering (DIS) -- or the results obtained for the renormalization relating dimensionally regularized quantities to $\overline{MS}$ (see, for instance, Eq.~(9) of \cite{Bertone:2022frx}). This does not come as a surprise since ${\mathfrak M}(\nu,z^2)$ is scheme-independent. Therefore, the SDF step-scaling and splitting functions are independent of the scheme used to compute them in perturbation theory, provided all orders are considered. Notably, scheme independence can be seen by recognizing that $\beta(\alpha_s) \,d/d\alpha_s$ is scheme-invariant and that the scheme dependence of the first term cancels that of the second term. However, any truncation of the perturbative expansion introduces scheme dependence.

\subsection{Performing the perturbative truncation}

Eqs.~\eqref{eq:vihcdsjak} and \eqref{eq:gheiufwjdlwkn} give, respectively, the SDF step-scaling function and the SDF splitting function at any desired order in perturbation theory. To derive the SDF perturbative evolution, we only need to find a consistent expression for ${\cal C}_0^{\otimes-1}$. For instance, if one works at order ${\cal O}(\alpha_s)$ for the matching kernel  \eqref{eq:bvekjcnsds}, one would be tempted to use:
\begin{equation}
{\cal C}_0^{\otimes-1} (z_0^2) = \left(1_{\otimes} - \frac{\alpha_s(\mu_0^2)}{2\pi}D\right)^{\otimes-1}\,,\label{eq:bvschknj}
\end{equation}
where $\mu_0^2 = (\lambda z_0^2)^{-1}$.
However, that this expression defines an actual distribution is doubtful. Working in Mellin space allows one to pinpoint the potential issue. The Mellin moments of the matching kernel are defined as:
\begin{align}
c_n(z^2) \equiv  \int_0^1 d\alpha\,\alpha^{n-1} {\cal C}_0\left(\alpha,z^2\right) \,.
\label{eq:wilsmom}
\end{align}
It is well-known and easy to verify that the Mellin transform converts the DGLAP convolution $\otimes$ into an ordinary product of Mellin moments, and that, provided ${\cal C}_0^{\otimes-1}$ exists, its Mellin moments are simply $1 / c_n$. An issue arises obviously if $c_n = 0$. We refer the interested reader to Appendix \ref{appendixa} for more details on the consequences of this pole in the Mellin moments of ${\cal C}_0^{\otimes-1}$. The intuitive conclusion of Appendix \ref{appendixa} is that one can obtain a workable definition of ${\cal C}_0^{\otimes-1}$ provided that one truncates it equally to the same perturbative order, as done \textit{e.g.} in \cite{Radyushkin:2019owq, Gao:2022uhg}. Here:
\begin{equation}
{\cal C}_0^{\otimes-1} (z_0^2) = 1_{\otimes} + \frac{\alpha_s(\mu_0^2)}{2\pi}D\,.\label{eq:ewfijpoo}
\end{equation}
Then the direct relation between the SDF and $\overline{MS}$ step-scaling functions \eqref{eq:vihcdsjak} yields:
\begin{equation}
\Sigma(z_0^2,z_1^2) = {\cal E}(\mu_0^2, \mu_1^2) \otimes \left(1_{\otimes} - \frac{\alpha_s(\mu_1^2)}{2\pi}D\right) \otimes \left(1_{\otimes} + \frac{\alpha_s(\mu_0^2)}{2\pi}D\right)\,.\label{eq:viebwjcwa}
\end{equation}
A contribution of order ${\cal O}(\alpha_s^2)$ results in that $\Sigma(z_0^2, z_0^2) \neq 1_\otimes$, so not applying any evolution does not correspond to the identity operator. To preserve $\Sigma(z_0^2, z_0^2) = 1_\otimes$, a further truncation is required yielding the final form of the SDF step-scaling function matched at fixed order ${\cal O}(\alpha_s)$. We will be using this form for our numerical applications:
\begin{equation}
\Sigma(z_0^2,z_1^2) = {\cal E}(\mu_0^2, \mu_1^2) \otimes \left(1_{\otimes} + \frac{\alpha_s(\mu_0^2)-\alpha_s(\mu_1^2)}{2\pi}D\right)\,.\label{eq:eqhiojwo}
\end{equation}

\subsubsection{Alternative truncation choices}

In the $\overline{MS}$ scheme a fixed-order truncation at the level of the step-scaling function is sub-optimal, and  a fixed-order truncation at the level of the splitting function is preferred. When integrating the differential equation with the truncated splitting function, one obtains a step-scaling function with renormalization group improvement -- that is the resummation of the dominant logarithms of the ratio of scales, classified as LL (leading logarithms resummed), NLL (next-to-leading logarithms), and so on. Eq.~\eqref{eq:eqhiojwo} presents a factorized form of the step-scaling function, with an evolution and a matching term. It is straightforward to implement the $\overline{MS}$ renormalization group improvement on ${\cal E}$. On the other hand, the truncation we have suggested is of fixed-order at the level of the matching term. Since we have computed the SDF splitting function, we could try to introduce a renormalization group improvement of the matching part as well, using Eq.~\eqref{eq:gheiufwjdlwkn}, instead of working directly with the step-scaling functions \eqref{eq:vihcdsjak}.

Applying our ${\cal O}(\alpha_s)$ truncated ${\cal C}_0$ and ${\cal C}^{\otimes-1}_0$ in Eq.~\eqref{eq:gheiufwjdlwkn} yields:
\begin{equation}
    \frac{d}{d\ln z_1^2}\Sigma(z_0^2,z_1^2) = \bigg\{-B(\alpha_s(\mu_1^2)) + \frac{\beta(\alpha_s(\mu_1^2))}{2\pi}D\otimes \left[1_{\otimes} + \frac{\alpha_s(\mu_1^2)}{2\pi}D\right]\bigg\} \otimes \Sigma(z_0^2,z_1^2)\,.\label{eq:bhkecnsm}
\end{equation}
However, this expression lends itself to multiple different perturbative truncations depending on the ingredients one chooses to use. Indeed, as $\beta(\alpha_s) = -\beta_0\alpha_s^2 + ...$, the one-loop direct matching kernel $D$ only enters at order ${\cal O}(\alpha_s^2)$ in Eq.~\eqref{eq:bhkecnsm}. The one-loop inverse matching kernel only enters at order ${\cal O}(\alpha_s^3)$. If one used a fixed-order truncation of Eq.~\eqref{eq:bhkecnsm} to order ${\cal O}(\alpha_s^3)$ for instance, one would be manipulating simultaneously the three-loop DGLAP splitting function, two-loop direct matching kernel and one-loop inverse matching kernel, in a curious mixture from a physical point of view. 

Instead, we propose to illustrate the similarities and differences between Eqs \eqref{eq:bhkecnsm} and \eqref{eq:eqhiojwo} by expressing both with the same physical ingredients, namely one-loop DGLAP splitting function (${\cal O}(\alpha_s)$), matching kernel (${\cal O}(\alpha_s)$), and beta function (${\cal O}(\alpha_s^2)$).

Let us first review the effect of renormalization group improvement at one-loop on the $\overline{MS}$ step-scaling function. The LL evolution operator, ${\cal E}_{LL}(\mu_0^2, \mu_1^2)$ is given by the solution of:
\begin{spreadlines}{1ex}
\begin{equation}
\begin{dcases}
\frac{d}{d \ln \mu_1^2} {\cal E}_{LL}(\mu_0^2, \mu_1^2) = \frac{\alpha_s(\mu_1^2)}{2\pi} B_1 \otimes {\cal E}_{LL}(\mu_0^2, \mu_1^2)\,, \\
\frac{d}{d \ln \mu_1^2} \alpha_s(\mu_1^2) = -\beta_0\alpha_s^2(\mu_1^2)\,.
\end{dcases}
\end{equation}
\end{spreadlines}
This system of differential equations yields:
\begin{equation}
    {\cal E}_{LL}(\mu_0^2, \mu_1^2) = \sum_{k=0}^{+\infty} \frac{1}{k!} B_1^{\otimes k} \left(\frac{1}{2\pi\beta_0}\ln\left(\frac{\alpha_s(\mu_0^2)}{\alpha_s(\mu_1^2)}\right)\right)^k\,. \label{eq:wsuhoilk}
\end{equation}
The result looks more familiar when expressed in terms of Mellin moments. If $\gamma_n$ and $e_n(\mu_0^2, \mu_1^2)$ are respectively the Mellin moments of $B_1$ and ${\cal E}_{LL}(\mu_0^2, \mu_1^2)$ defined analogous to Eq.~\eqref{eq:wilsmom}, then:
\begin{equation}
e_n(\mu_0^2, \mu_1^2) = \left(\frac{\alpha_s(\mu_0^2)}{\alpha_s(\mu_1^2)}\right)^{\gamma_n/(2\pi\beta_0)}\,.\label{eq:evfdvscvc}
\end{equation}
A crucial observation is that, although $B_1$ is mathematically a distribution defined as a plus-prescription according to Eq.~\eqref{eq:APdef}, ${\cal E}_{LL}(\mu_0^2, \mu_1^2)$ is an ordinary function as long as $\mu_1 > \mu_0$: the $\overline{MS}$ renormalization group improvement has the effect of removing the plus-prescription that appears in the splitting function. This effect is not evident as ${\cal E}_{LL}(\mu_0^2, \mu_1^2)$ is written as an infinite sum of convolutions of $B_1$, as seen in Eq.~\eqref{eq:wsuhoilk}. 

As we will show below, ${\cal E}_{LL}(\mu_0^2, \mu_1^2)$ is not only an ordinary function, but a positive one for $\mu_1 > \mu_0$. In the context of the probabilistic interpretation of the PDF $q(x, \mu^2)$ as the number density of partons carrying a momentum fraction $x$ at scale $\mu^2$, one could likewise interpret ${\cal E}_{LL}(\alpha, \mu_0^2, \mu_1^2)$ as the probability of finding a parton with momentum fraction $\alpha x$ at scale $\mu_1^2$ inside a parton with momentum fraction $x$ at scale $\mu_0^2$. Indeed, we remind that the probability density of the product of two independent probability densities $f$ and $g$ restricted to the interval $[0,1]$ is precisely $f \otimes g$. This picture has been used to derive the DGLAP evolution equation \cite{Collins:1988wj}. Note, however, that the probabilistic interpretation may appear problematic, as it is established that renormalization may spoil the positivity of PDFs at low $\overline{MS}$ scales \cite{Collins:2021vke}.

Aside from probabilistic considerations, the existence of evolution operators, whether in $\mu^2$ or $z^2$, can be tied to the validity of the OPE that defines the moments of (pseudo-) PDFs \cite{Karpie:2018zaz} at small $z^2$. Provided that the moments exist, in Mellin space the DGLAP convolution reduces to:
\begin{equation}
    q_n(\mu_1^2) = e_n(\mu_0^2, \mu_1^2) q_n(\mu_0^2)\,,\label{eq:r3efvd}
\end{equation}
where $q_n(\mu^2)$ are the Mellin moments of $q(x, \mu^2)$. Eq. \eqref{eq:r3efvd} defines the evolution operator straightforwardly as the inverse Mellin transform of $e_n(\mu_0^2, \mu_1^2) = q_n(\mu_1^2) / q_n(\mu_0^2)$\footnote{An enlightening parallel could be drawn with the Radon transform and the polynomiality property of generalized parton distributions (GPDs). The polynomiality property implies a sophisticated dependence of GPDs on their variables $(x, \xi)$ \cite{Ji:1998pc,Radyushkin:1998bz}. However, if one applies an inverse Radon transform to GPDs to express them under the form of ``double distributions'' \cite{Radyushkin:1997ki}, the polynomiality property becomes trivially verified \cite{Teryaev:2001qm, Chouika:2017dhe}.  In the end, the polynomiality property is converted into a requirement of well-definition of the inverse Radon transform with additional requirements on the support of the distributions. In our case, the DGLAP relation implies a sophisticated dependence of PDFs on their variables $(x, \mu^2)$ that can be converted into a requirement of a well-defined inverse Mellin transform with again some support properties. That the Mellin transform is applicable is ultimately guaranteed by the validity of the OPE.}. However, one still needs to demonstrate that the operator defined in this fashion is restricted to a support $\alpha \in [0,1]$, which can be done to all orders in perturbation theory thanks to the use of partonic on-shell states (see \textit{e.g.} \cite{collins_2011}). 

Let us mention another perspective, related to the arguments of growth of higher-twists and breaking of perturbation theory mentioned in the introduction, on why the evolution operators ${\cal E}$ and $\Sigma$ are limited to small $z^2$. The factorization scale of PDFs is akin to a cut-off in the $k_\perp$ integration of transverse momentum of partons inside the hadron. One can introduce a ``primordial'' (straight-link) transverse momentum distribution (TMD) ${\cal{F}}(x,k_\perp)$ \cite{Radyushkin:2017cyf,Radyushkin:2016hsy,Radyushkin:2018nbf} whose Fourier transforms give the pseudo-ITD:
\begin{equation}
\mathfrak{M}(\nu,z_\perp^2) = \int dx \int d^2 k_\perp e^{i\nu x + i k_\perp \cdot z_\perp}     {\cal{F}}(x, k_\perp^2)\,.
\end{equation}
With $z^2=0$, one can write schematically the PDF evolution equation as:
\begin{equation}
    \frac{d q(x, \mu^2)}{d\ln \mu^2} = \frac{d}{d\ln \mu^2} \int_0^{\mu^2} d^2 k_\perp {\cal{F}}(x, k_\perp^2)\,.\label{eq:fufvduijnzf}
\end{equation}
The large $k_\perp$ region of the TMD can be computed in perturbation theory -- in renormalizable theories, it varies as $1/k_\perp^2$, which integrated against $d^2 k_\perp$ gives the usual logarithmic scale dependence at large $\mu^2$ or small $z^2$ of the PDFs. On the other hand, the small $k_\perp$ (soft) regime of the TMD is deeply non-perturbative and tied to the confinement and finite size of the proton. Typical models of the soft TMD component have a very different form compared to the hard tail, such as a Gaussian dependence in $k_\perp$. The evolution equation is therefore fundamentally different at small $k_\perp$ or large $z$, and there are no first-principle grounds that the step-scaling definition we have used would hold at large $z$ -- except maybe for the hope that the cancellation of $z^2$ effects induced by the ratio renormalization results in a trivial evolution at large $z$. 

Finally, note that regardless of the probabilistic interpretation attributed to ${\cal E}_{LL}(\alpha, \mu_0^2, \mu_1^2)$, it is a positive ordinary function, which allows one to consider evolution to a higher scale as a simple reweighting of the parton distribution at an initial scale. This forms the basis of the study led in \cite{Dutrieux:2023qnz} which deals with the question ``which region of the parton distribution at initial scale contributes the most to which region at final scale?''.

We will only show that ${\cal E}_{LL}$ is an ordinary function in the main text, the positivity being demonstrated in Appendix \ref{positivity}. That ${\cal E}_{LL}$ does not present the same distribution-like features as $B_1$ can be broadly understood as a consequence of the fact that ${\cal E}_{LL}$ results from the integration of $B_1$, and therefore smooths out the singular behavior of the latter. The argument can be made rigorous in Mellin space, where the distribution-like features, which are localized at $\alpha = 1$, are mirrored in the large $n$ behavior of the Mellin moments. Indeed, the Mellin moments of an ordinary integrable function whose support is restricted to $[0,1]$ tend to 0 as $n \rightarrow +\infty$. A Dirac delta at $\alpha = 1$ induces Mellin moments which tend to a non-zero constant as $n \rightarrow +\infty$, whereas a plus-prescription ${\cal O}(1/(1-\alpha))_+$ induces moments diverging logarithmically at large $n$. That is the case for the Mellin moments of $B_1$, known as the leading order (LO) DGLAP anomalous dimensions, $\gamma_n$:
\begin{equation}
    \gamma_n  \  \mathrel{\overset{n\rightarrow+\infty}{\scalebox{1.5}[1]{$=$}}}\   -2C_F\ln(n) + C_F\left(\frac{3}{2}-2\gamma_E\right) + {\cal O}\left(\frac{1}{n}\right)\,. \label{eq:frbjknecfd}
\end{equation}
Therefore, the large $n$ behavior of the Mellin moments of 
${\cal E}_{LL}(\mu_0^2, \mu_1^2)$ reads:
\begin{equation}
e_n(\mu_0^2, \mu_1^2) = \left(\frac{\alpha_s(\mu_0^2)}{\alpha_s(\mu_1^2)}\right)^{\gamma_n / (2\pi \beta_0)}   \  \widesim[2]{n \rightarrow +\infty} \  A  n^{-\displaystyle \frac{C_F}{\pi\beta_0}\ln\left(\frac{\alpha_s(\mu_0^2)}{\alpha_s(\mu_1^2)}\right)}\,, 
\end{equation}
where
\begin{equation}
A = \left(\frac{\alpha_s(\mu_0^2)}{\alpha_s(\mu_1^2)}\right)^{C_F(3-4\gamma_E) / (4\pi\beta_0)}\,. \label{eq:refrihbjkenc}
\end{equation}
When $\mu_1 > \mu_0$, we have $\alpha_s(\mu_0^2)/\alpha_s(\mu_1^2) > 1$, so $e_n(\mu_0^2, \mu_1^2)$ falls off as a negative power of $n$. A straightforward inverse Mellin transform implies a behavior of ${\cal E}_{LL}(\alpha, \mu_0^2, \mu_1^2)$ for $\alpha \rightarrow 1$ dominated by the ordinary function:
\begin{equation}
    {\cal E}_{LL}(\alpha, \mu_0^2, \mu_1^2)  \  \widesim[2]{\alpha \rightarrow 1}\   \left(-\ln(\alpha)\right)^{\displaystyle -1+\frac{C_F}{\pi\beta_0}\ln\left(\frac{\alpha_s(\mu_0^2)}{\alpha_s(\mu_1^2)}\right)} \times \frac{A}{\displaystyle \Gamma\left(\frac{C_F}{\pi\beta_0}\ln\left(\frac{\alpha_s(\mu_0^2)}{\alpha_s(\mu_1^2)}\right)\right)} \,, \label{eq:bbjekvdcds}
\end{equation}
where $\Gamma$ is the ordinary Gamma function. The LL $\overline{MS}$ step-scaling function therefore only presents an integrable power divergence  at $\alpha = 1$ when $\mu_1 > \mu_0$. On the contrary, if $\mu_1 = \mu_0$, the step-scaling function reduces to the identity $1_\otimes$, which is a Dirac delta at $\alpha = 1$. If $\mu_1 < \mu_0$, then  $e_n(\mu_0^2, \mu_1^2)$ increases at large $n$ as a power of $n$. This shows that the LL backward DGLAP evolution operator presents distribution-like features in the vicinity of $\alpha = 1$, which explains the observation that backward evolution is considerably noisier than the forward one: renormalization group improvement does not always smooth out the distribution-like features.

The large $n$ behavior of the Mellin moments of the matching kernel, presented in more details in Eq.~\eqref{eq:bhevwcnjka}, diverges as a power of $\ln(n)$ as the matching kernel contains a plus-prescription:
\begin{equation}
    c_n(z^2) \  \widesim[2]{n \rightarrow +\infty} \ - \frac{\alpha_s(\mu^2)}{\pi} C_F \ln^2(n)\,.
\end{equation}
Therefore, the SDF step-scaling function derived in Eq.~\eqref{eq:eqhiojwo} has Mellin moments $\sigma_n(z_0^2, z_1^2)$ that satisfy:
\begin{equation}
   \sigma_n(z_0^2, z_1^2)    \  \widesim[2]{n \rightarrow +\infty}\  e_n\left(\mu_0^2, \mu_1^2\right)\frac{\alpha_s(\mu_0^2)-\alpha_s(\mu_1^2)}{\pi} C_F\ln^2(n)\,.\label{eq:brvbdkjcn}
\end{equation}
If $z_0^2 > z_1^2$, $e_n$ falls off like a negative power of $n$, and the $\ln^2(n)$ term added by the matching kernels does not change that general behavior: the SDF step-scaling function remains an ordinary function. An inverse Mellin transform gives that the large $\alpha$ behavior of the SDF step-scaling function is multiplied by $\ln^2(1-\alpha)$ compared to the LL $\overline{MS}$ step-scaling function.

On the other hand, Eq.~\eqref{eq:bhkecnsm} yields:
\begin{spreadlines}{1ex}
\begin{equation}
\begin{dcases}
\frac{d}{d\ln \mu_1^2}\Sigma(z_0^2,z_1^2) = \bigg\{\frac{\alpha_s(\mu_1^2)}{2\pi} B_1 + \beta_0\frac{\alpha_s^2(\mu_1^2)}{2\pi}D+\beta_0\frac{\alpha_s^3(\mu_1^2)}{(2\pi)^2}D^2\bigg\} \otimes \Sigma(z_0^2,z_1^2)\,, \\
\frac{d}{d \ln \mu_1^2} \alpha_s(\mu_1^2) = -\beta_0\alpha_s^2(\mu_1^2)\,.
\end{dcases}
\end{equation}
\end{spreadlines}
The solution, which one may verify by differentiating the result and observing that it obeys the previous system, is obtained as:
\begin{equation}
\Sigma(z_0^2,z_1^2) = {\cal E}_{LL}(\mu_0^2, \mu_1^2) \otimes \left(\sum_{k=0}^{+\infty} \frac{1}{k!} D^{\otimes k} \left(\frac{\alpha_s(\mu_0^2)-\alpha_s(\mu_1^2)}{2\pi}\right)^k\right)\otimes \left(\sum_{k=0}^{+\infty} \frac{1}{k!} D^{\otimes 2k} \left(\frac{\alpha_s^2(\mu_0^2)-\alpha_s^2(\mu_1^2)}{8\pi^2}\right)^k\right)\,, \label{eq:vbihajksn}
\end{equation}
or noting $d_n$ the Mellin moments of $D$:
\begin{equation}
\sigma_n(z_0^2,z_1^2) = \left(\frac{\alpha_s(\mu_0^2)}{\alpha_s(\mu_1^2)}\right)^{\gamma_n/(2\pi\beta_0)}\exp\left(\displaystyle d_n \frac{\alpha_s(\mu_0^2)-\alpha_s(\mu_1^2)}{2\pi}+d_n^2\frac{\alpha_s^2(\mu_0^2)-\alpha_s^2(\mu_1^2)}{8\pi^2}\right)\,.
\end{equation}
By expanding Eq.~\eqref{eq:vbihajksn} to order ${\cal O}(\alpha_s)$, we obtain:
\begin{equation}
\Sigma(z_0^2,z_1^2) = {\cal E}_{LL}(\mu_0^2, \mu_1^2) \otimes \left(1+\frac{\alpha_s(\mu_0^2)-\alpha_s(\mu_1^2)}{2\pi} D + {\cal O}(\alpha_s^2)\right)\,,
\end{equation}
which is exactly the form of Eq.~\eqref{eq:eqhiojwo}: one finds naturally that the fixed-order truncation of the matching gives the same result as the renormalization group improved one up to higher order corrections. However, the dominant term at large $n$ of the Mellin moments of the step-scaling function is now  (see Eq.~\eqref{eq:bhevwcnjka} for more details):
\begin{align}
    &e_n(\mu_0^2, \mu_1^2)\exp\left(\displaystyle d_n \frac{\alpha_s(\mu_0^2)-\alpha_s(\mu_1^2)}{2\pi}+d_n^2\frac{\alpha_s^2(\mu_0^2)-\alpha_s^2(\mu_1^2)}{8\pi^2}\right)\nonumber \\
    &\hspace{10pt}\  \mathrel{\overset{n\rightarrow+\infty}{\scalebox{1.5}[1]{$=$}}}\  e_n(\mu_0^2, \mu_1^2) \exp\left(C_F^2 \ln^4(n)\frac{\alpha_s^2(\mu_0^2)-\alpha_s^2(\mu_1^2)}{2\pi^2} + {\cal O}( \ln^3(n))\right)\,. \label{eq:rfdbhkecdj}
\end{align}
Here $e_n$ tends to zero only as $\exp(K\ln(n))$, with $K < 0$ when $n \rightarrow +\infty$ (see Eqs.~\eqref{eq:frbjknecfd} and \eqref{eq:refrihbjkenc}), which is not fast enough to counter the divergence of the last factor in Eq.~\eqref{eq:rfdbhkecdj}.  Therefore, the SDF step-scaling function derived through the splitting function has lost the appealing characteristic of being an ordinary function and the renormalization group improvement applied to the matching part enhances the distribution-like behavior (as in the case of the backward evolution in $\overline{MS}$). Therefore, we will stick in the rest of this work with the form expressed in Eq.~\eqref{eq:eqhiojwo}. 

Let us draw a short conclusion from this discussion. Dealing with several perturbative objects, namely the evolution and the matching terms either at the level of the step-scaling or splitting functions, the possibilities of perturbative truncation at a given order are multiple, and give different results due to the handling of higher order terms. We have confronted several possible truncations, and selected the one we found to have the most desirable properties. However, it must be clear that all methods are essentially valid, the difference between them vanishing in the limit $\alpha_s \rightarrow 0$, or when all orders are taken into account. We will show in the next section, devoted to a numerical application, that we are in neither scenario. The uncertainty introduced by the perturbative matching will motivate us to advocate for a new technique later in the document.

\subsection{A perturbative numerical application}

For this preliminary numerical exploration of PDF evolution in the SDF scheme, we will revisit the results of the isovector proton PDF published in \cite{Egerer:2021ymv}, obtained with a lattice spacing of $a = 0.094$ fm. The space-like separation $z$ is an integer multiple of the lattice spacing $a$. It is customary to exclude non-local operators separated by only one lattice spacing, $z = a$, for which discretization errors are expected to be large. The smallest distances we will consider are therefore $z_1 = 2a = 0.188$ fm and $z_0 = 3a = 0.282$ fm. To give an account of the uncertainty in the scale fixing procedure, we will consider scales given by $\mu^2 = (\lambda z^2)^{-1}$ with $\lambda \in [-2, -0.5]$\footnote{We have derived the formalism so far with $\lambda = -e^{2\gamma_E+1}/4$ which guarantees the simple form of the matching in Eq.~\eqref{eq:bvekjcnsds}. For arbitrary values of $\lambda$, $D$ should be replaced by $D + \ln\left(-\frac{e^{2\gamma_E+1}}{4\lambda}\right)B_1$ (see Appendix \ref{canonical_scale}).}, which yields:
\begin{equation}\mu_1 \in [0.74, 1.48] \textrm{ GeV\ \ \ and }\ \ \mu_0 \in [0.49, 0.99] \textrm{ GeV}\,.
\label{eq:frbeolnk}\end{equation}

\begin{figure}[h]
    \centering
    \includegraphics[scale=.7]{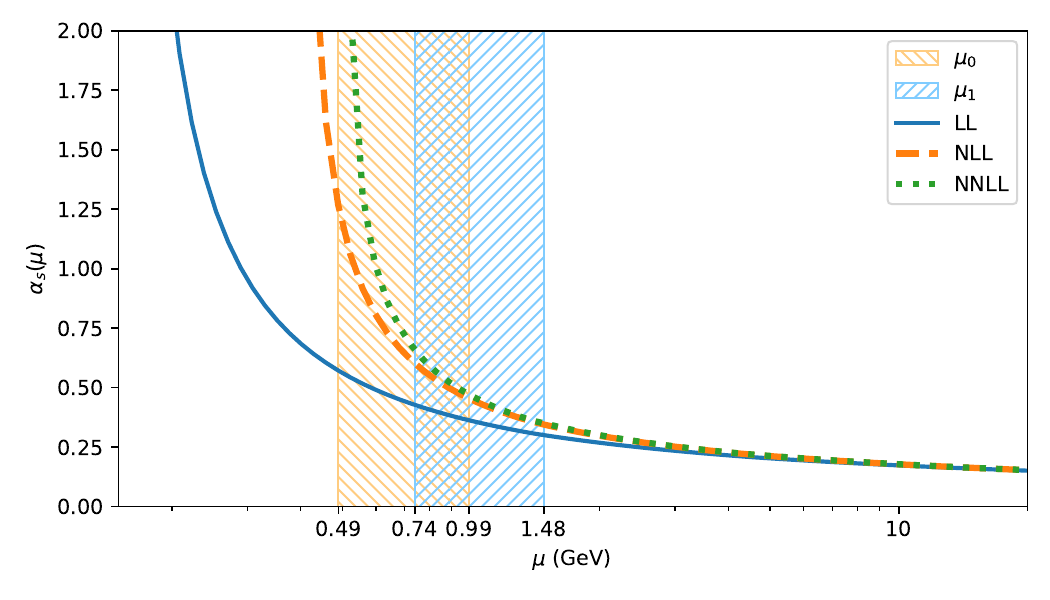}
    \caption{Running of the strong coupling as a function of the scale at LL, NLL and NNLL. }
    \label{fig4}
\end{figure}

The lower range of these $\overline{MS}$ scales is sufficiently low that instabilities in the perturbation theory could be expected. Indeed, using the PDG world average of $\alpha_s(M_Z = 91.19\ \textrm{GeV}) = 0.118$ \cite{PDG2023} in a variable flavor number scheme with threshold crossings at pole masses of the charm $m_c = 1.4$ GeV and bottom $m_b = 4.5$ GeV, the running of the strong coupling computed with the APFEL++ evolution code \cite{Bertone:2013vaa, Bertone:2017gds} is shown on Figure \ref{fig4}. This plot demonstrates that going to higher order in perturbation theory is not a tool to access lower scales, as the divergence of $\alpha_s$ appears earlier. That the renormalization group improvement makes SDF matching diverge as $z$ increases and gives a very different picture compared to the fixed-order truncation is well established, see \textit{e.g.}~Ref.~\cite{Ji:2022ezo} or \cite{Bhattacharya:2023ays}, where the renormalization group improvement can only be applied for $z \leq 3a$ for $a = 0.093$ fm. On the other hand, when the perturbative expansion converges (say above 1 GeV), going to higher order reduces the systematic uncertainty \cite{Li:2020xml}.

\begin{figure}[h]
    \centering
    \includegraphics[scale=0.7]{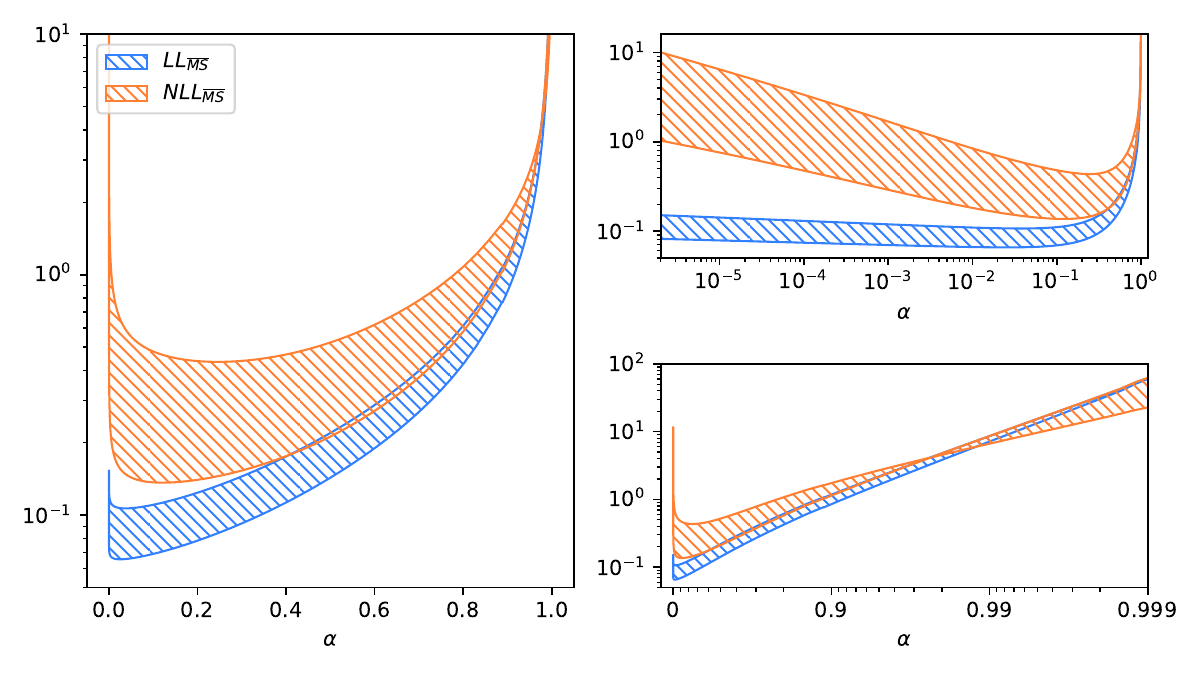}
    \caption{LL (hatched blue) and NLL (hatched orange) non-singlet $\overline{MS}$ step-scaling functions. The lower edge of the band around $\alpha = 0.5$ represents evolution with $\mu^2 = -2/z^2$, that is from $\mu_0 = 0.99$ GeV to $\mu_1 = 1.48$ GeV. The upper edge at $\alpha = 0.5$ represents evolution with $\mu^2 = -0.5/z^2$ (so stronger evolution effects as the scale is lower) from $\mu_0 = 0.49$ GeV to $\mu_1 = 0.74$ GeV. A zoom on the behavior at small $\alpha$ (upper right) and large $\alpha$ (lower right) is displayed.}
    \label{fig:fig5}
\end{figure}

The situation is critical and unreliable at the low scale $\mu_0$ at NNLL. We will therefore limit ourselves to a comparison of the LL and NLL results. We extract from APFEL++ the LL and NLL step-scaling functions in the $\overline{MS}$ scheme, between $\mu_0$ and $\mu_1$ in Figure \ref{fig:fig5}. The edges of the bands for perturbative quantities represent the results obtained for $\lambda = -0.5$ and $\lambda = -2$. At LL, there is only one non-singlet splitting function. In contrast, at NLL there are two components to the non-singlet splitting function, which are responsible for the independent evolution of the respective flavor asymmetries:
\begin{equation}
    (u - \bar{u}) - (d - \bar{d}) \textrm{\ \ and \ \ }(u + \bar{u}) - (d + \bar{d})\,.
\end{equation}
The first asymmetry corresponds to the real part of the isovector matrix element, whereas the second corresponds to the imaginary part. In Fig.~\ref{fig:fig5} we have plotted the NLL evolution operator of the real part of the isovector matrix element. The evolution operator of the imaginary part is essentially the same at this order in perturbation theory.

A few comments on Figure \ref{fig:fig5} are in order:
\begin{enumerate}
    \item We have demonstrated in the previous section that the (forward) LL $\overline{MS}$ step-scaling function is an ordinary positive function. We observe that the numerical extraction of the step-scaling function also converges towards an ordinary positive function at NLL, allowing us to present the results as simple curves in the figure.
    \item Both the LL and NLL $\overline{MS}$ step-scaling functions diverge at small and large $\alpha$. The NLL step-scaling function has a stronger divergence at small $\alpha$ compared to the LL, and inversely at large $\alpha$. This corresponds to the fact that the NLL step-scaling function deviates more than the LL one from the identity $1_\otimes = \delta(1-\alpha)$: NLL radiates more small momentum fraction partons than LL. At small $\alpha$, the difference is of several orders of magnitude, a testament to the poor perturbative convergence achieved at those scales. The scale uncertainty of the NLL curves is larger than that of the LL as a consequence of the stronger divergence of $\alpha_s$.
    \item We have already derived, in Eq.~\eqref{eq:bbjekvdcds}, the behavior of the LL $\overline{MS}$ step-scaling function at large $\alpha$ through an analysis of the large $n$ behavior of the anomalous dimensions $\gamma_n$. Applying the formula to the case $\lambda = -1$ (\textit{i.e.} $\mu_0 = 0.70$ GeV and $\mu_1 = 1.05$ GeV) yields:
    \begin{equation}
        {\cal E}_{LL}(\alpha, \mu_0^2, \mu_1^2) \  \widesim[2]{\alpha \rightarrow 1}\ 0.148(-\ln(\alpha))^{-0.865}\,.
    \end{equation}
    We can proceed in a similar way at small $\alpha$ by studying the Mellin moments in the limit where $n \rightarrow 0$. We find for the anomalous dimensions:
\begin{equation}
    \gamma_n \  \mathrel{\overset{n\rightarrow0}{\scalebox{1.5}[1]{$=$}}}\  C_F\left(\frac{1}{n} + \frac{1}{2}\right) + {\cal O}(n)\,.
\end{equation}
Hence
\begin{equation}
    e_n(\mu_0^2, \mu_1^2) = \left(\frac{\alpha_s(\mu_0^2)}{\alpha_s(\mu_1^2)}\right)^{\gamma_n / (2\pi \beta_0)}   \  \widesim[2]{n \rightarrow 0} \   B\left(\frac{\alpha_s(\mu_0^2)}{\alpha_s(\mu_1^2)}\right)^{C_F / (2n\pi \beta_0)}\,, 
\end{equation}
where
\begin{equation}
    B = \left(\frac{\alpha_s(\mu_0^2)}{\alpha_s(\mu_1^2)}\right)^{C_F / (4\pi\beta_0)}\,.\label{eq:vhcisjnk}
\end{equation}
The inverse Mellin transform of this term is not as straightforward as the one we performed for the large $\alpha$ behavior. It is useful to introduce the confluent hypergeometric limit function $_0F_1(;2;x)$ defined by its series expansion:
\begin{equation}
    _0F_1(;2;x) = \sum_{n = 0}^\infty \frac{x^n}{n!(n+1)!}\,,
\end{equation}
and to notice that the Mellin transform of $_0F_1(;2;-a \ln(\alpha)) \Theta(1-\alpha)$ gives:
\begin{equation}
    \frac{e^{a/n}-1}{a}\,,
\end{equation}
where $\Theta$ is the Heaviside step function. In the limit of $n \rightarrow 0$, we may neglect the term $-1/a$, and obtain therefore that the small $\alpha$ behavior of the LL $\overline{MS}$ step-scaling function is dominated by:
\begin{equation}
    {\cal E}_{LL}(\alpha, \mu_0^2, \mu_1^2) \  \widesim[2]{\alpha \rightarrow 0}\ B\frac{C_F}{2\pi\beta_0}\ln\left(\frac{\alpha_s(\mu_0^2)}{\alpha_s(\mu_1^2)}\right) \ _0F_1\left(;2;-\frac{C_F}{2\pi\beta_0}\ln(\alpha)\ln\left(\frac{\alpha_s(\mu_0^2)}{\alpha_s(\mu_1^2)}\right)\right)\,. \label{eq:wesdjkn}
\end{equation}
For the values of $\mu_0$ and $\mu_1$ we consider here, we obtain:
\begin{equation}
    {\cal E}_{LL}(\alpha, \mu_0^2, \mu_1^2) \  \widesim[2]{\alpha \rightarrow 0}\ 0.0701\ _0F_1\left(;2;-0.0677\ln(\alpha)\right)\,.
\end{equation}
\end{enumerate}

\begin{figure}[h]
    \centering
    \includegraphics[scale=0.7]{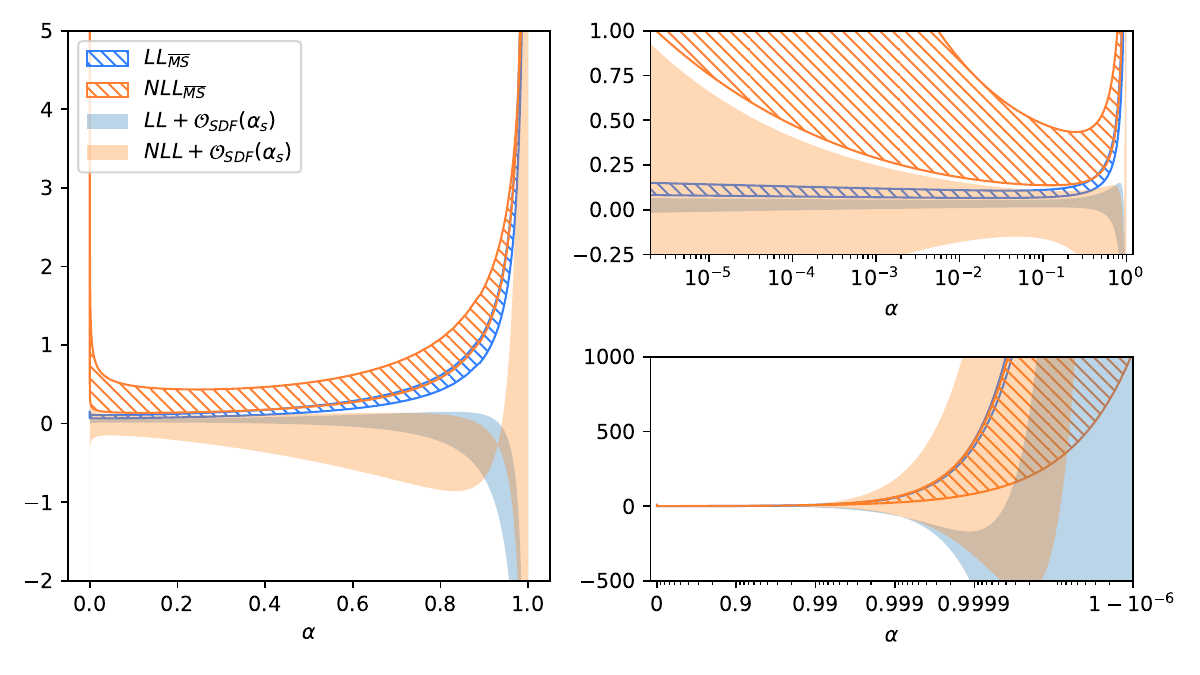}
    \caption{The same LL (hatched blue) and NLL (hatched orange) $\overline{MS}$ step-scaling functions as on Figure \ref{fig:fig5}. The uniformly colored bands are the ${\cal O}(\alpha_s)$ matching to the SDF scheme of both the LL and NLL step-scaling functions. The edges of the band are made of the result for $\mu^2 = -0.5/z^2$ and $\mu^2 = -2/z^2$.}
    \label{fig6}
\end{figure}

In Figure \ref{fig6} we represent the result of the matching at order ${\cal O}(\alpha_s)$ of both the LL and NLL $\overline{MS}$ step-scaling functions following Eq.~\eqref{eq:eqhiojwo}. A first consequence of the matching is a significant reduction of the contribution for $\alpha < 0.9$, to the point that the matched step-scaling becomes compatible with zero or slightly negative in this region. That evolution in the SDF scheme might not populate -- or at least considerably less than the $\overline{MS}$ evolution -- the small momentum fraction domain is an interesting finding. In particular, the reasoning developed in \cite{Dutrieux:2023qnz}, which used the fact that evolution ended up dominating the small $x$ behavior of the parton distributions to produce a perturbative modelling of this region, may be inapplicable in the SDF scheme.

More strikingly, for very large values of $\alpha$, the matched functions can dip to extremely negative values. The smaller the value of $|\lambda|$ (so the higher the perturbative scale), the stronger the negative dip. LL also exhibits a stronger dip than NLL. Only for values of $\alpha$ typically larger than $1-10^{-9}$ is the matched LL step-scaling function in line with the predicted asymptotic behavior -- namely $\ln^2(1-\alpha) C_F (\alpha_s(\mu_0^2)-\alpha_s(\mu_1^2)) / \pi$ times the LL $\overline{MS}$ large $\alpha$ behavior \eqref{eq:brvbdkjcn}.

To understand this very stark oscillation at very large $\alpha$, recall that the $\overline{MS}$ step-scaling functions are positive functions whose weight is mostly gathered around $\alpha = 1$. As such, one could consider them akin to a smeared Dirac delta function. The higher the perturbative scale, the less evolution effects are felt and the closer the $\overline{MS}$ step-scaling is to a Dirac delta function. We have noticed already that LL step-scaling functions are also closer to a Dirac delta than the NLL ones. The SDF step-scaling function \eqref{eq:eqhiojwo} is the convolution of this smeared Dirac delta with a term which contains a plus-prescription, the distribution $D$. A plus-prescription distribution convoluted with a smeared Dirac delta exhibits large oscillations at large $\alpha$, which become bigger as the smeared Dirac delta approaches an actual Dirac delta. In that limit, the convolution would become equal to the plus-prescription itself, which cannot be represented as an ordinary function since it would exhibit oscillations of infinite amplitude localized at $\alpha = 1$.

However spectacular this strong oscillation is, it is  barely relevant to our study for two main reasons. The first is that it is likely an artifact of the perturbative truncation. A fixed-order truncation of the $\overline{MS}$ step-scaling function presents a ``naked'' plus-prescription -- an infinitely narrow oscillation of infinite amplitude. One needs the resummation of the full leading logarithmic expansion to ensure the $\overline{MS}$ step-scaling function is an ordinary well-behaved function. Although we use renormalization group improvement for the $\overline{MS}$ step-scaling function, we used a fixed-order truncation of the non-logarithmic matching part (see \textit{e.g.} Eq.~\eqref{eq:eqhiojwo} and the discussion below). Renormalization group improvement of the non-logarithmic part through the SDF splitting function did not actually prove beneficial (see Eq.~\eqref{eq:rfdbhkecdj} and discussion). The discussion of the last section demonstrates that oscillating features at large $\alpha$ are very dependent on the handling of missing higher orders. Therefore, one could legitimately worry that there is little physical insight to draw from the pointwise value of the evolution operator, at least at large $\alpha$. This is indeed the case, but in practice, this is unlikely to cause a problem: what matters is only what can be measured, as we explain below. A resummation of threshold logarithms (see \textit{e.g.} \cite{Catani:1996yz, Gao:2021hxl}) could be envisioned, but the data we are working with is in fact not sensitive to these kind of effetcs.

\begin{figure}[h]
    \centering
    \includegraphics[scale=0.7]{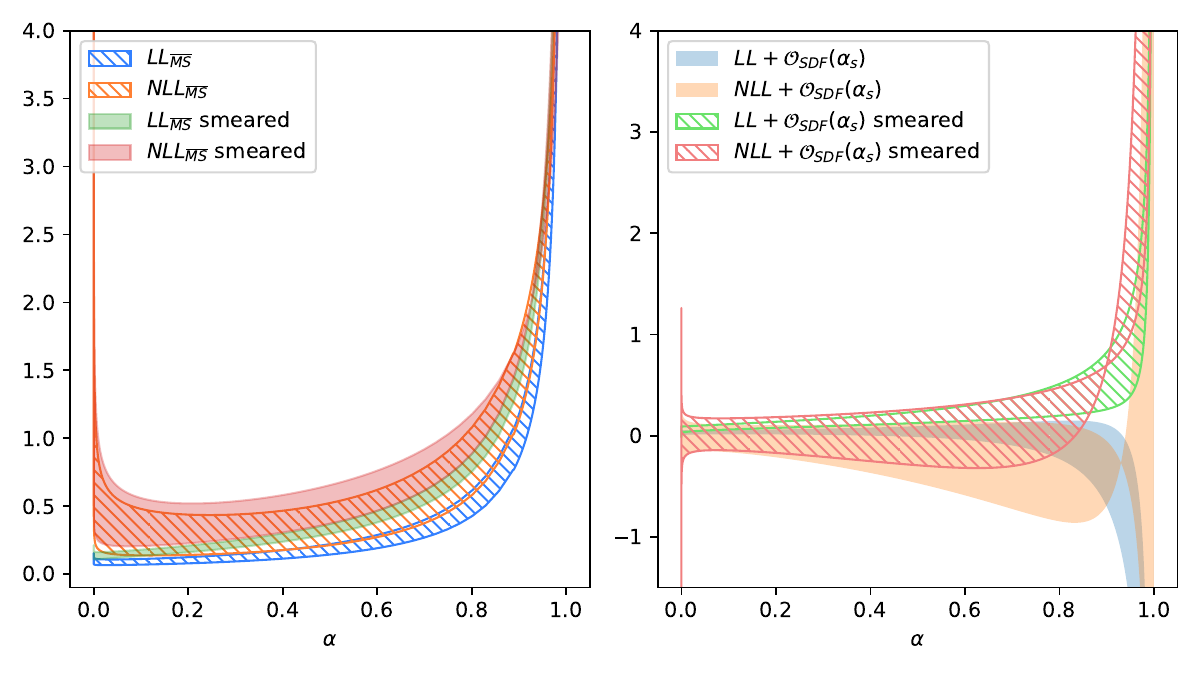}
    \caption{On the left, the effect of smearing the $\overline{MS}$ step-scaling functions with the kernel $S(\alpha)$ \eqref{eq:frfbdjknes}. There is little influence on the overall shape of the function. On the right, the smearing of the matched step-scaling functions. This time, the effect is dramatic, entirely erasing the sharp negative fluctuation at very large $\alpha$.}
    \label{fig7}
\end{figure}

More fundamentally, the second reason why these oscillations are mostly not relevant is that the step-scaling function is only ever accessed when convoluted with a PDF / ITD / matrix element, which are smooth functions. If the ITD only exhibits sizeable variations on a scale $\Delta \nu$, then the information content in the ITD over the interval $[0, \nu_{max}]$ can be roughly summarized by $\nu_{max} / \Delta \nu$ points. If one views the extraction of the step-scaling function as a linear system where we have about $\nu_{max} / \Delta \nu$ good measurements, then we can also only characterize the evolution operator on a similar number of points, so at a resolution level of $\Delta \nu / \nu_{max}$. 

This argument is of course only qualitative. One could argue that by measuring arbitrarily well the ITD on any interval in Ioffe-time, one could extract equally well the step-scaling regardless of the value of $\nu_{max}$ or the level of smoothness $\Delta \nu$. However, in practice, $\nu_{max} / \Delta \nu$ represents an empirical threshold between the resolution which can be reasonably reached, and the one which becomes computationally highly challenging.

In fact, current lattice data probably do not reach the threshold of $\Delta \nu / \nu_{max}$. If $\nu = z_3 \cdot P_z$, the closest $\Delta \nu$ we can probe at fixed value of $z^2$ is given by a variation of one lattice unit of the hadron momentum. Then the lattice proxy of $\Delta \nu / \nu_{max}$ is given by the inverse of the number of available hadron momenta, which is typically of the order of $\sim 0.1$. Some of the matched step-scaling functions in Figure \ref{fig6} show a strong negative dip up to $\alpha \sim 0.9999$, followed by an extremely sharp trend reversal. We would need a resolution of the order $10^{-4}$ to be effectively sensitive to this kind of feature of the step-scaling function.  Therefore, the data we collect do not allow us to access the step-scaling function in the full complexity of its $\alpha$-dependence, but rather its convolution with some unknown smearing kernel of width $\Delta \alpha \sim 0.1$. As an example, we depict on Figure \ref{fig7} the effect of convoluting all the step-scaling functions with an additional smearing kernel: 
\begin{equation}
    S(\alpha) = \frac{(-\ln(\alpha))^{-0.9}}{\Gamma(0.1)}\,.\label{eq:frfbdjknes}
\end{equation}

The smearing has the general effect of taming the behavior at large $\alpha$, and displacing some weight of the step-scaling function towards lower values of $\alpha$. The general shape of the $\overline{MS}$ step-scaling functions is only weakly altered by the smearing. On the other hand, the smearing has entirely suppressed the sharp negative oscillation of the SDF-matched functions at large $\alpha$. Therefore, we do not expect that these oscillations will play any role in evolving realistic data points.
The displacement of the curves when convoluted with the smearing kernel represents a typical systematic uncertainty we will be facing in any extraction of the step-scaling function due to the fact that we do not know precisely what the smearing kernel is aside from an estimate of its resolution.

\section{Fitting a non-perturbative step-scaling function to lattice data}\label{sec:fit}

In principle, the lattice matrix elements already published should give us a first hint of non-perturbative evolution in the SDF scheme, which we can compare to the perturbative application we have performed in the previous section. The leading order DGLAP evolution was first used in~\cite{Orginos:2017kos}, where numerical evidence suggested that the quenched lattice QCD results for $\mathfrak{M}(\nu,z^2)$ indeed follow this evolution in $z^2$. More generally, a wealth of numerical analyses have shown that for $z < 0.2 \sim 0.3$ fm, various truncations of the perturbative matching give rather similar results, mostly compatible with the phenomenological knowledge of unpolarized PDFs \cite{Zhang:2018nsy,Lin:2018pvv,Alexandrou:2018pbm,Joo:2019jct,Joo:2019bzr,Lin:2020ssv,Joo:2020spy,DelDebbio:2020rgv,Gao:2020ito,Alexandrou:2020zbe,Gao:2022uhg,Karpie:2021pap,Egerer:2021ymv,Bhat:2022zrw,HadStruc:2021qdf,HadStruc:2021wmh,HadStruc:2022yaw,HadStruc:2022nay,Delmar:2023agv,Karpie:2023nyg}.

\begin{figure}[h]
    \centering
    \includegraphics[scale=0.7]{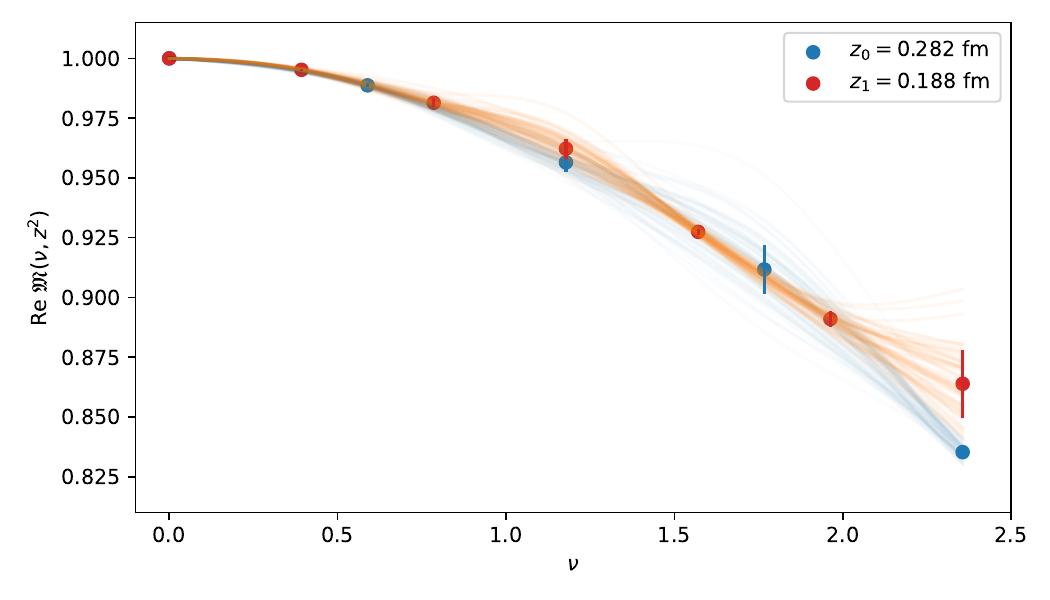}
    \caption{The real part of the isovector proton matrix element ${\frak M}(\nu, z^2)$ published in \cite{Egerer:2021ymv}. The light curves represent 60 cubic splines interpolations of jack-knife samples of the matrix elements.}
    \label{fig8}
\end{figure}

Data published in \cite{Egerer:2021ymv} with 2+1 flavors of Wilson clover quarks with a lattice spacing of $a = 0.094$ fm, volume of $32^3 \times 64$, pion mass of 358 MeV and 349 gauge configurations are shown on Figure \ref{fig8}. We present the two values $z_0 = 3a$ and $z_1 = 2a$, and the different Ioffe times are obtained by varying the momentum $P$ of the hadron. The small value of $z$ should guarantee a reduction of higher-twist effects and offers scales where we can compare to the perturbative results at LL and NLL. 

As the hadron momentum increases, the signal-to-noise ratio decreases. To counter this phenomenon, for $P \geq 4 \times \frac{2\pi}{L}$, the hadron interpolating operators are ``phased'' \cite{Bali:2016lva,Egerer:2020hnc}: their spatial extension is modulated by an oscillation which increases their coupling to the ground state of the boosted hadron. The effect is striking on Figure \ref{fig8}. The fifth red point, where the phasing kicks in, has a significantly smaller uncertainty than the fourth point as witnessed by the tightening of the red band. The same behavior is also visible for the fifth blue point (the rightmost one -- the first blue point is hidden at $\nu = 0$ by the first red one) compared to the fourth blue point. This gives however the unfortunate consequence that the distribution at the initial scale is considerably less constrained than at the final scale for $\nu \in [1.2, 2]$, and the opposite is true for $\nu \geq 2$, to some extent obfuscating the $z^2$ dependence of the matrix elements. 

\subsection{Methodology}\label{sec:fitting_method}

Our strategy to extract an empirical step-scaling function from the lattice data is first to propose a parametric form of $\Sigma(\alpha;z_0^2, z_1^2)$, and then to fit it to the lattice data:
\begin{equation}
    \frak{M}(\nu, z_1^2) = \int_0^1 d\alpha \,\Sigma(\alpha;z_0^2, z_1^2) \frak{M}(\alpha\nu; z_0^2)\,. \label{eq:evdojnlkme}
\end{equation}
As discussed in section III.B, the existence of $\Sigma$ can be inferred at small $z^2$ from the validity of the OPE and all order perturbative results. Let us note that if there are significant higher twist contributions in the lattice data, they will be encompassed in some way in the extracted $\Sigma$, whose existence or support we might not be able to guarantee. This limits us to only studying the small $z^2$ domain.

We only have at our disposal a discrete set of Ioffe-time values $(\nu_{i,0})_i$ at the initial scale $z_0^2$ to compute the integral of Eq.~\eqref{eq:evdojnlkme}. Therefore, we perform a cubic spline interpolation of the initial-scale matrix element samples. Denoting by $(\nu_{j,1})_j$ the set of Ioffe-time values at the final scale $z_1^2$, we can write schematically the fit as the optimization of a function $T_\Sigma$ of the initial-scale matrix elements: 
\begin{equation}
    \frak{M}(\nu_{j,1}, z_1^2) = T_\Sigma \bigg(\nu_{j,1}, z_1^2; \left(\frak{M}(\nu_{i,0}; z_0^2)\right)_i\bigg)\,,\label{eq:beisjknwc}
\end{equation}
where $T_\Sigma$ contains both the parametric dependence of the step-scaling function $\Sigma$, and the effect of the cubic spline  interpolation on the discrete set of matrix elements at initial scale $\left(\frak{M}(\nu_{i,0}; z_0^2)\right)_i$.

The presence of largely correlated uncertainties within the jack-knife samples of the matrix elements must be accounted for in the fit. At a fixed value of $P$, the points at $z_0^2$  and $z_1^2$ are correlated in excess of 97\%. On the other hand, at a fixed value of $z^2$, neighboring points in Ioffe time (so in $P$) are typically correlated at the level of 50\%. It drops to about 20\% with one more separation in Ioffe time, and to less than 10\% with one more.

Eq.~\eqref{eq:beisjknwc} represents a typical fit of the form $y(t) = f(t, x)$ where we want to learn the parameters of $f$, but there are uncertainties in the value of $x$ which are furthermore correlated with the uncertainties of $y$. For such an errors-in-variables model, we will use the total least square method. We assume that there exist true values $\frak{M}^*(\nu, z^2)$ of the matrix elements, and that the jack-knife samples arise from a measurement of $\frak{M}^*(\nu, z^2)$ with \textit{classical} errors. This means that although there is correlation between the measurement of $\frak{M}(\nu, z^2)$ at various kinematics, there is no correlation assumed between the uncertainty of the measurement and the true value $\frak{M}^*(\nu, z^2)$. We form a vector of fit residuals defined by:
\begin{equation}
    r = \begin{pmatrix}
        \bigg(\frak{M}^*(\nu_{i,0}, z_0^2) - \frak{M}(\nu_{i,0}, z_0^2)\bigg)_i\\
        \bigg(T_\Sigma \bigg(\nu_{j,1}, z_1^2; \left(\frak{M}^*(\nu_{i,0}; z_0^2)\right)_i\bigg) - \frak{M}(\nu_{j,1}, z_1^2)\bigg)_j
    \end{pmatrix}\,.
\end{equation}
The vector of residuals contains both the discrepancy of the true values to the measured ones, and quantifies the ability of the model to reproduce the data. Then introducing the empirical covariance matrix evaluated from the full set of jack-knife samples:
\begin{equation}
    \Omega = \begin{pmatrix} \textrm{Cov}[\frak{M}(\nu_{i,0}, z_0^2), \frak{M}(\nu_{j,0}, z_0^2)]_{i,j} & \textrm{Cov}[\frak{M}(\nu_{i,0}, z_0^2), \frak{M}(\nu_{j,1}, z_1^2)]_{i,j}  \\  \textrm{Cov}[\frak{M}(\nu_{i,0}, z_0^2), \frak{M}(\nu_{j,1}, z_1^2)]_{i,j} & \textrm{Cov}[\frak{M}(\nu_{i,1}, z_1^2), \frak{M}(\nu_{j,1}, z_1^2)]_{i,j}
    \end{pmatrix}\,,
\label{eq:fwcvshjbkca}\end{equation}
we form a goodness-of-fit measure:
\begin{equation}
    \chi^2 = r^T \Omega^{-1} r\,,
\label{eq:gvjkjk}\end{equation}
where $r^T$ denotes the transpose of $r$. To give an account of uncertainty, we study the distribution of parameters in the vicinity of the minimal $\chi^2$. This distribution reflects the aleatoric uncertainty of the data, but not the epistemic uncertainty introduced by the specific choice of the parametric form.

In practice we face an ill-defined inverse problem. The step-scaling function might be in full generality a distribution with an arbitrary dependence on $\alpha$. However, as we have discussed in the previous section, since we only have a restricted kinematic access to the PDFs which phenomenology tells us are smooth functions, we are only sensitive to the step-scaling function smeared on some resolution of the order of $\Delta \alpha \sim 0.1$. The fact that we can really only extract the smeared step-scaling function is actually a blessing to some extent, as demonstrated by our toy smearing on Figure \ref{fig7}: there, the smeared perturbative SDF step-scaling function exhibits a fairly generic shape similar to its $\overline{MS}$ counterparts, \textit{i.e.} an ordinary function with a simple divergence at $\alpha = 0$ and $\alpha = 1$ once the large oscillation at very large $\alpha$ is erased by smearing. At first, we hope to circumvent the ill-posedness by finding a generic parameterization of the smeared step-scaling function that we can fully constrain from the data, but it is clear that this object can only be used to evolve the PDF at the same level of resolution as the one used for its extraction. That we have done a good job at extracting the smeared step-scaling function within the uncertainties of the data will be identified by the fact that we obtain a good reproduction of the $z^2$ dependence of the data, that is a $\chi^2$ \eqref{eq:gvjkjk} per fitted points close to one (called $\chi^2$ per $n_{pts}$ in the following). We will be considering another, less model-dependent approach later on.

\begin{figure}[h]
    \centering
    \includegraphics[scale=0.7]{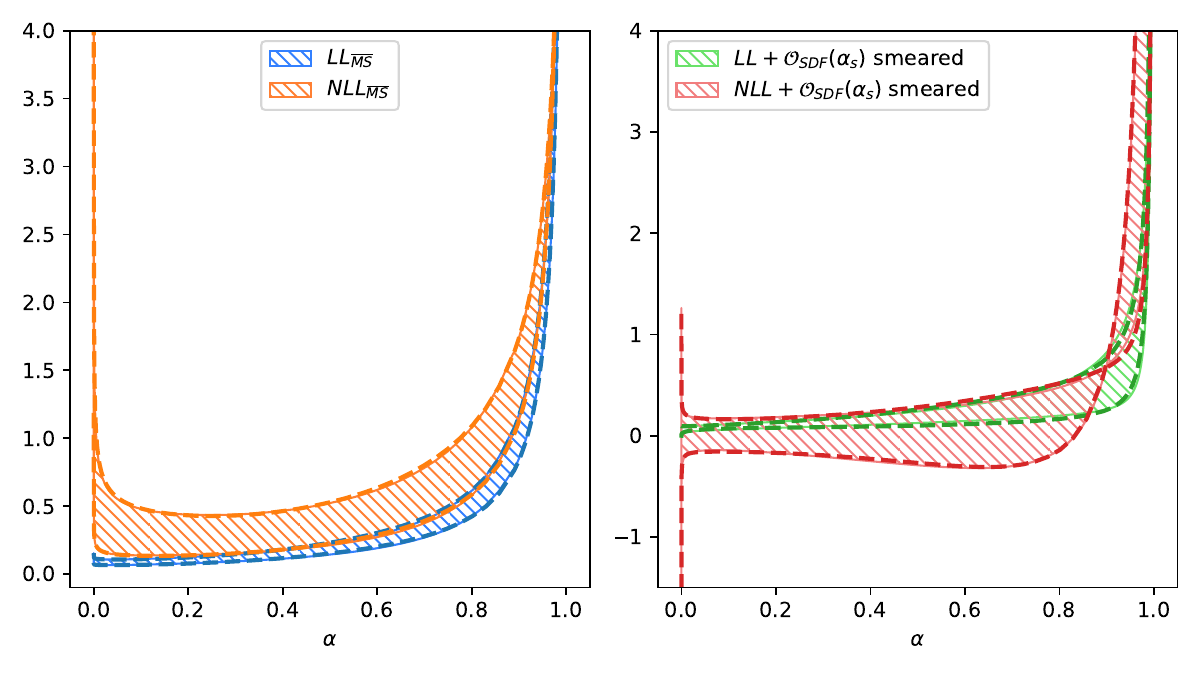}
    \caption{The bands are the same as in Figure \ref{fig7}. The dotted lines represent fits with the functional form of Eq.~\eqref{eq:efvbdibuska} on the edges of the bands. The agreement is very satisfactory. Since the asymptotic behaviors are not the same, going to extremely small or large values of $\alpha$ would show discrepancies, but those are not meaningful numerically due to the resolution smearing induced by the convolution.}
    \label{fig9}
\end{figure}

We propose the following parameterization of the step-scaling function:
\begin{equation}
\Sigma(\alpha; z_0^2, z_1^2) = A \alpha^{-\delta} (1+r\alpha) + b (-\ln(\alpha))^{-\eta} \ln^2(1-\alpha) + \sigma \alpha(1-\alpha)\,. \label{eq:efvbdibuska}
\end{equation}
There are six parameters: $A, \delta, r, b, \eta, \sigma$. The first term produces a divergence when $\alpha \rightarrow 0$, the second term when $\alpha \rightarrow 1$, and the third term only contributes to the intermediate $\alpha$ range. One will notice that the asymptotic behaviors of Eq.~\eqref{eq:efvbdibuska} when $\alpha \rightarrow 0$ and 1 do not correspond to those we have analytically computed for the LL $\overline{MS}$ case (see Eqs.~\eqref{eq:wesdjkn} and \eqref{eq:bbjekvdcds}). The large $\alpha$ behavior of the LL $\overline{MS}$ step-scaling function only diverges as $b (-\ln(\alpha))^{-\eta}$, and the small $\alpha$ behavior was determined to be of the general form $A\,_0F_1(;2;-\delta\ln(\alpha))$, which is less divergent than any term $A\alpha^{-\delta}$ where $\delta > 0$. But although Eq.~\eqref{eq:efvbdibuska} does not have the same asymptotical behavior than the true LL $\overline{MS}$ step-scaling function, it is able to produce a very satisfactory account on a compact interval $[\alpha_0, \alpha_1] \subset (0,1)$. We demonstrate in Figure \ref{fig9} that our parametric form is able to reproduce satisfactorily the various perturbative step-scalings we have derived before on a large range of $\alpha$ when we use the smearing of the extremely sharp variations of the SDF-matched step-scaling functions.

\subsection{Test of the methodology}

\begin{figure}[h]
    \centering
    \includegraphics[scale=0.7]{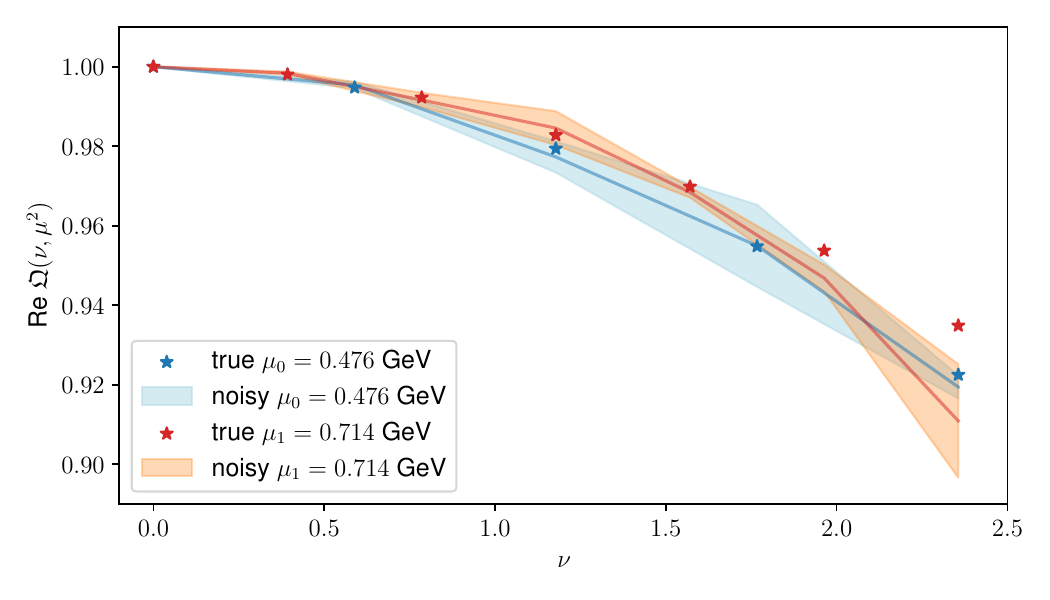}
    \caption{Real part of the Ioffe-time distribution associated to the PDF \eqref{eq:esbkjnklwcs}. The stars represent the true values of the Ioffe-time distribution, and the bands the result of their reshuffling with the covariance matrix of the actual lattice data shown of Figure \ref{fig8}.}
    \label{fig10}
\end{figure}

To exemplify our fitting methodology, we first work in a controlled setting with synthetic data. We start with a simple toy valence PDF at $\mu_0 = 0.49$ GeV, normalized so that its integral from 0 to 1 equals 1:
\begin{equation}
    q(x, \mu_0^2) = \frac{35}{32} x^{-0.5} (1-x)^3\,.\label{eq:esbkjnklwcs}
\end{equation}
We evolve this PDF to the scale $\mu_1 = 0.74$ GeV using the exact LL DGLAP evolution, and extract the values of the ITD  \eqref{eq:ahonlvz} on the kinematics shown by the datapoints on Figure \ref{fig8}. Then we fluctuate the true data according to the covariance matrix of the lattice data \eqref{eq:fwcvshjbkca}. The result is depicted in Figure \ref{fig10}. Although the true values of the ITD at final scale (red stars) are strictly above their values at initial scale (blue stars), the uncertainty makes it far less visible at the level of the noisy data. Nonetheless, how well can we  extract the LL $\overline{MS}$ step-scaling function from this noisy dataset?

Exploring the 10-parameter space (6 parameters for the step-scaling function and 4 parameters for the unknown true values of the ITD at initial scale) with a non-linear $\chi^2$ proves rather challenging. We start from 2000 random initial sets of widely distributed parameters and run independent stochastic gradient optimizations to find good fit candidates. In fact, this will provide a series of local minima, and hopefully the global one among them. We achieve a smallest $\chi^2$ per $n_{pts}$ of 0.82\footnote{Since the data at Ioffe time $\nu = 0$ is exactly 1 with no uncertainty, to practically perform a fit, we have to grant it a standard deviation -- or enforce analytically the reproduction of that point. We choose the precise uncertainty of $0.0001$, which ensures that the integral of the step-scaling function is equal to $1 \pm 0.0001$.}, which demonstrates an excellent ability to explain the data with the parametric form of the step-scaling function and within the uncertainty. 

\begin{figure}[h]
    \centering
    \includegraphics[scale=0.7]{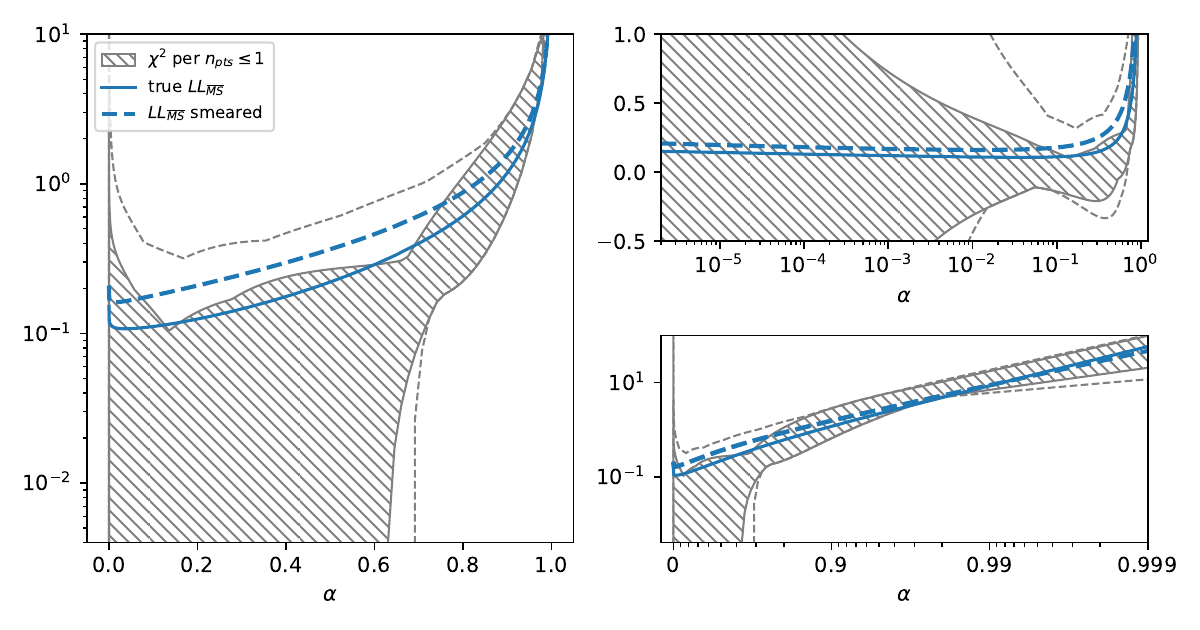}
    \caption{The hatched band results from an exploration of the fits whose $\chi^2$ per $n_{pts}$ is less than 1. The dotted grey lines represent the uncertainty with a $\chi^2$ per $n_{pts} \leq 3$. This extraction of the step-scaling function is in good agreement with the actual LL $\overline{MS}$ step-scaling function which has been used to produce the data, and also mostly compatible with its smeared version.}
    \label{fig12}
\end{figure}

To produce an account of uncertainty, we select first the 8 local minima with a $\chi^2$ per $n_{pts}$ below 1, and then the 32 ones with a $\chi^2$ per $n_{pts}$ below 3. We explore the vicinity of these local minima to produce as large a sampling as possible of the distribution of parameters which offers either a $\chi^2$ per $n_{pts}$ below 1 or 3\footnote{For this part of the study, we present uncertainties computed from a fixed threshold in $\chi^2$ per fitted points. Linear propagation of uncertainty invites to use one unit of the full $\chi^2$ (not divided by the number of points or degrees of freedom) around its (unique) minimum as the standard measure of uncertainty. However, in the highly non-linear case we are facing, we find this uncertainty quantification both overly optimistic and difficult to interpret. We prefer to use the intuitive absolute goodness of fit as a criterion. In the later results with a quadratic $\chi^2$, we will use the standard result of linear propagation of uncertainty.}. The result with a threshold of 1 is shown as the hashed band on Figure \ref{fig12}, whereas the result for a threshold of 3 is depicted as the dotted grey lines. In fact, exploring the $\chi^2$ in the vicinity of the local minima is rather useless. The plot would be exactly the same if we had only plotted the distribution of local minima: the $\chi^2$ varies so abruptly that a change of the parameters by less than 0.1\% is often enough to see the $\chi^2$ increase by several units. Therefore, the crucial part of the evaluation of the uncertainty is to attempt to determine the largest possible amount of local minima.

\begin{figure}[h]
    \centering
    \includegraphics[scale=0.65]{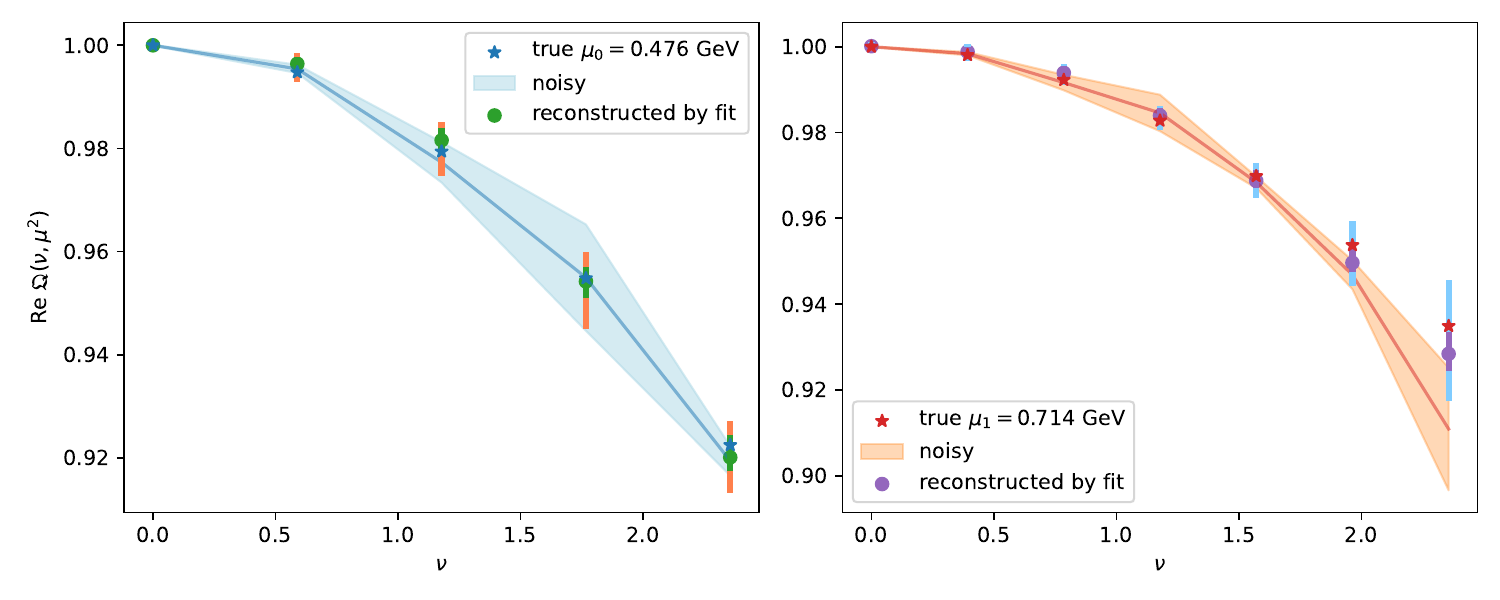}
    \caption{The stars and bands are the same as on Figure \ref{fig10}. The green points with error bars on the left represent the fitted result of the true ITD value with a $\chi^2$ per d.o.f. threshold of 1 at initial scale ${\cal Q}^*(\nu, \mu_0^2)$. The larger orange error bars stands for a threshold of 3. The purple points with purple and light blue error bars on the right represent the convolution of the fitted step-scaling function with the fitted true ITD values at initial scale $T_\Sigma(\nu, \mu_1^2, ({\cal Q}^*(\nu, \mu_0^2)))$.}
    \label{fig11}
\end{figure}

In spite of the significant noise in the data, the fit allows us to reconstruct the LL $\overline{MS}$ step-scaling function in good agreement with its true value. The small $\alpha$ domain is poorly constrained as one would expect considering the limited range in Ioffe time where the data are available. The smeared LL $\overline{MS}$ step-scaling function that we have introduced in Figure \ref{fig7} is also mostly compatible with the extraction, a fortiori if a threshold on $\chi^2$ per $n_{pts}$ of 3 is used. We expected this result as we have explained that we only have an access to the step-scaling function with a limited resolution.

In Figure \ref{fig11}, we show the ``true'' values of the ITD obtained thanks to the fit, with both the thresholds in $\chi^2$ per $n_{pts}$ of 1 and 3. The correction induced by the fit reduces significantly the uncertainty and discrepancy introduced by the noise. For instance, the last point at the higher scale $\mu_1$ is correctly displaced by more than $1\sigma$ compared to its noisy value, and its uncertainty is considerably reduced compared to the original data for a threshold of 1. The second to last point at initial scale sees its uncertainty reduced by a factor 3 compared to the original data. As we will see later on, it is the requirement of self-consistent evolution in $z^2$, encompassed in the form of the convolution of Eq.~\eqref{eq:evdojnlkme}, that is responsible for the correction of aberrations in the data, more than the parametric form chosen for $\Sigma$. These corrections would otherwise require significant computational time to obtain from better lattice calculations. This underlines how crucial a fine understanding of the evolution effect is to the effort of extraction of parton distributions from lattice QCD. 

\subsection{Extraction from actual lattice data with a parametric form}\label{sec:ss_fit_real_data}

\begin{figure}[h]
    \centering
    \includegraphics[scale=0.7]{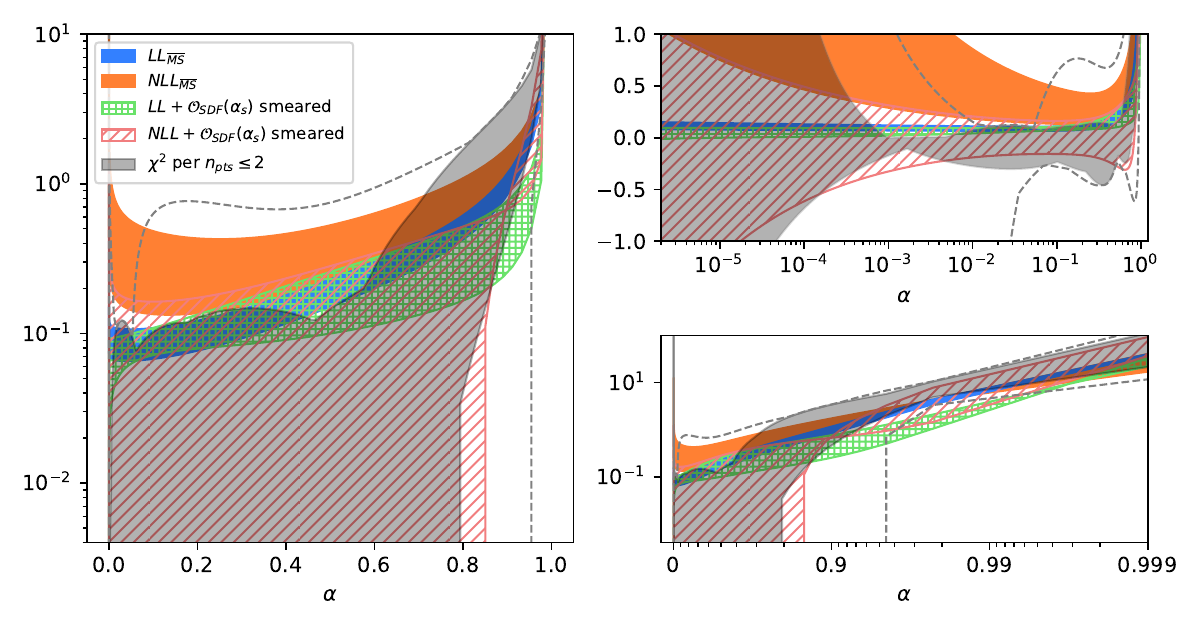}
    \caption{Empirical extraction of the SDF step-scaling function from the real part of the lattice data. The shadow band contains all fit results whose $\chi^2$ per $n_{pts}$ is less than 2, the dotted grey lines less than 3.}
    \label{fig13}
\end{figure}

\begin{figure}[h]
    \centering
    \includegraphics[scale=0.7]{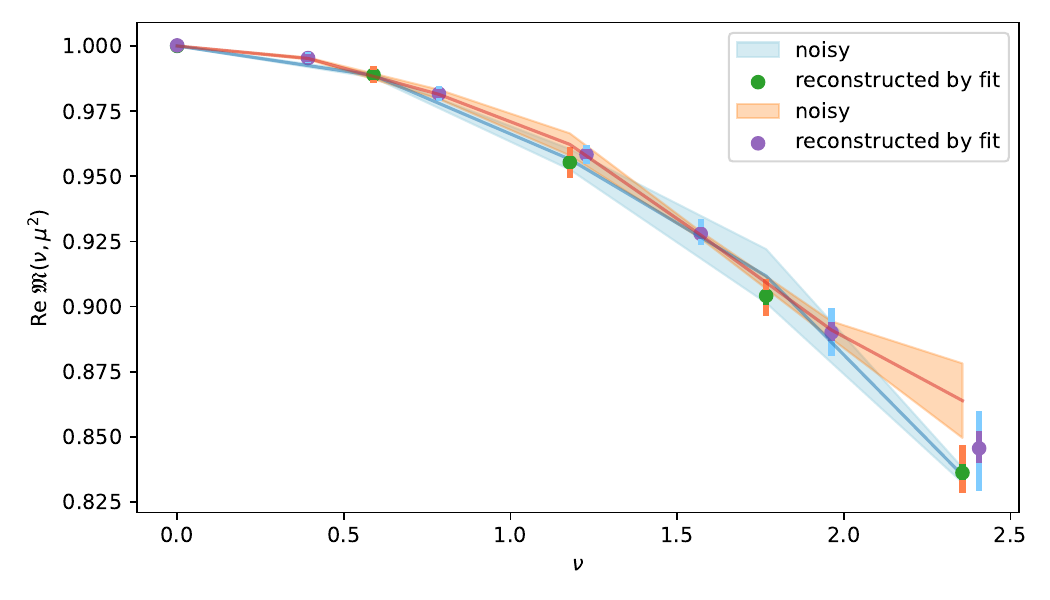}
    \caption{Real part of the isovector proton matrix element ${\frak M}(\nu, z^2)$ displaced by our fitting methodology. The two colored error bars stand for a tolerance parameter of 2 and 3. The fourth and seventh purple points have been slightly displaced to avoid collision with the green points. The fit correction produces data with a smaller evolution effect than the -- slightly erratic -- effect present in the original data.}
    \label{fig14}
\end{figure}

The success of the previous test motivates us to apply the same technique to the actual lattice data. The results are shown in Figures \ref{fig13} and \ref{fig14} for the real part. The smallest $\chi^2$ per $n_{pts}$ that we have found for the real part is 1.65, which shows a satisfactory level of explanation of the data. This time, we will therefore compare a threshold of 2 (9 local minima found) and 3 (32 local minima found).

Overall, the extracted SDF step-scaling function is highly compatible with most inputs from perturbation theory, except the NLL $\overline{MS}$ evolution if one uses the most stringent threshold in $\chi^2$. The SDF step-scaling function extracted directly from the lattice data shows compatibility with zero as soon as $\alpha < 0.8$ as the matched and smeared NLL calculation, and a very similar large $\alpha$ behavior to all perturbative derivations. One should remember that the uncertainty depicted for perturbative derivations is the theoretical uncertainty tied to scale fixing, whereas the uncertainty of the lattice SDF step-scaling function is tied to the uncertainty of the lattice data and the choice of parametric form. That both coincide so well for the matched and smeared NLL term is therefore rather coincidental. The ``true'' values of the matrix elements are shown in Figure \ref{fig14}. The matrix elements corrected by the requirement of self-consistent $z^2$ evolution show a much reduced effect of evolution compared to that typically produced by evolution in the $\overline{MS}$ scheme, in line with the conclusions we drew from the study of the perturbative matching. We observe that, as in our toy model, the last point at higher scale has been significantly moved, and the uncertainty of some data points largely reduced by our fitting procedure.  

\begin{figure}[h]
    \centering
    \includegraphics[scale=0.7]{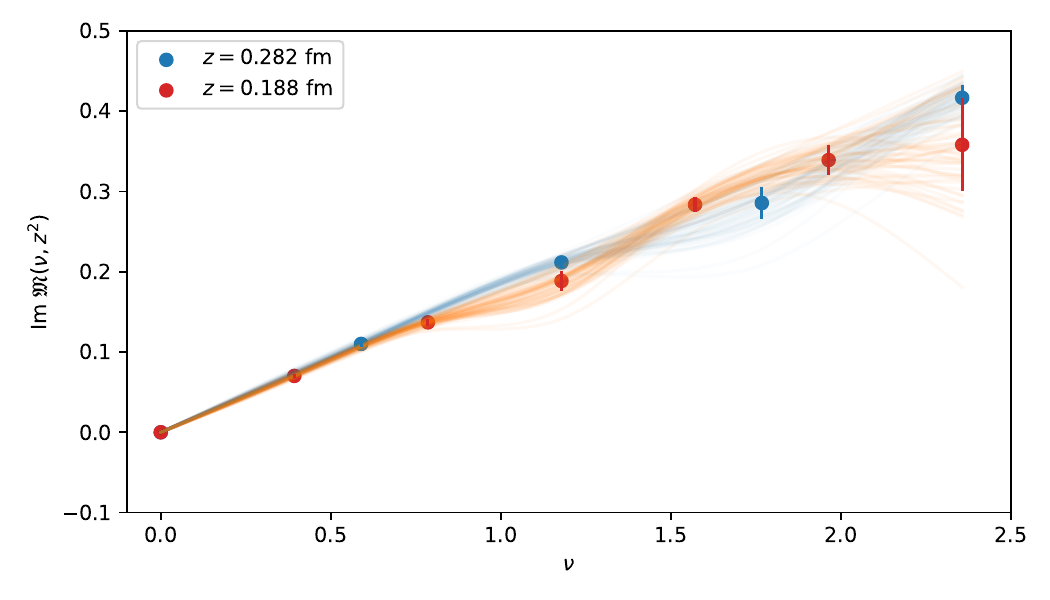}
    \caption{Imaginary part of the isovector proton matrix element ${\frak M}(\nu, z^2)$ published in \cite{Egerer:2021ymv}.}
    \label{fig15}
\end{figure}

We conduct a similar study for the imaginary part of the lattice data at the same scales. At $\nu = 0$, the imaginary part of the isovector data is zero, due to the $x$ symmetries of this non-singlet distribution. This means that we lose one constraint on the step-scaling function, namely that its integral over $\alpha$ should be equal to one. The data is depicted on Figure \ref{fig15}. The effect of $z^2$ evolution is here again fairly difficult to discern clearly, due to the noise in the data.

\begin{figure}[h]
    \centering
    \includegraphics[scale=0.7]{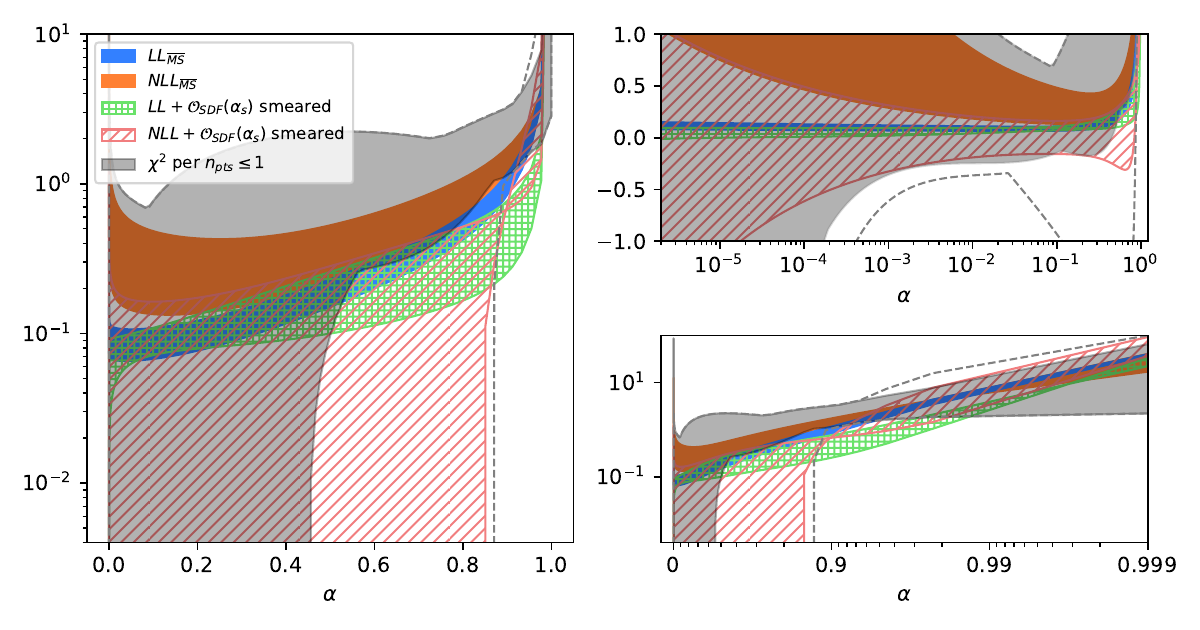}
    \caption{Empirical extraction of the SDF step-scaling function from the imaginary part of the lattice data.}
    \label{fig16}
\end{figure}

\begin{figure}[h]
    \centering
    \includegraphics[scale=0.7]{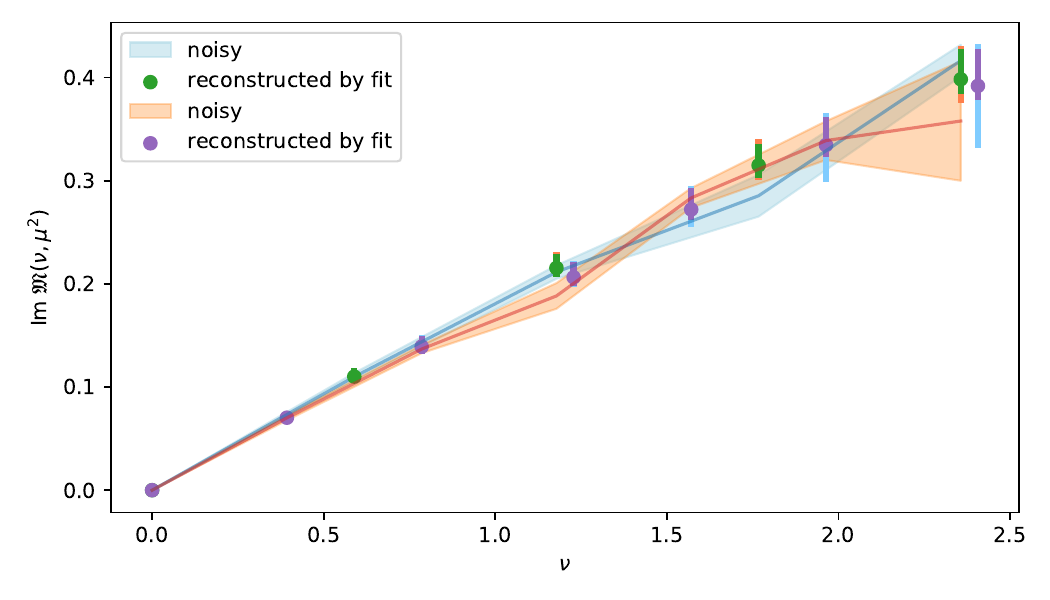}
    \caption{Imaginary part of the isovector proton matrix element ${\frak M}(\nu, z^2)$ displaced by our fitting methodology. The error bars are noticeably dissymmetric as the non-linear fitting procedure produces non-Gaussian uncertainty distributions. The dot corresponds to the mean of the distribution, whereas the error bar to all values below the threshold in $\chi^2$.}
    \label{fig17}
\end{figure}

The smallest $\chi^2$ per $n_{pts}$ we find is 0.54, that is significantly smaller that with both our previous studies of the real part (ideal LL $\overline{MS}$ case and actual lattice data). The data is easier to explain, and the $\chi^2$ not such an abrupt function as for the real case, to the point that all 22 local minima we find correspond to a $\chi^2$ per $n_{pts} \leq 1$. We will use a tolerance parameter of 1 and 3. The result is shown on figures \ref{fig16} and \ref{fig17}.

Overall, the imaginary data is less constrained by evolution, can be fit much more easily and shows larger uncertainties both at small and large $\alpha$. Less stark behavior at large $\alpha$ are more favored by the fit, but also much larger contributions at small $\alpha$. The results of the ``corrected'' lattice data on Figure \ref{fig17} show very little evolution effect. 

Notice that we have theoretically no expectation that the real and imaginary parts of the isovector matrix elements evolve in a similar way. However, there is no difference at LL in $\overline{MS}$, and the difference at NLL is so minute that it is not visible on the plot. Unfortunately, the larger uncertainties of the step-scaling function extracted from the imaginary part do not allow any conclusive statement on a difference with the evolution of the real part.

\subsection{Bayesian reconstruction of the step-scaling function}

As we have already mentioned, determining the SDF step-scaling function from lattice data is an ill-posed integral inverse problem, which tends to have significant modeling bias when studying fixed functional forms. There exists an alternative class of solutions to the parametric forms studied in the previous sections to this sort of inverse problem based upon Bayes's theorem, such as the Maximum Entropy Method~\cite{MEM} and Bayesian Reconstruction~\cite{BR}. These approaches are similar in theme, but vary in the specific choice of prior distribution given via Bayes's Theorem. In work on lattice calculation of PDFs, these approaches have been used to determine the PDF from lattice matrix elements~\cite{Karpie:2019eiq,Liang:2019frk}.

The goal of each of these approaches is to determine the most probable function whose integrals correspond to a given set of data and some prior information. The most probable function is given by the expectation value:
\begin{equation}
    \langle\Sigma(\alpha;z_0^2,z_1^2)\rangle = \int D\Big[\Sigma(\alpha;z_0^2,z_1^2)\Big] \,\Sigma(\alpha;z_0^2,z_1^2) \, P\Big[\Sigma(\alpha;z_0^2,z_1^2)| \mathfrak{M}(z_0^2), \mathfrak{M}(z_1^2), I\Big]\,,
\end{equation}
where the posterior distribution $P\Big[\Sigma(\alpha;z_0^2,z_1^2)| \mathfrak{M}(z_0^2), \mathfrak{M}(z_1^2), I\Big]$ is defined by Bayes's theorem as:
\begin{equation}
    P\Big[\Sigma(\alpha;z_0^2,z_1^2)| \mathfrak{M}(z_0^2), \mathfrak{M}(z_1^2), I\Big] \propto \exp\left[-\frac{\chi^2}2 + u S\big[\Sigma(\alpha;z_0^2,z_1^2)]\right]\,.
\end{equation}
The $\chi^2$ term represents the likelihood of the observed data knowing the value of the step-scaling function, and is defined in the same manner as in Eq.~\eqref{eq:gvjkjk} including the interpolation of $\mathfrak{M}^\ast(z_0^2)$ with a cubic spline. The $S\big[\Sigma\big]$ term represents the prior enforced on the step-scaling function, which distinguishes the various approaches of this type. The hyperparameter $u$ controls the relative importance of the goodness of fit and the prior. The hyperparameter can be varied to study its effect or can be integrated away. The values of the function $\Sigma$ on a discrete grid in $\alpha$ become the model parameters and the integral whose inverse is desired and those in $S$ are evaluated numerically. In this work, the trapezoid method is used on various grids. Due to the fact that no evolution corresponds to a step-scaling which is a Dirac delta at $\alpha = 1$, high resolution in the large $\alpha$ region is required to accurately reproduce $\Sigma$. The choice of this grid, the integration method, and the choice of $S$ will collectively define a model with some specific bias. By varying the choices, the model dependence can be studied. 

The goal of the prior information is to bias the result away from unphysical results. In the case of spectral functions, known to be positive definite, it is desirable to require the function to be positive. This feature was therefore baked into the Bayesian Reconstruction prior information in \cite{BR}. Without physical guidance many ``reasonable'' choices must be studied, though as always with Bayesian methods the choice of prior information can dramatically change the result. Hence what is ``reasonable'' may end up debatable.

In this work, we shall use the Quadratic Difference Ratio (QDR) which is defined by the prior function:
\begin{equation}
   \log\Big[ P\big[\Sigma | I\big] \Big] = u S_{\rm QDR}\big[\Sigma\big] = -u \int_0^1 d\alpha\, \frac{\big(\Sigma(\alpha)- h(\alpha)\big)^2}{2\sigma^2(\alpha)} \label{eq:vsinjqvwq}
\end{equation}
where $h$ is a prior model and $\sigma$ a prior weight function. This prior will attempt to push the step-scaling function towards the prior model, but through the prior weight it allows $\Sigma$ in different regions of $\alpha$ to differ from the prior model by a varying amount. If the hyperparameter $u$ is integrated away, then the magnitude of the weight function becomes trivialized and only its shape dictates which regions of $\alpha$ must agree more or less than other regions. Without integrating $u$, its size as well as $\sigma$ determine how restrictive this prior distribution is. 

\begin{figure}
    \centering
    \includegraphics[width=0.45\textwidth]{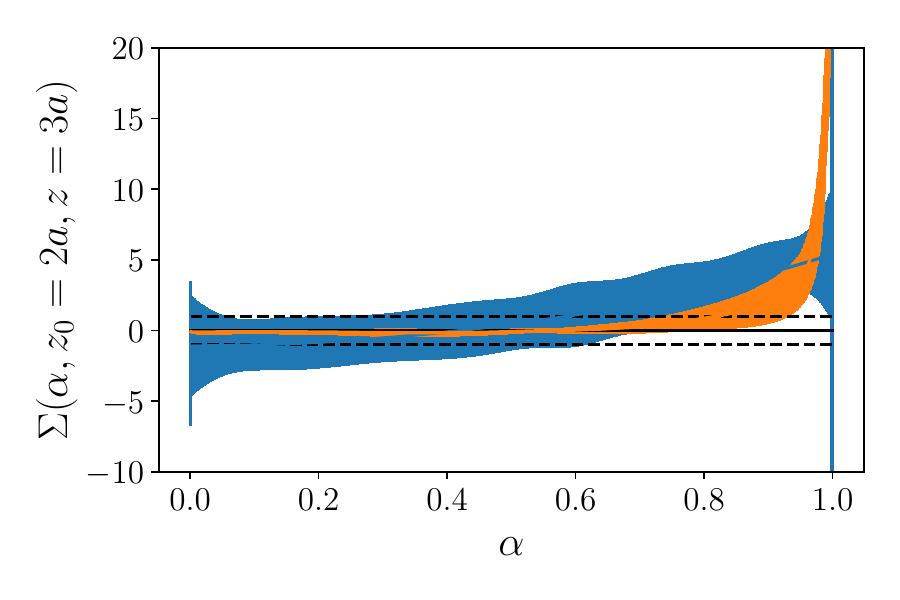}
    \includegraphics[width=0.45\textwidth]{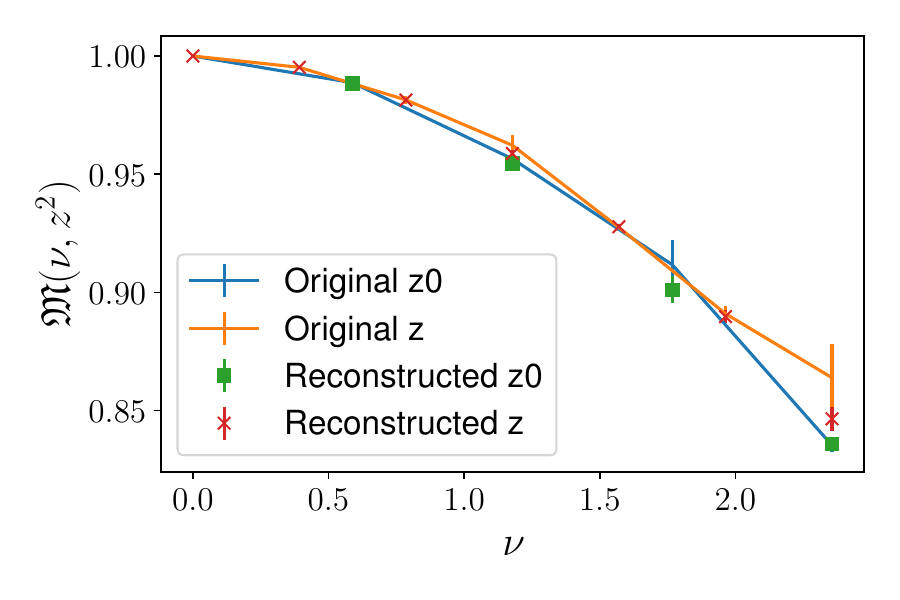}
    \caption{(Left) The step-scaling function from Bayesian Reconstruction (blue) and from the analysis of Sec.~\ref{sec:ss_fit_real_data} (orange). The grid consists of 1000 evenly spaced points in $\alpha$. The prior distribution is defined by $u=1$, $h(\alpha)=0$, and $\sigma(\alpha)=1$. The prior model (solid) and width (dashed) are shown in black. (Right) The reproduction of the data sets.}
    \label{fig:br_first_even}
\end{figure}

For the first example of this approach, we take the seemingly simple choices of: 
\begin{equation}
    u=1 \,\qquad h(\alpha)=0 \,\qquad \sigma(\alpha)=1
\end{equation}
for the prior distribution. The function is parameterized on a grid of 1000 equally spaced points. The set of $\Sigma_i$ on that grid that maximizes the posterior is shown in Figure~\ref{fig:br_first_even}. The minimum of twice the negative log posterior is given by $\chi^2=3.76$ and $2uS=9.20$. It should be noted that the $\chi^2$ reported has not been reduced by the number of datapoints, so this value represents significant agreement with the data. The errors are determined by the inverse of the Hessian near the minimum since both $S$ and $\chi^2$ are quadratic in $\Sigma_i$ and $\mathfrak{M}^\ast(z_0^2)$. Some level of tension with the prior was expected, as for instance the prior does not respect the basic requirement that the integral of the step-scaling function should be close to 1. On the other hand, the $\chi^2$ is significantly less than the number of data points, so the data is well reproduced. The reconstructed step-scaling function is globally compatible with the parametric fit result presented in Figure \ref{fig13}, except at very large $\alpha$ where the flat prior penalizes an attempt at divergence. Although the uncertainty obtained thanks to this Bayesian reconstruction is typically quite larger than that of the parametric fit, the correction effect on the matrix elements shown on the right panel of Figure \ref{fig:br_first_even} is similar to the one derived with the most stringent statistical tolerance in the parametric fit of Figure \ref{fig14}. 

\begin{figure}
    \centering
    \includegraphics[width=0.45\textwidth]{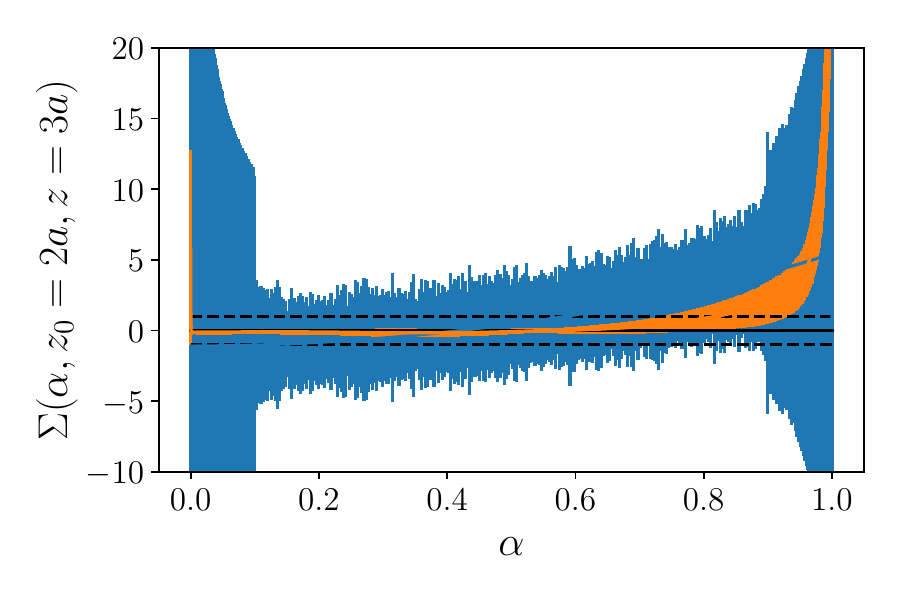}
    \includegraphics[width=0.45\textwidth]{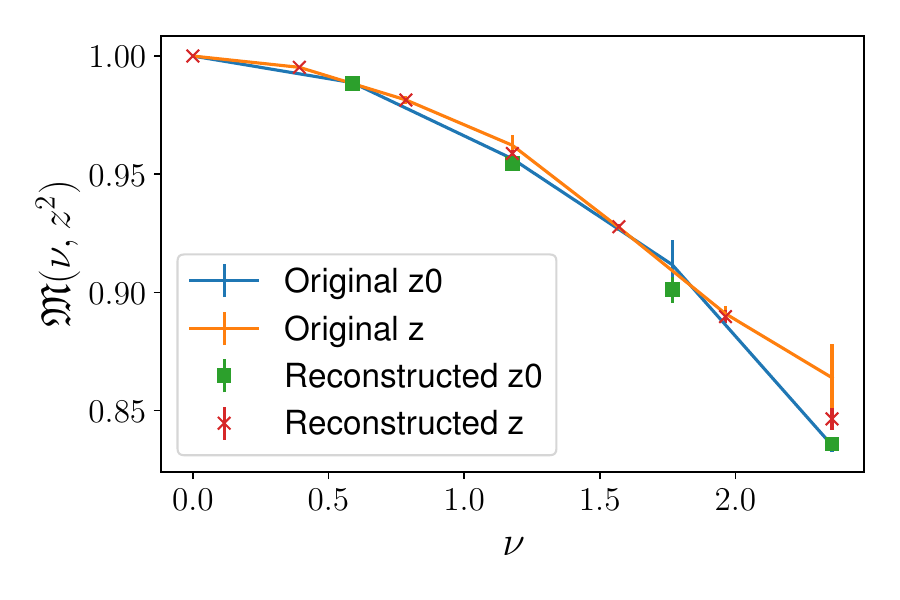}
    \caption{(Left) The step-scaling function from Bayesian Reconstruction (blue) and from the analysis of Sec.~\ref{sec:ss_fit_real_data} (orange). The grid consists of 1000 unevenly spaced points in $\alpha$. The prior distribution is defined by $u=1$, $h(\alpha)=0$, and $\sigma(\alpha)=1$. The prior model (solid) and width (dashed) are shown in black. (Right) The reproduction of the data sets. }
    \label{fig:br_first_uneven}
\end{figure}

Using the same flat prior, we introduce another grid in $\alpha$ consisting of 1000 unevenly spaced points. Specifically, the points between $[10^{-4},0.1]$ and $[0.9,1-10^{-12}]$ are evenly spaced logarithmically, while those in the middle are evenly spaced linearly. There are twice as many points in the upper $\alpha$ region than the other two. The results are shown in Figure~\ref{fig:br_first_uneven}. The minimum of twice the negative log posterior is very similar: $\chi^2=3.76$ and $2uS=9.24$. Since the grid is denser for $\alpha \leq0.1$ and $\alpha \geq 0.9$ compared to the previous case, more fluctuations away from the prior are allowed for the same prior penalty in Eq.~\eqref{eq:vsinjqvwq} as these fluctuations occur on a shorter range in $\alpha$. This generates a much larger uncertainty in the extraction of the step-scaling function as the smoothness of the reconstructed function decreases. Remarkably, this has very little effect on the correction of the matrix elements. Indeed, the matrix elements result from a convolution with the spline-interpolated data, so as we have already argued, they are not sensitive to the short-range fluctuations of the extraction of the step-scaling function.

\begin{figure}
    \centering
    \includegraphics[width=0.45\textwidth]{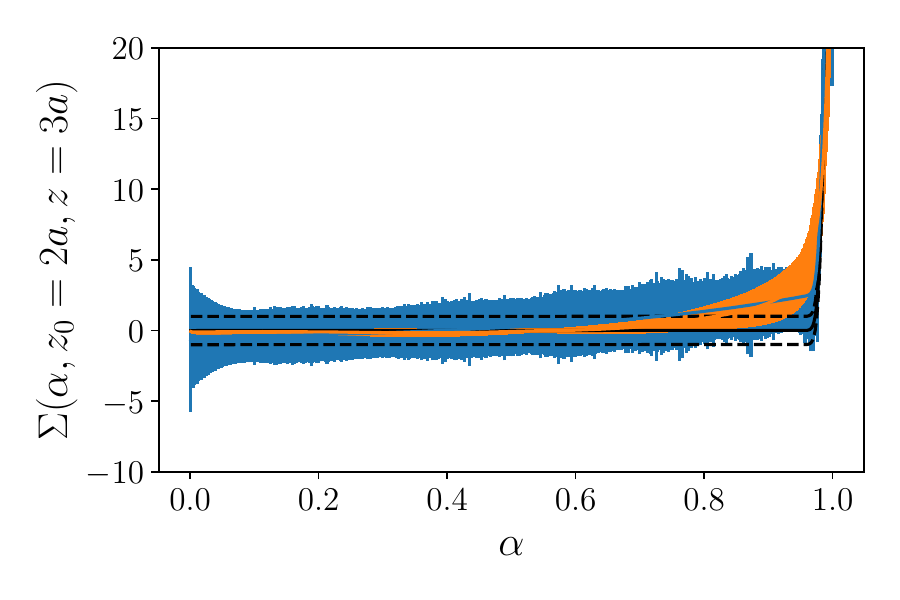}
    \includegraphics[width=0.45\textwidth]{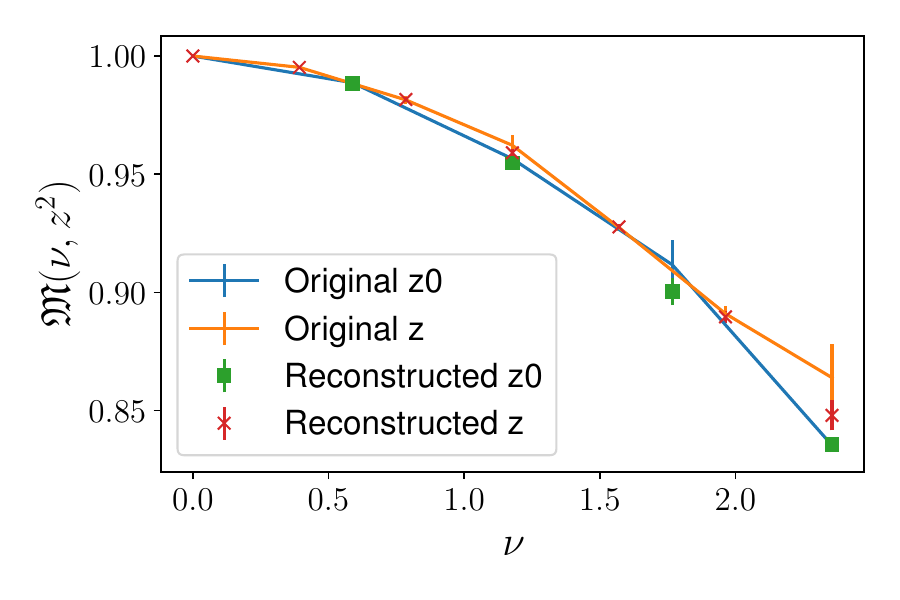}
    \caption{(Left) The step-scaling function from Bayesian Reconstruction (blue) and from the analysis of Sec.~\ref{sec:ss_fit_real_data} (orange). The grid consists of 1000 evenly spaced points in $\alpha$. The prior distribution is defined by $u=1$, $h(\alpha)=N(1-\alpha,w=0.01)$, and $\sigma(\alpha)=1$. The prior model (solid) and width (dashed) are shown in black. (Right) The reproduction of the data sets. }
    \label{fig:br_gauss_prior_1e-2_even}
\end{figure}

\begin{figure}
    \centering
    \includegraphics[width=0.45\textwidth]{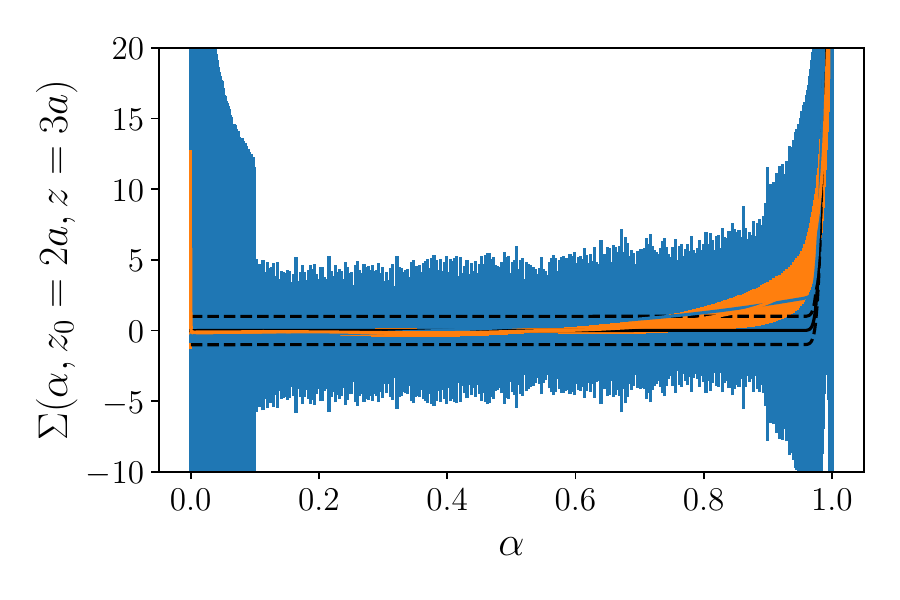}
    \includegraphics[width=0.45\textwidth]{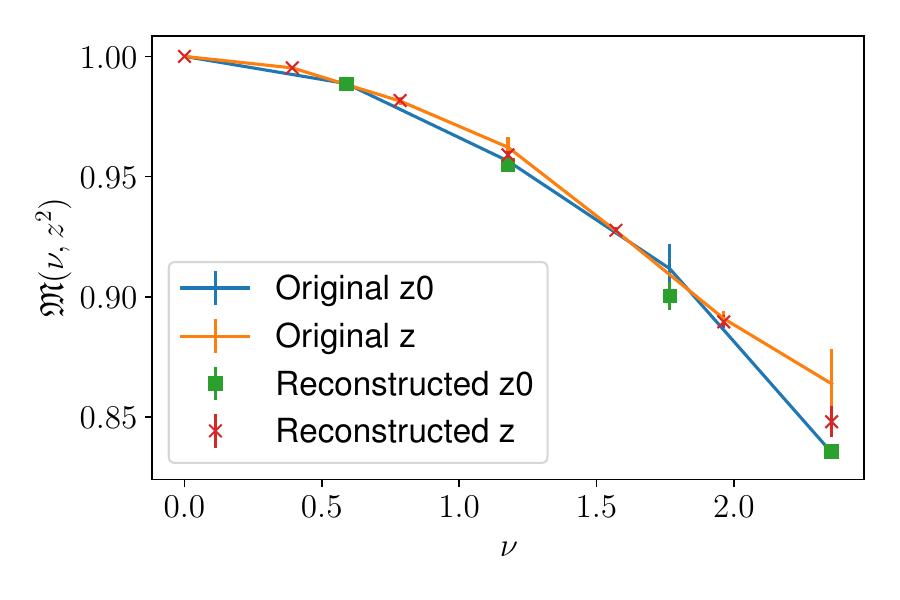}
    \caption{(Left) The step-scaling function from Bayesian Reconstruction (blue) and from the analysis of Sec.~\ref{sec:ss_fit_real_data} (orange). The grid consists of 1000 unevenly spaced points in $\alpha$. The prior distribution is defined by $u=1$, $h(\alpha)=N(1-\alpha,w=0.01)$, and $\sigma(\alpha)=1$. The prior model (solid) and width (dashed) are shown in black. (Right) The reproduction of the data sets. }
    \label{fig:br_gauss_prior_1e-2_uneven}
\end{figure}

There is little reason to believe that the prior model $h=0$ will generate a useful bias to obtain the physics we are interested in. For instance, it contains potentially incorrect limits for $\alpha=0,1$ and does not produce an overall integral close to 1. In order to reproduce the Dirac-like behavior as $\alpha\to1$, the prior distribution could be a sharply peaked Gaussian 
\begin{equation}
    h(\alpha) =N(1-\alpha,w)= \frac{\exp\Big[-\frac{(1-\alpha)^2}{2w^2}\Big]}{w\sqrt{2\pi}}
\end{equation}
which simulates $\frac12\delta(1-\alpha)$ for small $w$. 
Figures~\ref{fig:br_gauss_prior_1e-2_even} and~\ref{fig:br_gauss_prior_1e-2_uneven} show the results of this analysis with $w=0.01$. The same grids in $\alpha$ were used again. 
In both scenarios, the sharply peaked Gaussian prior pulls upward the large $\alpha$ region. The minima of these fits are $\chi^2=3.28$ and $2uS=2.94$ for the even grid and $\chi^2=3.24$ and $2uS=1.88$ for the uneven grid. The data is slightly better reproduced than with the previous prior model while the prior term has decreased significantly. The correction of the matrix elements is once again unbothered by the change from even to uneven grid, and very close to the results obtained with the parametric fit or the flat prior.

\begin{figure}
    \centering
    \includegraphics[width=0.45\textwidth]{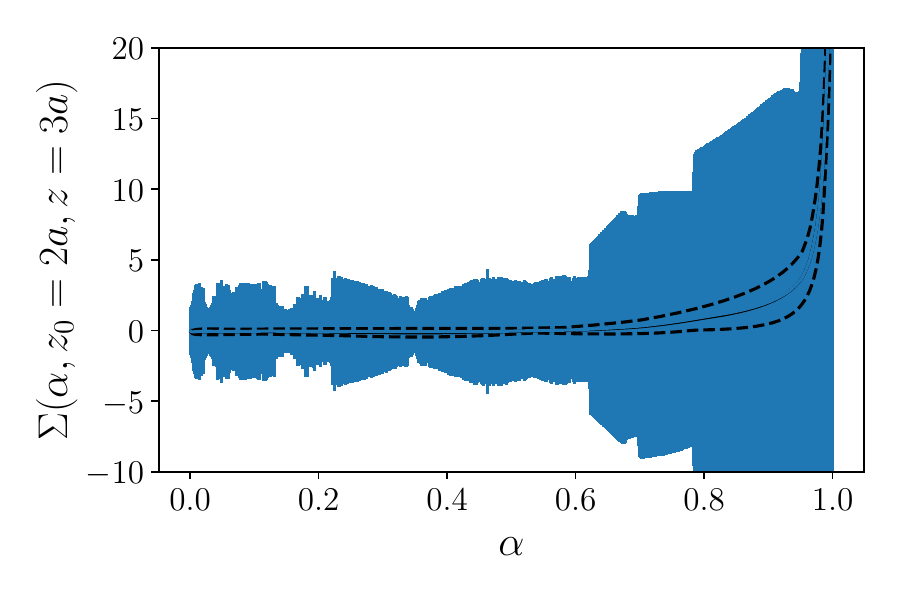}
    \includegraphics[width=0.45\textwidth]{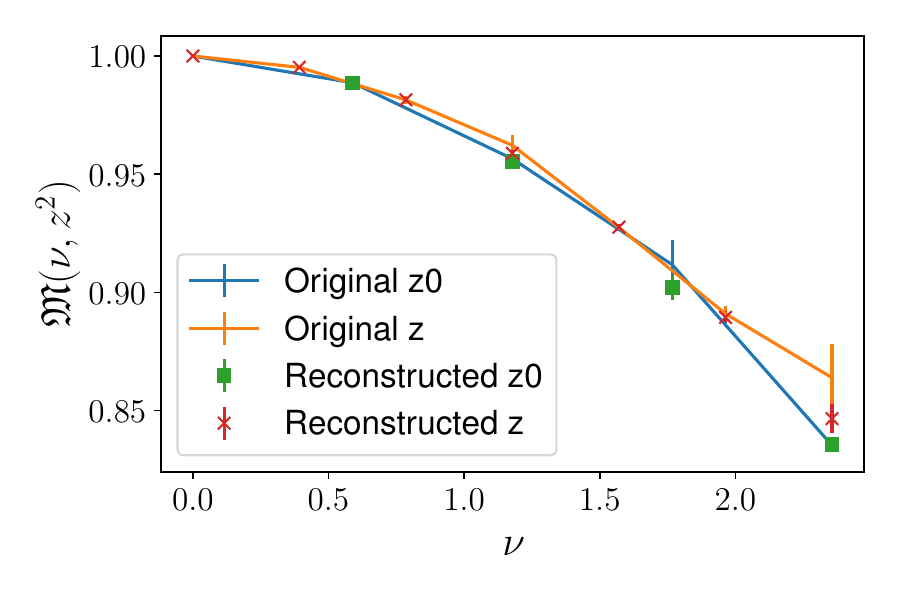}
    \caption{(Left) The step-scaling function from Bayesian Reconstruction (blue). The grid consists of 1000 evenly spaced points in $\alpha$. The prior distribution is defined by $u=1$, and the fit results of Sec.~\ref{sec:ss_fit_real_data}. The prior model (solid) and width (dashed) are shown in black. (Right) The reproduction of the data sets. }
    \label{fig:br_model_prior_model_even}
\end{figure}

A final prior model and width to test would be the results from the parameterized model in Sec.~\ref{sec:ss_fit_real_data}. The model is set to the central value and the width is given by the larger of the upper or lower bands in the $\chi^2$ per $n_{pts} <2$ error estimation. The results are shown in Figure~\ref{fig:br_model_prior_model_even}. This analysis has again a slightly better agreement to the data and a much better agreement to the prior model. For the even grid, $\chi^2=3.06$ and $2uS=0.046$ and for the uneven grid, $\chi^2=3.04$ and $2uS=0.044$. Since this prior model was itself obtained through a careful fit on the data, it is unsurprising that the prior distribution is well satisfied.

\begin{figure}
    \centering
    \includegraphics[width=0.45\textwidth]{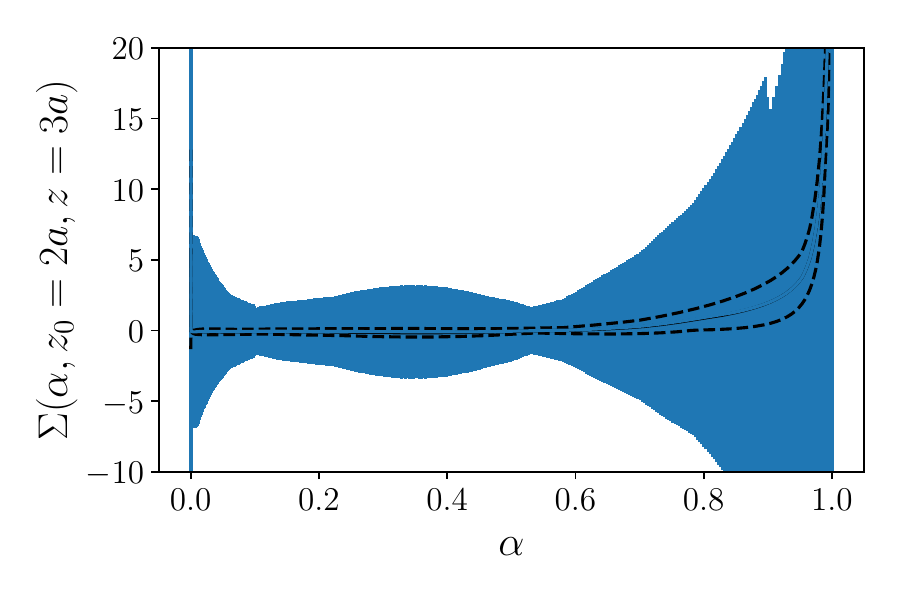}
    \includegraphics[width=0.45\textwidth]{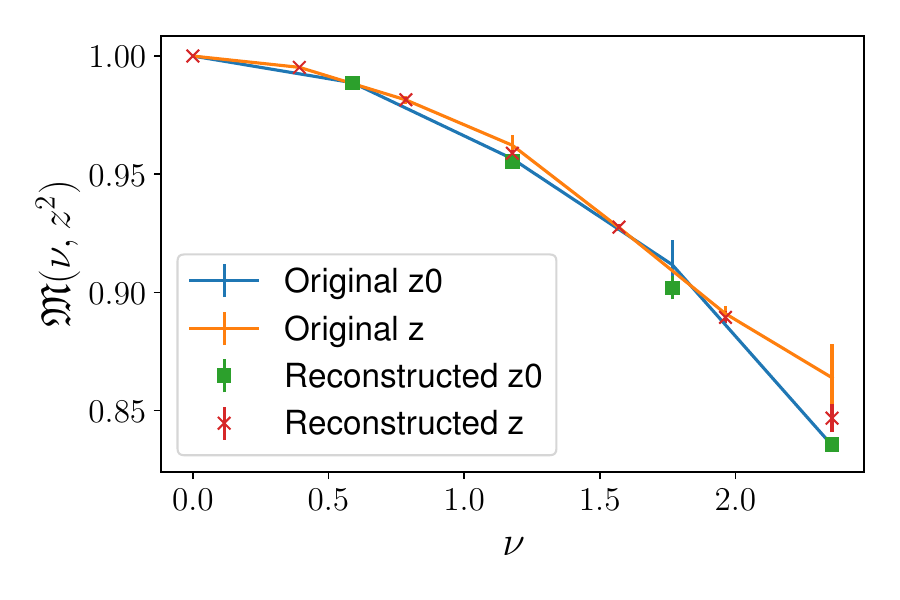}
    \caption{(Left) The step-scaling function from Bayesian Reconstruction (blue). The grid consists of 1000 unevenly spaced points in $\alpha$. The prior distribution is defined by $u=1$, and the fit results of Sec.~\ref{sec:ss_fit_real_data}. The prior model (solid) and width (dashed) are shown in black. (Right) The reproduction of the data sets. }
    \label{fig:br_model_prior_model_uneven}
\end{figure}

\begin{figure}
    \centering
    \includegraphics[width=0.45\textwidth]{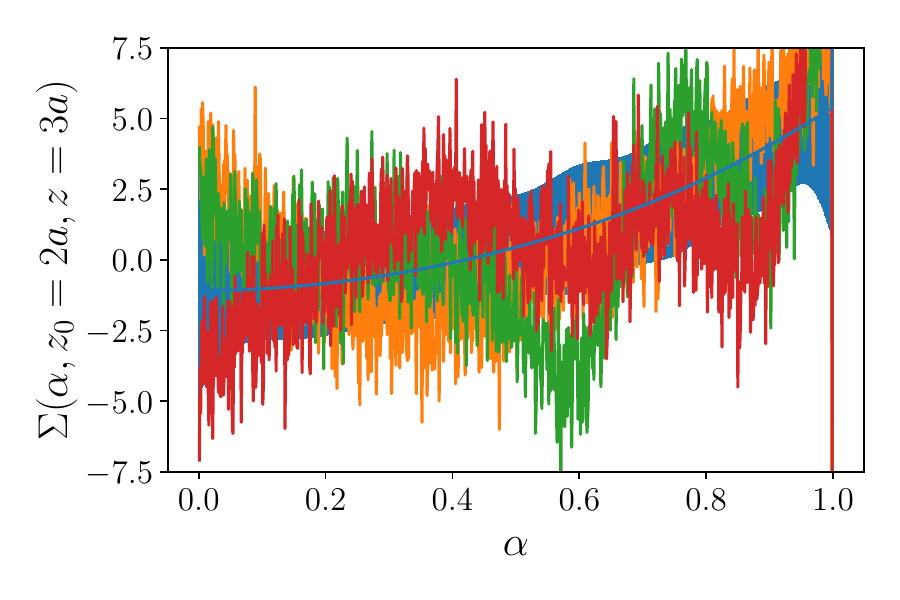}
    \includegraphics[width=0.45\textwidth]{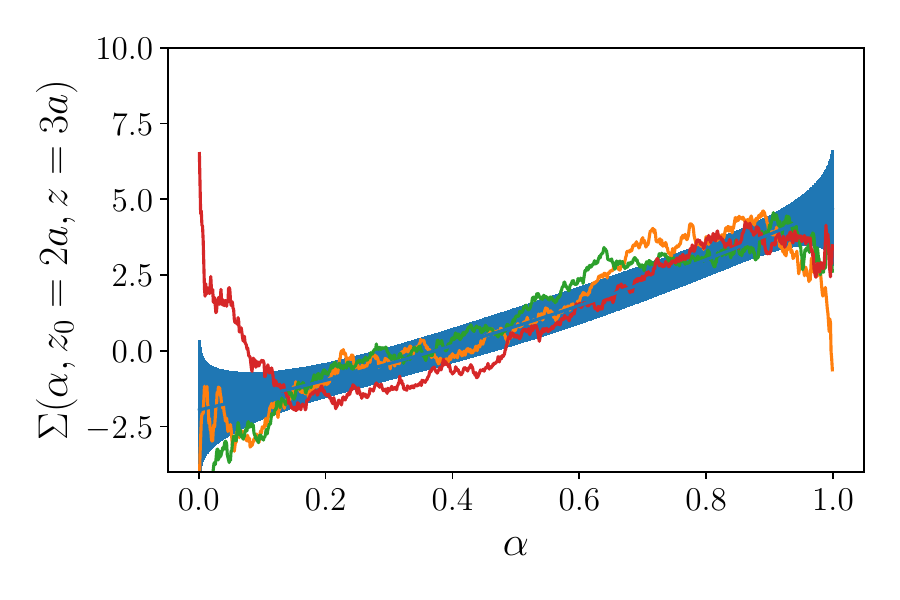}
    \caption{Probable step-scaling functions (red/orange/green) are sampled from a multivariate normal distribution with the full central value and covariance (blue). (Left) Results from the QDR prior with $h=0$ and $\sigma=1$. The function is allowed to vary wildly from one point to the next as represented by the large error band. (Right) Results from the smoothness prior. The functions vary significantly less from one point to the next resulting in the smaller variance.}
    \label{fig:variance}
\end{figure}

In all these cases, the errors have been quite large on the resulting step-scaling function. This is due to the fact that there is no requirement of smoothness in the QDR prior defined in Eq.~\eqref{eq:vsinjqvwq}: the reconstructed step-scaling function can vary abruptly from one value of the grid to the next. Using the inverse Hessian of the first fit with $h = 0$ and $\sigma = 1$, we produce samples of the step-scaling function from the multivariate Gaussian distribution. The samples fluctuate quite wildly as shown in Figure~\ref{fig:variance}. Instead one could consider a prior which does not bias the central value at all but instead biases the derivatives to be smooth. 

\begin{figure}
    \centering
    \includegraphics[width=0.45\textwidth]{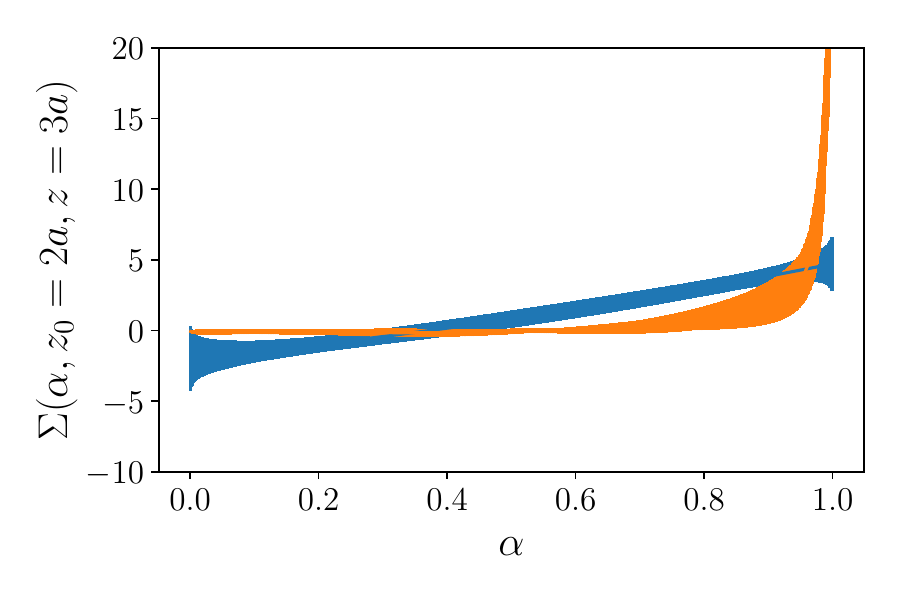}
    \includegraphics[width=0.45\textwidth]{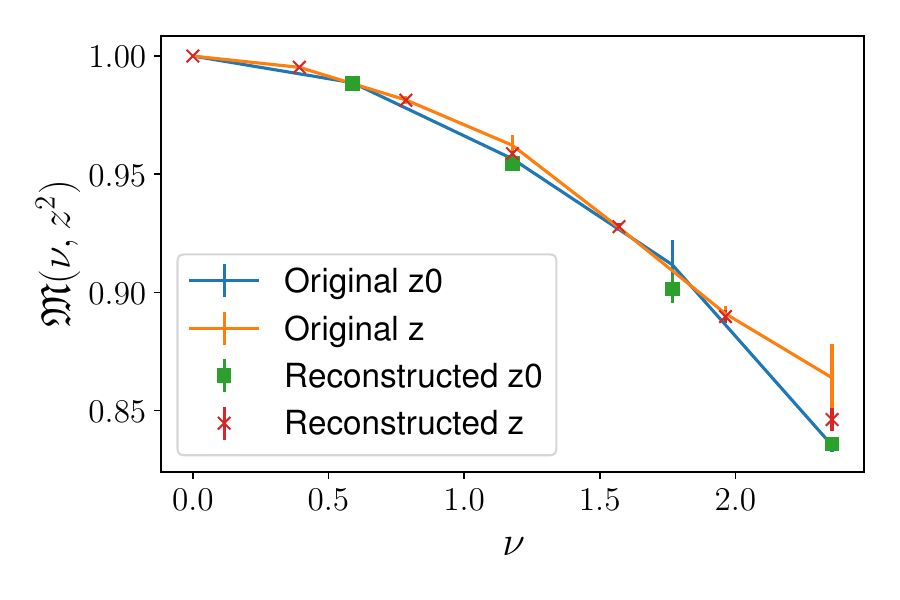}
    \caption{(Left) The step-scaling function from Bayesian Reconstruction (blue) and from the analysis of Sec.~\ref{sec:ss_fit_real_data} (orange). The grid consists of 1000 evenly spaced points in $\alpha$. The prior distribution is defined by $u=1$ and the smoothness function in Eq.~\eqref{eq:smooth}.  (Right) The reproduction of the data sets. }
    \label{fig:br_smooth_even}
\end{figure}

Consider instead a smoothness based prior such as:
\begin{equation}
    S = \int_0^1 d\alpha\, \frac{\alpha(1-\alpha)}2 \left( \Sigma'(\alpha) \right)^2 = \sum_i \frac{\alpha_i(1-\alpha_i)}2 \frac{(\Sigma(\alpha_{i+1}) - \Sigma(\alpha_i))^2}{\alpha_{i+1}-\alpha_i}
\end{equation}\label{eq:smooth}
This prior is intended to keep the function to be relatively smooth in the middle $\alpha$ region while not disfavoring the potential divergences are $\alpha=0$ or 1.
The fit results are shown in Figure~\ref{fig:br_smooth_even} where the variance is smaller than the QDR case. As can be seen in Figure~\ref{fig:variance} this corresponds to sampled step-scaling functions that are noticeably smoother than the QDR case. The posterior at the minimum value is given by $\chi^2=3.82$ and $2uS=  7.56$. While the size of the prior terms are not directly comparable, the $\chi^2$ is comparable to the previous cases, and the effect of correction on the matrix elements again very similar.

We have shown that the Bayesian reconstruction methods can be utilized to determine the step-scaling function from lattice QCD data. With each of our different choices of prior distribution, the data are well reproduced by step-scaling functions that share the same general features, also common with the parametric fit. Among those, a small $\alpha$ contribution mostly compatible with zero, and an increase at large $\alpha$. Although we have exhibited a large sensitivity of the reconstruction to the value of hyperparameters such as the grid density or the prior model, the correction of the matrix elements induced by the fit has shown a great stability. This motivates our claim that it is really the requirement of existence of a step-scaling function along the convolution of Eq.~\eqref{eq:evdojnlkme} at small $z^2$, and the self-consistency that it imposes on the $z^2$ dependence of the lattice data, that has been the main driver of this correction. Implementing such a correction to the lattice data allows a systematic improvement at the cost of a cheap post-processing with the possibility to limit model dependence considerably compared to using a perturbative matching to achieve the same result. 

Finally, this discussion of Bayesian reconstruction has brought again the central aspect of the regularization of the inverse problem. We have addressed it through the angle of the smoothness and smearing throughout this paper. An alternative could be to study many choices of prior distributions to create a Bayesian Model Averaged~\cite{Jay:2020jkz} final result for the step-scaling functions to alleviate the biases from any individual choice. The model averaging could also be augmented by adding results not just from Bayesian Reconstruction but also from parametric models. To accurately and precisely determine the step-scaling function, one must perform this task on a series of ensembles, as we outline in the following section.

\section{Lattice calculations of parton distributions in small volumes}\label{sec:dedicated_proposal}

At small $z^2$ the extraction of a step-scaling function in $z^2$ is possible by imposing self-consistency  of the $z^2$ dependence, which results in a determination of corrected  matrix elements with sizable corrections in some cases.  A much better improvement could likely be achieved if, instead of using an internal consistency of the lattice data, one used a separate dedicated lattice computation of the step-scaling function. In particular, the control of higher-twist effects -- which we have assumed to be negligible here, but could still be present at large Ioffe time for $z \gtrapprox 0.2$ fm -- could be understood by performing the extraction of the step-scaling function from multiple external states accessible from lattice QCD, such as a pion, a kaon, and even a quark state in a fixed gauge. Then, a range of validity of a truly universal leading-twist step-scaling operator could be determined and used to hunt down higher twist contamination that would appear as discrepancies in the $z^2$ dependence of the data compared to the expected leading-twist step-scaling function. 

\begin{figure}[h]
    \centering
    \includegraphics[scale=.35]{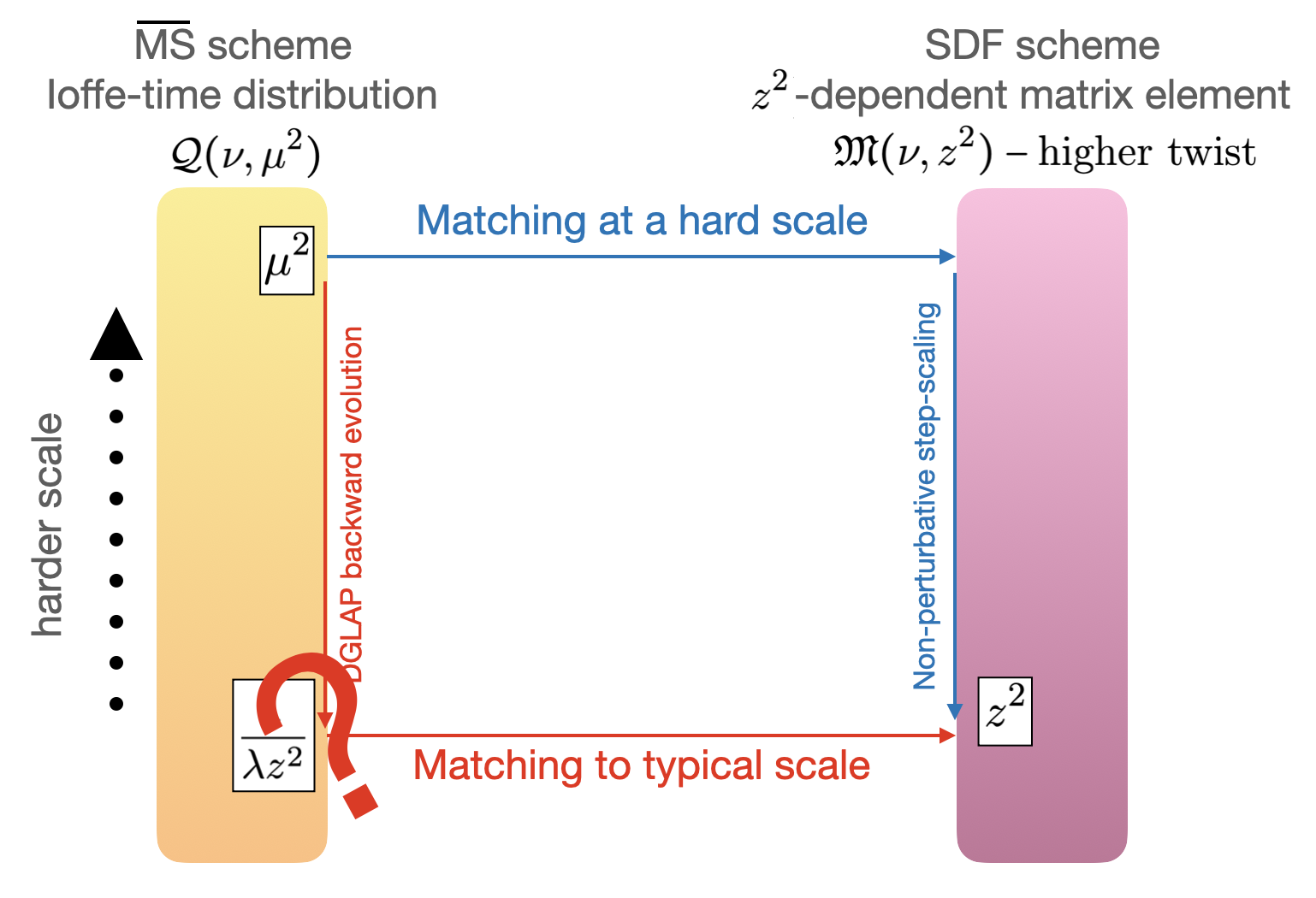}
    \caption{Instead of performing the matching at a low scale $\mu^2\lambda z^2 = 1$ where $\alpha_s$ is nearly divergent and perturbation theory uncertain, we evolve in a non-perturbative way the matrix element to a smaller value of $z^2$, where ordinary matching can now be performed safely.}
    \label{fignew_prop_}
\end{figure}

This non-perturbative evolution operator could be used to bring lattice calculations to higher scales more amenable to perturbation theory in a self-consistent way, thereby reducing the size of theoretical uncertainty induced by the perturbative matching as illustrated on Figure \ref{fignew_prop_}. Additionally if short enough distance scales can be reached, one could invert Eq.~\eqref{eq:vihcdsjak} to obtain the $\overline{MS}$ step scaling function from a non-perturbative calculation of the pseudo-PDF step scaling function. This evolution can be used in global analyses of datasets whose hard scales are perhaps too low for accurate pertubative evolution. Small volume calculations, as we will explore below, could be used to reach much smaller values of $z^2$, at which point one could hope to factorize directly experimental processes without any explicit reference to matching at all. To some extent, the use of non-perturbative evolution to reduce the uncertainty in the matching could be compared to the approach advocated in~\cite{Karthik:2021sbj}. There, the authors suggest to determine the Mellin moments of the matching kernel by comparing experimental estimates of moments of the PDF and lattice estimates of the moments of pseudo-PDF in the small $z^2$ region where the matching is universal. This non-perturbative matching kernel could be used to derive a non-perturbative evolution operator, although the main ingredient is quite different: \cite{Karthik:2021sbj} uses experimental data as a point of reference whereas we suggest to use a dedicated lattice computation fully inscribed within the theoretical framework of the Standard Model which allows for an independent prediction required for BSM studies.

One less obvious use of the discussion of section IV is however its connection to finite volume effects, and the perspectives it opens for a radical new proposal of computations of parton distributions in small volumes. Before highlighting this connection, let us present why we believe that small volume calculations could present a crucial perspective for the program of parton physics on the lattice. Demanding lattice computations by current standards could typically use the following settings with a near-physical pion mass: $a = 0.05$ fm and $L = 128a = 6.4$ fm. Then the largest accessible momentum allowed by the Brillouin zone of the lattice is of the order of $P = \frac{2\pi}{L} \times \frac{L}{6a} = \pi / (3a) \approx 4.2$ GeV. This happens to also be a momentum at which the issue of signal-to-noise ratio becomes quite critical. The break down of renormalization group improved perturbation theory and the growth of higher twist effects precludes the use of non-local matrix elements with a separation larger than 0.3 fm. If one uses a separation of $z = 0.1$ fm in the hope that it is small enough to effectively remove the theoretical uncertainties in the partonic interpretation of the matrix element, then the largest accessible Ioffe times $z \cdot P$ are only of about 2. This range in Ioffe time (coincidentally, it is roughly the one which we have been working with in this paper -- although we worked at smaller momentum and larger $z$) is quite disappointing considering the efforts required to handle such a fine lattice. Of course, if we do not want to increase $z$, the only margin of improvement is to go to ever larger momenta and therefore finer lattices. However, both aspects are a serious issue. Just to obtain a range in Ioffe time of about 4 would require momenta of the order of 8 GeV, and lattice spacing of the order of 0.03 fm. The computational cost due to the issue of ``critical slowing-down'' \cite{Luscher:2010iy} and the degrading signal-to-noise ratio make this situation unrealistic for what is additionally still a very limited range in Ioffe time. The challenge of accessing large momenta occurs in rather similar terms in the LaMET formalism. For instance, Ref.~\cite{Fan:2020nzz} notes that if the validity of the LaMET perturbative matching is limited to $z \approx 0.2$ fm, then reliable calculations down to $x = 0.1$ would require momenta in excess of 10 GeV. Efforts to probe larger momenta at fixed lattice size are underway \cite{Gao:2023lny}, but cannot yet achieve the large momenta necessary.

Requiring very fine lattices, while preserving a volume (much) larger than the hadronic scale (two-scale problem),  has been a challenge in several domains of lattice QCD, notably for B-meson physics calculations~\cite{Guagnelli:2002jd,Guazzini:2007ja,ETM:2009sed}, where the large quark mass plays an analogous role as our requirement of large momentum $P$. Such calculations are performed by dropping the requirement of infinite volume limit, which allows very fine lattices in ever smaller volumes, while keeping the number of sites constant. This solves both the question of the largest available momentum and the issue of signal-to-noise ratio. In sub-hadronic lattice volumes, hadrons no longer exist, and the uncomfortably light pion no longer dictates the rate of increase of the noise. Instead, it is the scale tied to the finite volume $1/L$ that determines all spectral gaps and the scaling of noise. Therefore, as the volume decreases in physical units while keeping it fixed in lattice units the spectral gaps and the noise scaling will remain roughly constant allowing us to access increasing momentum in physical units. 

Every time the volume and spacing of the lattice are divided by two while keeping the physical value of $z$ constant, the accessible range in Ioffe time is multiplied by two. For a relatively similar computational cost as the $a = 0.05$ fm, $L = 128a = 6.4$ fm lattice that we have mentioned before, we could therefore carry out a computation with $a = 0.0063$ fm and $L = 128a = 0.8$ fm. This would give access to a range of Ioffe time of about 16 for $z = 0.1$ fm! The catch, of course, is that in this small volume the matrix element is no longer that of a hadron, and its interpretation in terms of PDFs is impossible without further work. Therefore, one must find a way to understand the nature of the finite volume effects to connect the matrix element computed in small volume to its hadronic counterpart in infinite volume. This is precisely the traditional role of the step-scaling function whose potential in B-physics calculations  has been demonstrated in~\cite{Guagnelli:2002jd,Guazzini:2007ja,ETM:2009sed}.

The possibility to perform small volume computations of the first DGLAP anomalous dimension has been demonstrated in the quenched approximation using the Schr\"odinger functional method and local matrix elements \cite{Guagnelli:1998ve,Guagnelli:1999gu}. We believe that the success of this study, which reproduces the perturbative results at large scales, is a sign that even for non-local matrix elements, with $z \ll L$, the finite volume effects can be propagated through the form of the standard DGLAP convolution. More precisely, we have been working so far in this paper implicitly in the infinite volume and continuum limit, and only focused on the $\alpha$ dependence of the step-scaling function that evolves from one scale $z^2$ to another. But for $z \ll \Lambda_{QCD}^{-1}, L$, we could likewise introduce a generalized step-scaling function accounting for changes in volume: 
\begin{equation}
    \Sigma(\alpha;z_0^2, z_1^2; L_0, L_1)\,.
\end{equation}
The general principle of the computation of the PDF in small volume could then be the following:
\begin{enumerate}
    \item Compute the matrix element at $z = 0.1$ fm in a small volume lattice $L = 128a = 0.8$ fm. The accessible Ioffe time range could be up to 16.
    \item Compute the matrix element in a larger volume by a factor 2. The accessible Ioffe time range is reduced to 8. By using a similar procedure to that explained in section IV of this paper, compute the generalized step-scaling function $\Sigma(\alpha; z_0 = z_1 = 0.1 \textrm{ fm}; L_0 = 0.8 \textrm{ fm}, L_1 = 1.6 \textrm{ fm})$ using the data in both volumes on the Ioffe time range up to 8. 
    \item Repeat the previous step by doubling the volume up to $L=6.4$ fm, each time computing the step-scaling function on a shrinking Ioffe time range.
    \item Use the computed step-scaling functions to evolve the initial computation in the full glory of its Ioffe time range up to 16 all the way to the large volume.
\end{enumerate}
One will notice that the main ``trick'' is that the step-scaling function is independent of the Ioffe-time range if $z^2$ is small enough, and can therefore be computed even though the range in Ioffe time decreases as one approaches the infinite volume limit. One could argue that since the step-scaling functions to the largest volumes are computed on much reduced Ioffe time ranges, they will be less precise and will add systematic uncertainty when evolving the large Ioffe time range of the small volume matrix elements. We believe, however, that a control of the precision of the step-scaling uncertainty is possible for two main reasons. First, as the volume increases, the effects of small volume should be reduced and the step-scaling function should contain less information -- it will be more akin to a simple Dirac delta. Second, if we find the quality of the step-scaling function to be disappointing -- and we have tools with reduced model dependence to evaluate that as we have presented in this paper -- then we can always increase the density of available Ioffe times within the same limited range thanks to twisted boundary conditions, and therefore obtain a more precise extraction of the step-scaling function.

\section{Conclusions}
In this work, we have studied the evolution of the flavor non-singlet unpolarized pseudo-distribution with respect to the scale $z^2$ and outlined the nature of the step scaling function, which evolves matrix elements from one scale to another. At small enough $z^2$, where the OPE can be established, we have shown that the existence of this step-scaling function implies a self-consistent effect of $z^2$ dependence among the lattice data points which can lead to sizeable corrections of fluctuations in the data. The ability to evolve matrix elements from longer to shorter distance scales is fundamentally important for the analysis of parton distributions from lattice QCD. By controlling this evolution properly, one can reach scales where perturbative comparison to the light-cone $\overline{MS}$ parton distributions can be done with reduced perturbative systematic errors. As has been shown this evolution is intimately connected to the standard DGLAP evolution in the $\overline{MS}$ scheme and the matching kernels relating the pseudo-distributions to the $\overline{MS}$ scheme. 

Unfortunately the perturbative expansions, and the application of the DGLAP evolution in the $\overline{MS}$ scheme, are potentially insufficient at the scales utilized for modern lattice QCD calculations. We have outlined a proposal to access a larger kinematic range on the lattice, while preserving good control of the matching accuracy. In analogy to approaches used in heavy quark physics in lattice QCD, the step scaling functions, both in scale $z^2$ and in volume $L$, can be modelled using the techniques outlined in Sec.~\ref{sec:fit} on a series of ensembles. By scaling in volume as well as in the invariant scale, the step scaling function can be computed with Schr\"odinger functional techniques with small enough lattice spacings that perturbative matching to the $\overline{MS}$ scheme can be done with minimal error. 

We study the extraction of the step scaling functions to determine the feasibility of our proposal. The step scaling function fit from real lattice data was modelled by a parameterized form. The result reproduced a function quite similar to the known perturbative expansions. To study the model dependence, a non-parameteric form, Bayesian reconstruction, was used to reproduce the step scaling function. The function is determined by its value on a grid over its domain and some interpolation rules. The ill-posed inverse problem is regulated by a choice of prior distribution, several of which were studied. A dedicated study of optimal choices of prior distributions can be performed in the future using mock or real lattice data. Such tests will be key to understanding the model dependence of the final step scaling functions.

Finally, this work can be extended to other parton distributions and their step scaling can be studied similarly. Comparing to the results of a pion matrix element analysis is critical to understanding the systematic errors and range of validity in this approach. Furthermore, the evolution of the nucleon GPD and TMD for example, would be of great interest to current exploratory lattice calculations. Each of these distributions are significantly less well-known from experimental results alone. If lattice QCD data for PDFs are to be compared directly to experimental results, as done in~\cite{Bringewatt:2020ixn,JeffersonLabAngularMomentumJAM:2022aix}, then the evolution of the lattice data from long distance scales must be done properly.

\appendix

\section{Technical appendix on the matching relations}
\label{canonical_scale}

\subsection{Perturbative expansion of the matching kernel}

Provided a perturbative expansion of the matching kernel exists, it is possible to express it from any $z^2$ to the $\overline{MS}$ scale $(\lambda z^2)^{-1}$ as:
\begin{equation}
    {\cal C}_0(z^2) = \sum_{k=0}^\infty \left(\frac{\alpha_s((\lambda z^2)^{-1})}{2\pi}\right)^k c_{k,0}(\lambda)\,,
\end{equation}
where $c_{k,0}(\lambda)$ are distributions which depend on the choice of $\lambda$. This expression singles out a specific $\overline{MS}$ scale $(\lambda z^2)^{-1}$. The general matching to any scale can be derived exactly following Eq.~\eqref{eq:efvdvnjke} as the convolution of ${\cal C}_0$ with the backward $\overline{MS}$ step-scaling function. This yields the ordinary presentation of the matching kernel in the literature as a double expansion in terms of $\alpha_s(\mu^2)$ and $\ln(\mu^2z^2\lambda)$. If all orders are considered, the intermediate scale $(\lambda z^2)^{-1}$ does not matter. If a perturbative truncation is performed, then this scale becomes a first choice with a simpler matching devoid of logarithms.

If $\lambda = -e^{2\gamma_E+1}/4$, then Eq.~\eqref{eq:bvjeknks} teaches us that:
\begin{equation}
    c_{0,0}(\lambda) = 1_\otimes\,,\ \ c_{1,0}(\lambda) = -D\,, ...
\end{equation}
In general, one has:
\begin{align}
    {\cal C}(z^2\mu^2, \alpha_s(\mu^2)) &= 1_\otimes - \frac{\alpha_s(\mu^2)}{2\pi}\left[\ln\left(-z^2\mu^2\frac{e^{2\gamma_E+1}}{4}\right)B_1 + D\right] + {\cal O}(\alpha_s^2) \,,\nonumber\\
    &= 1_\otimes - \frac{\alpha_s(\mu^2)}{2\pi}\left[\ln\left(z^2\mu^2\lambda\right)B_1 + \ln\left(-\frac{e^{2\gamma_E+1}}{4\lambda}\right)B_1 + D\right] + {\cal O}(\alpha_s^2)\,,
\end{align}
so
\begin{equation}
    c_{0,0}(\lambda) = 1_\otimes\,,\ \ c_{1,0}(\lambda) = -D - \ln\left(-\frac{e^{2\gamma_E+1}}{4\lambda}\right)B_1\,, ...
    \label{eq:e3fedviyhbkws}
\end{equation}
For instance, the derivation is performed with a value of $\lambda = -e^{2\gamma_E}/4$ in \cite{Li:2020xml}. One can verify that their matching relation gives indeed (see the next section for more details):
\begin{equation}
    c_{0,0}(\lambda) = 1_\otimes\,,\ \ c_{1,0}(\lambda) = -D - B_1\,, ...
\end{equation}
Let us demonstrate how the double expansion in terms of $\alpha_s(\mu^2)$ and $\ln(\mu^2z^2\lambda)$ arises, and how the coefficients in front of the logarithms can be directly obtained as functions of the DGLAP splitting function and the $c_{k,0}(\lambda)$ distributions. 

A Taylor expansion of the  $\overline{MS}$ step-scaling function in terms of $\ln(\mu^2)$ reads:
\begin{equation}
    {\cal E}(\mu_0^2, \mu^2) = \sum_{k = 0}^{+\infty} \frac{1}{k!}\ln^k\left(\frac{\mu^2}{\mu_0^2}\right)\frac{d^k {\cal E}(\mu_0^2, \mu^2)}{d\ln^k(\mu^2)}\bigg|_{\mu^2 = \mu_0^2} \,,
\end{equation}
where the DGLAP differential equation \eqref{eq:grvbdsisjqs} and the running of the strong coupling read:
\begin{spreadlines}{1ex}
\begin{equation}
\begin{dcases}
\frac{d}{d\ln \mu^2}{\cal E}(\mu_0^2,\mu^2) = \bigg\{\frac{\alpha_s(\mu^2)}{2\pi} B_1 + \left(\frac{\alpha_s(\mu^2)}{2\pi}\right)^2 B_2 + ...\bigg\} \otimes {\cal E}(\mu_0^2,\mu^2)\,, \\
\frac{d}{d \ln \mu^2} \alpha_s(\mu^2) = -\beta_0\alpha_s^2(\mu^2) -\beta_1\alpha_s^3(\mu^2) - ...\,.
\end{dcases}
\end{equation}
\end{spreadlines}
It is easy to see that the differential of order $k$ of ${\cal E}(\mu_0^2,\mu^2)$ starts at order ${\cal O}(\alpha_s^k)$. For instance:
\begin{equation}
    \frac{d^2}{d\ln^2(\mu^2)}{\cal E}(\mu_0^2,\mu^2)\bigg|_{\mu^2 = \mu_0^2} = \left(\frac{\alpha_s(\mu_0^2)}{2\pi}\right)^2 (B_1^{\otimes 2} - 2\pi\beta_0 B_1) + {\cal O}(\alpha_s^3)\,.
\end{equation}
Straightforward calculations then gives that the $\overline{MS}$ step-scaling function can be written as:
\begin{equation}
{\cal E}(\mu_0^2, \mu^2) = 1_\otimes + \sum_{n=1}^{\infty} \left(\frac{\alpha_s(\mu_0^2)}{2\pi}\right)^n\sum_{k=1}^n \ln^k\left(\frac{\mu^2}{\mu_0^2}\right) {\cal E}_{n,k}\,,
\end{equation}
where for instance: 
\begin{align}
    {\cal E}_{1,1} = B_1\,,\ \ {\cal E}_{2,1} = B_2\,,\ \ 
    {\cal E}_{2,2} = \frac{1}{2}(B_1^{\otimes 2} - 2\pi\beta_0 B_1)\,.
\end{align}
An important observation is that the terms ${\cal E}_{n,n-k}$ which constitute the $k$-th leading logarithmic expansion (N$^k$LL) are only expressed as functions of $B_{1+j}$ and $\beta_{j}$ such that $0 \leq j \leq k$.

In fact, we are interested in the backward $\overline{MS}$ step-scaling function in Eq.~\eqref{eq:efvdvnjke}, so simple calculations give:
\begin{equation}
{\cal E}(\mu^2, \mu_0^2) = 1_\otimes + \sum_{n=1}^{\infty} \left(\frac{\alpha_s(\mu_0^2)}{2\pi}\right)^n\sum_{k=1}^n \ln^k\left(\frac{\mu^2}{\mu_0^2}\right) \widetilde{\cal E}_{n,k}\,, \label{eq:ehcijaxo}
\end{equation}
where:
\begin{align}
    \widetilde{\cal E}_{1,1} = -B_1\,,\ \ \widetilde{\cal E}_{2,1} = -B_2\,,\ \ 
    \widetilde{\cal E}_{2,2} = \frac{1}{2}(B_1^{\otimes 2} + 2\pi\beta_0 B_1)\,.\label{eq:efbeviscoi}\end{align}
Evaluating Eq.~\eqref{eq:ehcijaxo} for $\mu_0^2 = (\lambda z^2)^{-1}$ and convoluting with the expansion of ${\cal C}_0(z^2)$ yields:
\begin{equation}
    {\cal C}( z^2\mu^2,\alpha_s(\mu^2)) = 1_\otimes + \sum_{n=1}^{\infty} \left(\frac{\alpha_s(\mu_0^2)}{2\pi}\right)^n\sum_{k=0}^n \ln^k\left(\mu^2 z^2 \lambda\right) {\cal F}_{n,k}\,,
\end{equation}
where:
\begin{align}
    {\cal F}_{n,0} = c_{n,0}(\lambda)\,,\ \ {\cal F}_{1,1} = -B_1\,,\ \ {\cal F}_{2,1} = - c_{1,0}(\lambda)\otimes B_1 -B_2\,,\ \ 
    {\cal F}_{2,2} = \frac{1}{2}(B_1^{\otimes 2} + 2\pi\beta_0 B_1)\,.
\end{align}
Finally, we expand $\alpha_s(\mu_0^2)$ as a series in $\alpha_s(\mu^2)$ and $\log(\mu^2 / \mu_0^2)$:
\begin{align}
    \alpha_s(\mu_0^2) &= \sum_{j = 0}^\infty \frac{(-1)^j}{j!}\ln^j\left(\frac{\mu^2}{ \mu_0^2}\right)\frac{d^j \alpha_s(\mu^2)}{d\log^j(\mu^2)} = \alpha_s(\mu^2) + \beta_0\alpha^2_s(\mu^2)\ln\left(\frac{\mu^2}{ \mu_0^2}\right) +  \mathcal{O}(\alpha_s^3)\,,
\end{align}
and we obtain the general perturbative expansion of the matching kernel ${\cal C}$: 
\begin{equation}
    {\cal C}( z^2\mu^2,\alpha_s(\mu^2)) = 1_\otimes + \sum_{n=1}^{\infty} \left(\frac{\alpha_s(\mu^2)}{2\pi}\right)^n\sum_{k=0}^n \ln^k(\mu^2z^2\lambda ) c_{n,k}(\lambda)\,,
    \label{eq:ebfhvjocijwpo}
\end{equation}
where:
\begin{align}
    c_{1,1}(\lambda) = -B_1\,,\ \ 
    c_{2,1}(\lambda) = c_{1,0}(\lambda) \otimes(2\pi\beta_0-B_1) - B_2\,,\ \  
    c_{2,2}(\lambda) = \frac{1}{2}\left(B_1^{\otimes 2}-2\pi\beta_0 B_1\right)\,.\label{eq:vrbhecioxsijo}
\end{align} 
This gives a very simple understanding of the structure of the matching kernel whereas the analytical expressions of $c_{2,1}$ and $c_{2,2}$ already are seriously cumbersome. We check explicitly the validity of the formula for $c_{2,2}(\lambda)$ thanks to the matching kernel at order ${\cal O}(\alpha_s^2)$ derived in \cite{Li:2020xml} in the next section. The terms $c_{n,n-k}(\lambda)$ which constitute the N$^k$LL expansion are only expressed as functions of $c_{j, 0}(\lambda)$, $B_{1+j}$ and $\beta_j$ such that $0 \leq j \leq k$. In particular, since $c_{0,0}(\lambda) = 1_\otimes$, the LL expansion is independent of $\lambda$. 

\begin{figure}
    \centering
    \includegraphics[width=0.38\textwidth]{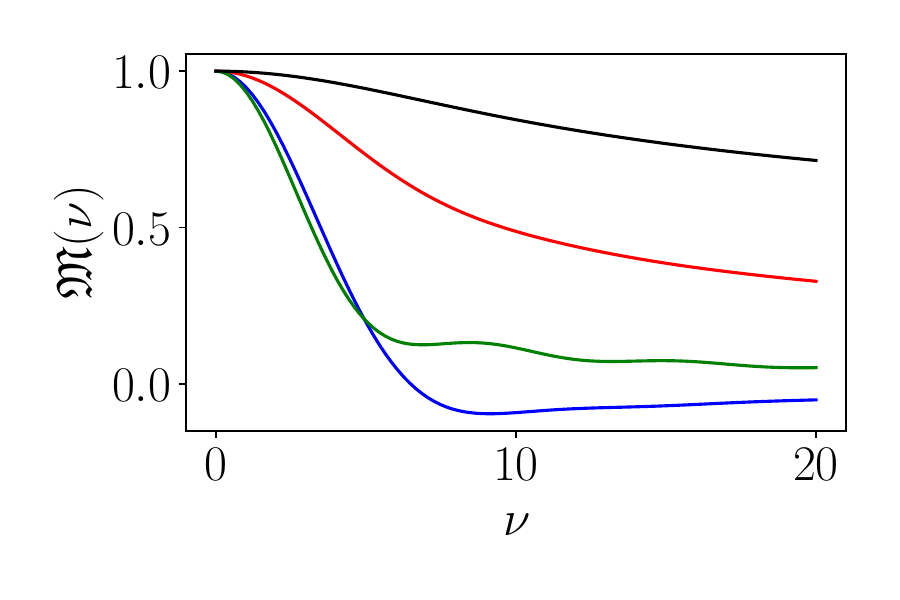}
    \includegraphics[width=0.38\textwidth]{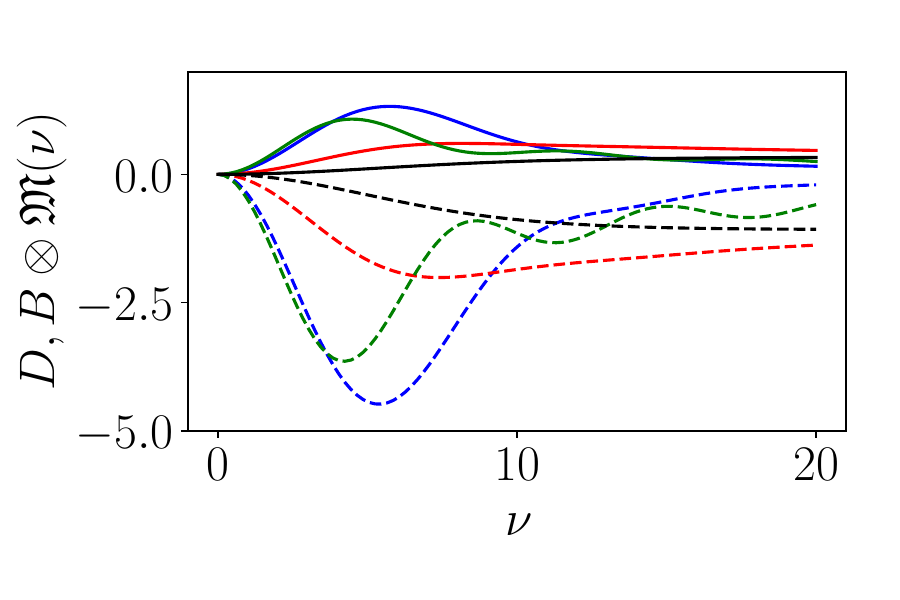}
    \includegraphics[width=0.38\textwidth]{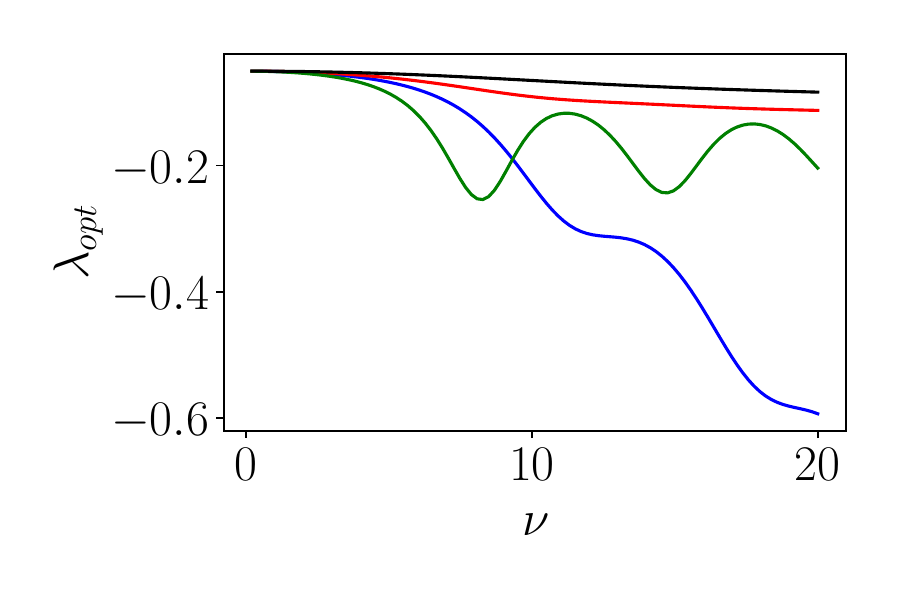}
    \includegraphics[width=0.38\textwidth]{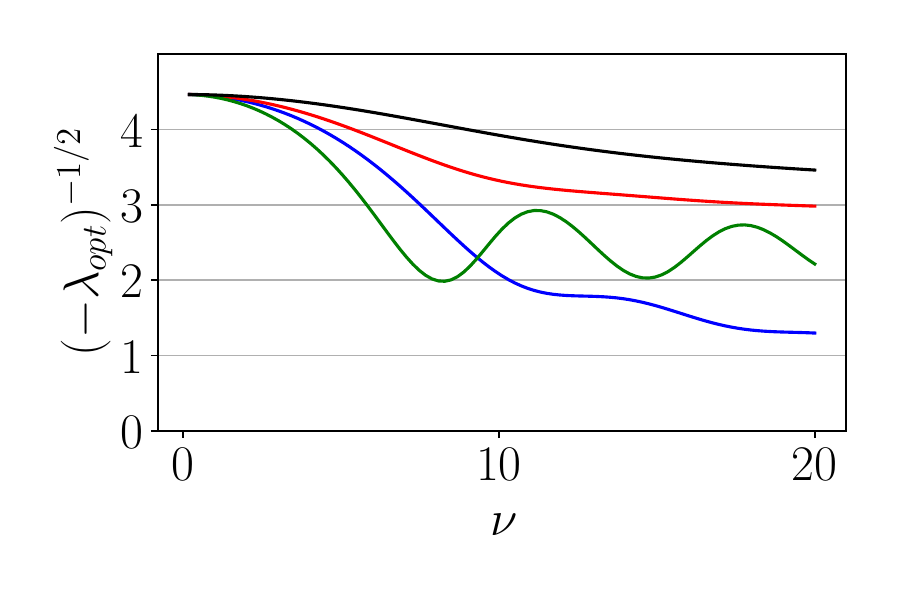}
    \caption{(Upper Left) Models of ITDs similar to those arising from the lattice. (Upper Right) Convolutions of those models with $B_1$ (solid) and $D$ (dashed). The two convolutions differ in sign and magnitude. (Lower Left) The value of $\lambda_{opt}$ which will make the convolution with $c_{1,0}$ to be 0. (Lower Right) $(-\lambda_{opt})^{-1/2}$ represents how much larger $\mu$ should be than $1/z$ in order to simultaneously cancel the logarithms of those scales and $c_{1,0}$. }
    \label{fig:model_optimization}
\end{figure}

\begin{figure}
    \centering
    \includegraphics[width=0.38\textwidth]{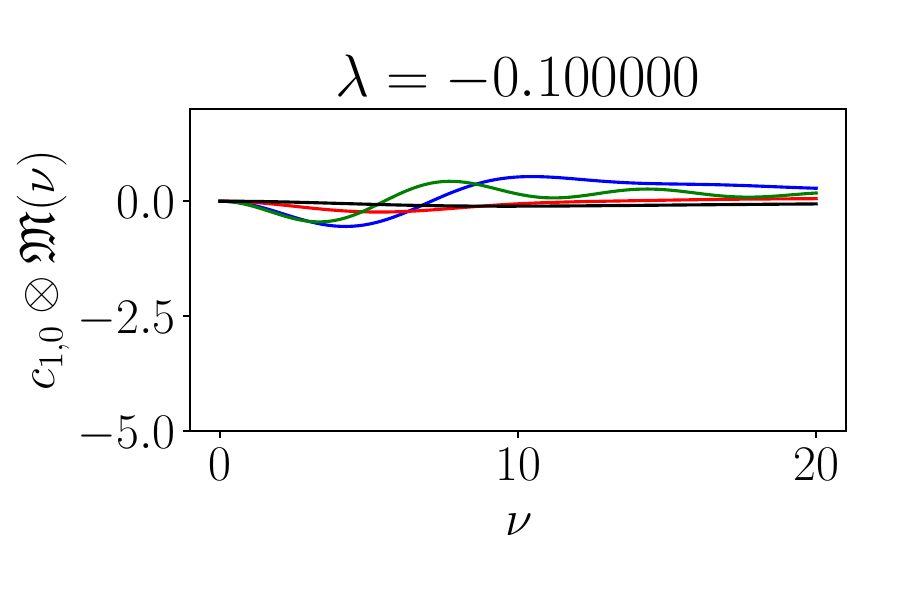}
    \includegraphics[width=0.38\textwidth]{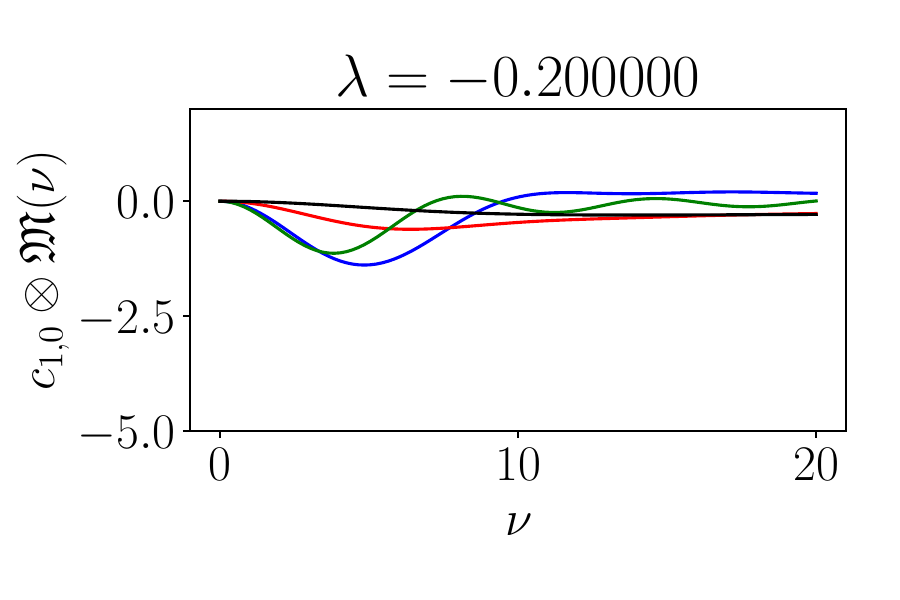}
    \includegraphics[width=0.38\textwidth]{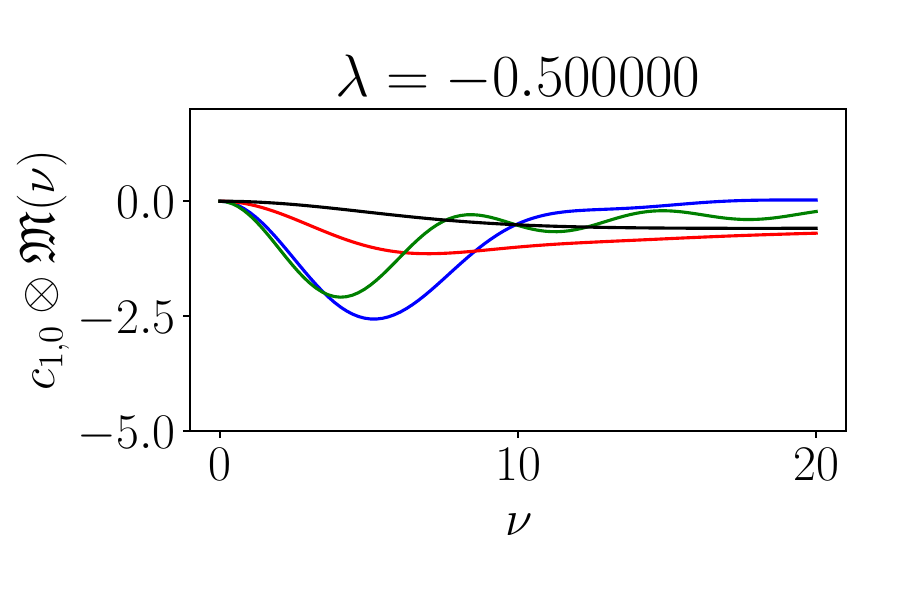}
    \includegraphics[width=0.38\textwidth]{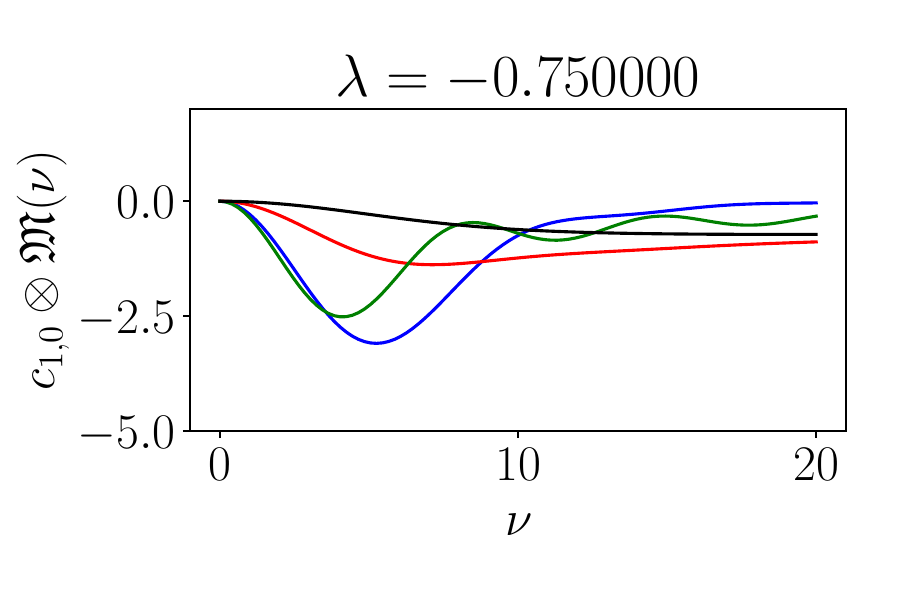}
    \caption{The convolution of the model functions with $c_{1,0}$ for variance $\lambda$. This can be compared to $\lambda=-e^{2\gamma_E+1}/4 \approx 2.15$ which are the dashed lines in upper right of Figure~\ref{fig:model_optimization}.}
    \label{fig:model_lambdas}
\end{figure}

The conventions of Eq.~\eqref{eq:bvjeknks} use $-\lambda = e^{2\gamma_E+1}/4 \approx 2.16$. On the other hand, \cite{Radyushkin:2018cvn, Li:2020xml} uses $-\lambda = e^{2\gamma_E}/4 \approx 0.79$. Considering that a given choice of $\lambda$ implies a simplified expression of the matching relation to the scale $\mu^2 = (\lambda z^2)^{-1}$, one could wonder if there is a natural choice to select the value of $\lambda$ with which to present the results. A similar question has been raised in \cite{Radyushkin:2018cvn} where a value of $\mu^2 \sim -16/z^2$ ($\lambda = -0.06$) has been argued. A natural way to address the question could be to try and minimize the convolution of a typical ITD with $c_{1,0}(\lambda)$, so as to single out a scale which brings minimal corrections at fixed order ${\cal O}(\alpha_s)$. It has been noted in the literature \cite{Joo:2019jct} that the terms $B_1$ and $D$ which intervene in the matching at order ${\cal O}(\alpha_s)$ produce typically contributions of opposite signs when convoluted to realistic ITDs.
There is an optimal value $\lambda_{opt}$ for which the convolution of the ITD with $c_{1,0}$ at a given $\nu$ will be 0. Figure~\ref{fig:model_optimization} shows 4 models of ITDs obtained from the cosine transform of $x^a(1-x)^b/B(a+1,b+1)$ where $B$ is the Beta function with varying $(a,b)$. These ITDs are convoluted with $B_1$ and $D$. In the range of Ioffe times available to modern lattice QCD, $|\lambda_{opt}| < 0.4$ for all the models given. This value corresponds to scales $\mu$ greater than $1/z$ by the factor $(-\lambda_{opt})^{-1/2}$ which is between [2,4.5] for $\nu<12$. Figure~\ref{fig:model_lambdas} shows the convolutions of $c_{1,0}$ with $\lambda=-0.1$ and -0.2. These convolutions have reduced magnitude compared to the convolutions of $D$ in Figure~\ref{fig:model_optimization}, which corresponds to $\lambda = -e^{2\gamma_E+1}/4$. There does appear to be a reduction in the magnitude, specifically for the model with the largest convolutions. 

Although this choice of scale reduces the typical size of the contribution of order ${\cal O}(\alpha_s)$, it does not offer much information on the size of further terms in the expansion. The trade-off is a potentially increase in $c_{2,0}$ which contains terms proportional to $\ln^2\left(-\frac{e^{2\gamma_E+1}}{4\lambda}\right)\approx 9.41$ for $\lambda=-0.1$. Even so a choice of $\lambda > -e^{2 \gamma_E+1}/4$ could still improve the perturbative expansion without larger contaminations of neglected terms. For $\lambda=-0.5$ the $c_{1,0}$ convolution's peak is reduced nearly a factor of 2 for some models while the higher order terms increase by the smaller coefficient of $\ln^2\left(-\frac{e^{2\gamma_E+1}}{4\lambda}\right)\approx2.16$. While the full optimization of $\lambda$ for NLO and NNLO effects is beyond the extent of this study, it appears clear that varying significantly for $\lambda > -e^{2 \gamma_E+1}/4$ has the potential to reduce perturbative effects.

\FloatBarrier
\subsection{Observations on the matching relations at order ${\cal O}(\alpha_s^2)$}

Modifying slightly the matching relation in \cite{Li:2020xml} for our needs, we can write:
\begin{equation}
    \langle P | \bar{\psi}(z) \gamma^\nu { \hat E} (0,z; A) \psi(0) | P\rangle = Z^{\overline{MS}}(z^2, \mu^2) \int_{-1}^1 dx\, q(x, \mu^2) K^\nu(x\nu, z^2, \mu^2)\,. \label{eq:rdonktefd}
\end{equation}
On the other hand, we have used in this document the definition, neglecting the higher-twist contributions:
\begin{equation}
    {\frak M}(\nu, z^2) \equiv \frac{\langle P | \bar{\psi}(z) \gamma^0 { \hat E} (0,z; A) \psi(0) | P\rangle}{\langle 0 | \bar{\psi}(z) \gamma^0 { \hat E} (0,z; A) \psi(0) | 0\rangle} \frac{\langle 0 | \bar{\psi}(0) \gamma^0 { \hat E} (0,z; A) \psi(0) | 0\rangle}{\langle P | \bar{\psi}(0) \gamma^0 { \hat E} (0,z; A) \psi(0) | P\rangle} = \int_{0}^1 d\alpha\, {\cal C}(\alpha, z^2, \mu^2) {\cal Q}(\alpha\nu, \mu^2)\,. \label{eq:rdbjkecdoi}
\end{equation}
From Eq.~\eqref{eq:rdonktefd}, we find that:
\begin{equation}
    {\frak M}(\nu, z^2) = \frac{1}{AK^0(0, z^2, \mu^2)}\int_{-1}^1 dx\, q(x, \mu^2) K^0(x\nu, z^2, \mu^2)\,, \label{eq:wsdfeidjknds}   
\end{equation}
where $A = \int_{-1}^1 dx\,q(x,\mu^2)$. Remembering that we have defined the Ioffe-time distribution through Eq.~\eqref{eq:ahonlvz}:
\begin{equation}
     {\cal Q}(\nu,\mu^2) \equiv \frac{1}{A}\int_{-1}^1 dx\, e^{i\nu x} q(x, \mu^2)\,,
\end{equation}
we transform Eq.~\eqref{eq:rdbjkecdoi} into:
\begin{align}
    {\frak M}(\nu, z^2) = \frac{1}{A}\int_{-1}^1 dx\, q(x, \mu^2) \int_0^1 d\alpha\,e^{ix\alpha\nu} {\cal C}(\alpha, z^2, \mu^2)\,,\label{eq:ddbhiknds}
\end{align}
and the comparison between Eqs.~\eqref{eq:wsdfeidjknds} and \eqref{eq:ddbhiknds} gives:
\begin{equation}
    {\cal C}(\alpha, z^2, \mu^2) = \frac{1}{K^0(0,z^2, \mu^2)} \int_{-\infty}^{+\infty} \frac{dy}{2 \pi} e^{-iy\alpha} K^0(y, z^2, \mu^2)\,. \label{eq:fevdbbijkns}
\end{equation}
From \cite{Li:2020xml}, we obtain that
\begin{align}
    K^0(0,z^2, \mu^2) &= 2 + \frac{\alpha_sC_F}{\pi}\left(\frac{3}{2}L+1\right) + {\cal O}(\alpha_s^2)\,,
\end{align}
where $L = \ln(-\mu^2z^2 e^{2\gamma_E+1}/4)$. On the other hand,
\begin{equation}
    \int_{-\infty}^{+\infty} \frac{dy}{2 \pi} e^{-iy\alpha} K^0(y, z^2, \mu^2) = K^0(0, z^2, \mu^2)\delta(1-\alpha) - \frac{\alpha_sC_F}{\pi}\left(L\left[\frac{1+\alpha^2}{1-\alpha}\right]_++\left[\frac{4 \ln(1-\alpha)}{1-\alpha} - 2(1-\alpha)\right]_+ \right) + {\cal O}(\alpha_s^2)\,.
\end{equation}
The ratio with $K^0(0,z^2,\mu^2)$ gives the expected result \eqref{eq:bvjeknks} up to the factor $A$ which arises due to the lost normalization of the PDF in the ratio. We also obtain a correction of order ${\cal O}(\alpha_s^2)$ which is a non-trivial consequence of the ratio:
\begin{align}
    {\cal C}(\alpha, z^2, \mu^2) &= \delta(1-\alpha) - \frac{\alpha_sC_F}{2\pi}\left(L\left[\frac{1+\alpha^2}{1-\alpha}\right]_++\left[\frac{4 \ln(1-\alpha)}{1-\alpha} - 2(1-\alpha)\right]_+ \right) \nonumber \\
    &\hspace{-30pt} + \left(\frac{\alpha_sC_F}{2\pi}\right)^2\left(\frac{3}{2}L^2\left[\frac{1+\alpha^2}{1-\alpha}\right]_++L\left[\frac{1+\alpha^2}{1-\alpha}+\frac{6 \ln(1-\alpha)}{1-\alpha} - 3(1-\alpha)\right]_++\left[\frac{4 \ln(1-\alpha)}{1-\alpha} - 2(1-\alpha)\right]_+ \right) +{\cal O}(\alpha_s^2)\,.\label{eq:fedijknec}
\end{align}
The missing term of order ${\cal O}(\alpha_s^2)$ is the genuine 2-loop contribution which is particularly cumbersome to express. The leading logarithmic term gives:
\begin{align}
&\frac{\alpha_s^2C_F}{2\pi^2}L^2\bigg[-C_F\bigg\{\frac{1-\alpha}{2}+\frac{(1+3\alpha^2)\ln(\alpha)}{4(1-\alpha)}-\frac{(1+\alpha^2)\ln(1-\alpha)}{1-\alpha}\bigg\}-\frac{11C_A(1+\alpha^2)}{24(1-\alpha)}+\frac{n_fT_F(1+\alpha^2)}{6(1-\alpha)}\bigg]_+\,.
\end{align}
According to the discussion of the previous section in Eq.~\eqref{eq:vrbhecioxsijo}, the sum of that term and the corresponding correction of order ${\cal O}(\alpha_s^2)$ derived in Eq.~\eqref{eq:fedijknec} should yield exactly $c_{2,2}(\lambda) = \left(B_1^{\otimes 2}-2\pi\beta_0 B_1\right)/2$. The sum writes:
\begin{equation}
    c_{2,2}(\lambda) = 2C_F\bigg[-C_F\bigg\{\frac{1-\alpha}{2}+\frac{(1+3\alpha^2)\ln(\alpha)}{4(1-\alpha)}-\frac{(1+\alpha^2)\ln(1-\alpha)}{1-\alpha}-\frac{3(1+\alpha^2)}{4(1-\alpha)}\bigg\}-\frac{11C_A(1+\alpha^2)}{24(1-\alpha)}+\frac{n_fT_F(1+\alpha^2)}{6(1-\alpha)}\bigg]_+\,. \label{eq:acbkzjcaz}
\end{equation}
It is easy to recognize in the last two terms $-\pi\beta_0B_1$. The first term is more sophisticated, and corresponds to $B_1^{\otimes 2}/2$, as can be observed by comparing the Mellin moments for instance.

\section{Existence and properties of the inverse for the DGLAP convolution}
\label{appendixa}

At order ${\cal O}(\alpha_s)$, the Mellin moments of the matching kernel defined in Eq.~\eqref{eq:wilsmom} read:
\begin{equation}
c_n\left(z^2\right)\ = 1 - \frac{\alpha_s(\mu^2)}{2\pi}C_F\left(2\psi^{(0)}(n)[2\gamma_E+\psi^{(0)}(n)] - 2\psi^{(1)}(n)+1+2\gamma_E^2+\frac{\pi^2}{3}-\frac{2}{n}+\frac{2}{(n+1)}\right) \,,
\label{eq:refdbjed}
\end{equation}
where $\psi^{(k)}(n)$ is the polygamma function, and $\mu^2 = (\lambda z^2)^{-1}$. The asymptotical behavior of the moments for $n \rightarrow 0$ and $n \rightarrow +\infty$ yields:
\begin{equation}
c_n\left(z^2\right)\ \  \mathrel{\overset{n\rightarrow 0}{\scalebox{1.5}[1]{$=$}}}\  \  1 - \frac{\alpha_s(\mu^2)}{2\pi}C_F\left(-\frac{2}{n}+3-\frac{2\pi^2}{3}+{\cal O}(n)\right) \rightarrow +\infty \,,\label{eq:vejdbln}
\end{equation}
\begin{equation}
c_n\left(z^2\right)\ \  \mathrel{\overset{n\rightarrow+\infty}{\scalebox{1.5}[1]{$=$}}}\  \  1 - \frac{\alpha_s(\mu^2)}{2\pi}C_F\left(2\ln(n)[\ln(n)+2\gamma_E]+1+2\gamma_E^2+\frac{\pi^2}{3}+{\cal O}\left(\frac{1}{n}\right)\right) \rightarrow -\infty\,.\label{eq:bhevwcnjka}
\end{equation}
Furthermore, we have demonstrated earlier \eqref{eq:kernel_norm} that, to all orders, 
\begin{equation}
    c_1(z^2) = 1\,.
\end{equation}
The derivative of Eq.~\eqref{eq:refdbjed} is strictly negative for $n \in [1, +\infty)$. The technical demonstration, without interest for the rest of the discussion, is produced at the end of this Appendix. 
As a consequence, Eq.~\eqref{eq:refdbjed}, which spans $(-\infty, 1]$ for $n \in [1, +\infty)$ is bound to vanish for a unique (non-integer) value which we denote $n_0(\alpha_s) > 1$. Therefore, if $c_n$ is truncated to order ${\cal O}(\alpha_s)$, $1/c_n$ exhibits a pole at $n = n_0(\alpha_s) > 1$. 

\begin{figure}[h]
    \centering
    \includegraphics[scale=.7]{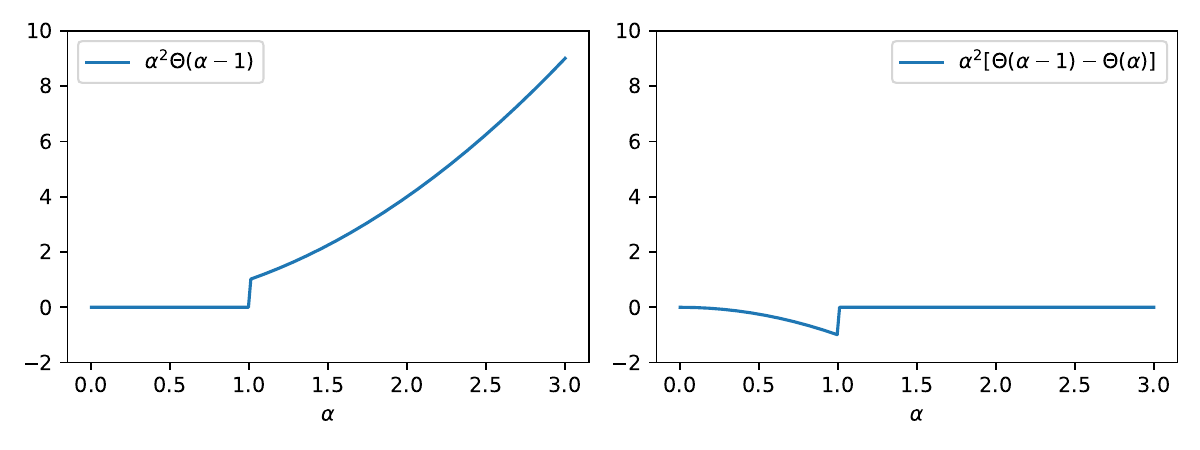}
    \caption{Two distributions whose Mellin moments are $-1/(n+2)$, but defined on different holomorphy strips: (left) $(-\infty, -2) + i\mathbb{R}$, (right) $(-2, +\infty)+ i\mathbb{R}$.}
    \label{fig3}
\end{figure}

In general, the Mellin transform \eqref{eq:wilsmom} can be evaluated for complex values of $n$. It is known classically that the integral converges in ``holomorphy strips''  \cite{Mellintransform} defined in the complex plane generically by $(n_1, n_2) + i\mathbb{R}$, where $n_1, n_2 \in \mathbb{R}\cup \{\pm\infty\}$. The Mellin transform of a distribution is really the pair consisting of the Mellin moments and the holomorphy strip on which they are defined. Let us briefly observe an example borrowed from \cite{Mellintransform} where the Mellin moments diverge at some value of $n$ (here $n=-2$):
\begin{equation}
    c_n = -\frac{1}{n+2}\,. \label{eq:rbvesn}
\end{equation}
There are two possible holomorphy strips, $(-\infty, -2) + i\mathbb{R}$ and $(-2, +\infty)+ i\mathbb{R}$. In fact, one can verify easily that the distribution giving rise to the Mellin moments \eqref{eq:rbvesn} in the first holomorphy strip is $\alpha^2 \Theta(\alpha-1)$ where $\Theta$ is the Heaviside step function:
\begin{equation}
    \int_0^\infty d\alpha\,\alpha^{n-1}\,\alpha^2 \Theta(\alpha-1) = \int_{1}^\infty d\alpha\,\alpha^{n+1} = -\frac{1}{n+2}\ \ \textrm{if Re}(n) \leq -2\,.
\end{equation}
On the other hand, the distribution giving rise to the same Mellin moments in the second holomorphy strip is $\alpha^2 [\Theta(\alpha-1)-\Theta(\alpha)]$:
\begin{equation}
    \int_0^\infty d\alpha\,\alpha^{n-1}\,\alpha^2 [\Theta(\alpha-1)-\Theta(\alpha)] = -\int_0^{1} d\alpha\,\alpha^{n+1} = -\frac{1}{n+2}\ \ \textrm{if Re}(n) \geq -2\,.
\end{equation}
The two distributions which give rise to the same general expression for the Mellin moments, but on two different holomorphy strips are shown on Figure \ref{fig3}. It appears clearly that the choice of holomorphy strip changes crucially the properties of the distribution, notably its support. 

In the case of the inverse matching kernel, we face a similar issue. In fact, there are infinitely many holomorphic strips available to reconstruct the inverse matching kernel from $1/c_n$, spanning the entire real values of $n$. But only one of those strips produces an integrable distribution: by definition, if the distribution is integrable, its moment for $n = 1$ is finite. Therefore, the holomorphy strip which we will use is $(n', n_0(\alpha_s))+i\mathbb{R}$, where $n' < 1 < n_0(\alpha_s)$\footnote{In fact, $n' = 0$ as a consequence of the (non-demonstrated) fact that the derivative of Eq.~\eqref{eq:refdbjed} is strictly negative on $n \in (0, +\infty)$, and not just on $n \in [1, +\infty)$.}. If the inverse matching kernel exists, then its Mellin moments in the vicinity of $n_0(\alpha_s)$ are:
\begin{equation}
    \frac{1}{c_{n}}\  \mathrel{\overset{n\rightarrow n_0(\alpha_s)}{\scalebox{1.5}[1]{$=$}}}\    \frac{1}{n-n_0(\alpha_s)}\left(\frac{d}{dn}c_{n_0}\right)^{-1} + {\cal O}(1)\,.\label{eq:jevbdn}
\end{equation}
An inverse Mellin transform of the right-hand side of \eqref{eq:jevbdn} on the strip $(n', n_0(\alpha_s))+i\mathbb{R}$ gives that, \textit{provided the inverse of the matching kernel exists and is integrable}, its general behavior at large $\alpha$ has to be dominated by: 
\begin{equation}
    {\cal C}^{\otimes-1}(\alpha)  \  \widesim[2]{\alpha \rightarrow +\infty} \  \left(\frac{d}{dn}c_{n_0}\right)^{-1} \alpha^{-n_0(\alpha_s)}\,.
\end{equation}
Therefore, the only integrable inverse matching kernel we can plausibly reconstruct from $1/c_n$ has a support which extends to arbitrarily large $\alpha$, unlike the direct matching kernel whose support is limited to $\alpha \in [0,1]$. This unwanted large $\alpha$ behavior decreases as a power function whose exponent depends on the value of $\alpha_s$, and vanishes in the limit where $\alpha_s = 0$ which implies $n_0(\alpha_s) = +\infty$.

Schematically speaking, the Mellin moments of the direct matching kernel have the form:
\begin{equation}
    c_n = 1 - \alpha_s h(n)\,. \label{eq:gbrvckhaj}
\end{equation}
Hence: \begin{equation}
    \frac{1}{c_n} = 1 + \alpha_s h(n) + \alpha_s^2 h^2(n) + ...
\end{equation}
The terms of order $\alpha_s^2$ and beyond are incomplete, since they will receive unknown corrections from higher orders in the perturbative expansion  of $c_n$. Therefore, the issue of determining whether the inverse of the matching kernel truncated at order ${\cal O}(\alpha_s)$ exists, and what its support is, is not particularly worrisome \textit{per se}. It should serve however as a warning of potentially complicated effects linked to the perturbative truncation. To recover a well-defined object that can serve as an inverse matching kernel and whose support is restricted to $[0,1]$, it is enough to truncate ${\cal C}_0^{\otimes-1}$ (or equivalently $1/c_n$) to order ${\cal O}(\alpha_s)$ too, thereby ignoring altogether the higher-order contributions we partially resummed when we considered the exact value of $1/c_n$:
\begin{equation}
    {\cal C}_0^{\otimes-1}(z^2) = 1_{\otimes} + \frac{\alpha_s(\mu^2)}{2\pi}D + {\cal O}(\alpha_s^2)\,.\label{eq:gbevwocjna}
\end{equation}

\subsection*{A technical demonstration}

\label{technique}
We want to demonstrate that the derivative of the Mellin moments of the matching kernel truncated at order ${\cal O}(\alpha_s)$ \eqref{eq:refdbjed} is strictly negative for $n \in [1, +\infty)$. In fact, that is also true for $n \in (0, +\infty)$ although the demonstration is more sophisticated in the latter case and of no interest for this discussion. The polygamma function may be defined as:
\begin{equation}
    \psi^{(k)}(n) \equiv \frac{d^{k+1}}{dn^{k+1}} \ln \Gamma(n)\,,\label{eq:vrinjfevfv}
\end{equation}
where $\Gamma$ is the ordinary Gamma function. Then it follows immediately that:
\begin{equation}
    \frac{d}{dn}\psi^{(k)}(n) = \psi^{(k+1)}(n)\,, \label{eq:evjsfdnrsd}
\end{equation}
and the derivative of Eq.~\eqref{eq:refdbjed} gives:
\begin{equation}
    \frac{d}{dn} c_n = - \frac{\alpha_s(\mu^2)}{2\pi} C_F\bigg( 4\psi^{(1)}(n)(\gamma_E+\psi^{(0)}(n)) - 2\psi^{(2)}(n) + \frac{2}{n^2}-\frac{2}{(n+1)^2}\bigg)\,. \label{eq:edjncscsd}
\end{equation}
In parallel, it is well-known that the polygamma function admits the following integral representation for $k > 0$:
\begin{equation}
    \psi^{(k)}(n) = -\int_0^1 dt\frac{t^{n-1}}{1-t} \ln^k (t)\,, \label{eq:einjczcex}
\end{equation}
which gives trivially that $\psi^{(1)}(n) > 0$ and $\psi^{(2)}(n) < 0$ for $n > 0$. Therefore, $\psi^{(0)}(n)$ is a strictly increasing function, and since $\psi^{(0)}(1) = -\gamma_E$, we find for $n \geq 1$,
\begin{equation}
    \gamma_E + \psi^{(0)}(n) \geq 0\,.
\end{equation}
This demonstrates that the derivative \eqref{eq:edjncscsd} is strictly negative for $n \geq 1$ and gives the expected result. The result is also true for $n \in (0,1)$, for instance since
\begin{equation}
    \psi^{(0)}(n) - \psi^{(0)}(1) > \frac{\psi^{(2)}(n)}{2\psi^{(1)}(n)}\,.
\end{equation}
This statement is particularly obvious if $n \geq 1$ since the l.h.s. is positive and the r.h.s. strictly negative, but is more subtle if $n \in (0,1)$. We will not seek a further demonstration.

\section{Positivity of the LL $\overline{MS}$ step-scaling function}
\label{positivity}

To demonstrate the positivity of the LL $\overline{MS}$ step-scaling function, we can show that it satisfies the Hausdorff moment problem \cite{Hausdorff}. The latter establishes the necessary and sufficient condition so that a sequence of integer Mellin moments for $n = 1$ to infinity corresponds to a unique Borel measure supported on the interval $[0, 1]$. The condition is that the sequence of Mellin moments $(e_n)_n$ is completely monotonic, that is for all $n \geq 1$ and $k \geq 0$:
\begin{equation}
    (-1)^k \Delta^k e_n \geq 0\,, \label{eq:rfnjkiorf}
\end{equation}
where $\Delta^k$ is the difference operator applied $k$-times:
\begin{equation}
    \Delta e_n = e_{n+1}-e_n\,,\ \ \Delta^2 e_n = e_{n+2}-2e_{n+1}+e_n\,, ...
\end{equation}
We remind that \eqref{eq:evfdvscvc}:
\begin{equation}
e_n(\mu_0^2, \mu_1^2) = \left(\frac{\alpha_s(\mu_0^2)}{\alpha_s(\mu_1^2)}\right)^{\gamma_n/(2\pi\beta_0)}\,,
\end{equation}
and the anomalous dimensions $\gamma_n$ are given by:
\begin{equation}
    \gamma_n = -2\psi^{(0)}(n)-2\gamma_E+\frac{3}{2}-\frac{1}{n}-\frac{1}{n+1}\,, \label{eq:rhuinjecdjn}
\end{equation}
where $\psi^{(0)}$ is the digamma function defined in Eq.~\eqref{eq:vrinjfevfv}.

The theorem 11d in Chapter 4 of \cite{widder2015laplace} gives that the sequence $(e_n)_n$ is completely monotonic if the function $e_n$ of $n$ is a completely monotonic function for $n \geq 1$, a related property expressed in terms of derivatives instead of difference operators:
\begin{equation}
    (-1)^k \frac{d^k}{dn^k} e_n \geq 0\,.\label{eq:cdnjkrvdv}
\end{equation}
The demonstration is a straightforward application of the mean value theorem. As $e_n \geq 0$, the property is obviously satisfied for $k = 0$. Differentiating repeatedly $e_n$ as a composite function allows to express its derivatives with respect to those of $\gamma_n$ for $k \geq 1$:
\begin{equation}
    \frac{d^k}{dn^k} e_n = e_n \sum_{m = 1}^k \frac{1}{(2\pi\beta_0)^m}\ln^{m}\left(\frac{\alpha_s(\mu_0^2)}{\alpha_s(\mu_1^2)}\right)\sum_{\substack{A = \{i_1,...,i_m\} \\ i_j \geq 1\,, \sum i_j = k}} c_A \prod_{i_j \in A} \frac{d^{i_j}}{dn^{i_j}} \gamma_n\,, \label{eq:eiojekedc}
\end{equation}
where
\begin{equation}
c_{\{m\}} = 1\,,\ \ c_{\{1,...,1\}} = 1\,,\ \ c_{\{2,1\}} = 3\,,\ \ c_{\{2,1,1\}} = 6\,,\ \ c_{\{2,2\}} = 3\,,\ \ c_{\{3,1\}} = 4\,, ...
\end{equation}
In spite of its apparent complicated form, Eq.~\eqref{eq:eiojekedc} is simply a Leibniz-rule kind of formula where all possible combinations of the derivatives of $\gamma_n$ with correct order enter. The precise value of the coefficients $c_A$ is a complicated combinatorial expression which does not matter in this discussion except for the observation that it is a positive number, since no negative signs are ever involved in the differentiation formula. Then, let us express the derivatives of Eq.~\eqref{eq:rhuinjecdjn} for $k \geq 1$:
\begin{equation}
    \frac{d^k}{d n^k} \gamma_n = -2\psi^{(k)}(n) - (-1)^k k!\left[\frac{1}{n^{k+1}}+\frac{1}{(n+1)^{k+1}}\right] \,,
\end{equation}
where we have used Eq.~\eqref{eq:evjsfdnrsd}. We can rewrite that relation in an integral form inspired from Eq.~\eqref{eq:einjczcex}:
\begin{equation}
    (-1)^k \frac{d^k}{d n^k} \gamma_n = 2\int_0^1 dt\,t^{n-1} (-\ln(t))^k\left[\frac{1}{1-t}-\frac{1}{2}-\frac{t}{2}\right]\,,
\end{equation}
and it is easy to see that the last term is positive for $t \in [0,1]$ (for instance by derivating it yet again) so that $(-1)^k \frac{d^k}{d n^k} \gamma_n \geq 0$ if $k \geq 1$ and $n \geq 0$.

Therefore, if $A = \{i_1, ..., i_m\}$ is a set of strictly positive integers such that $\sum i_j = k$, then
\begin{equation}
    (-1)^k \prod_{i_j \in A} \frac{d^{i_j}}{dn^{i_j}} \gamma_n \geq 0\,.
\end{equation}
Using this relation in conjunction with Eq.~\eqref{eq:eiojekedc}
gives the expected property of Eq.~\eqref{eq:cdnjkrvdv} provided that $\alpha_s(\mu_0^2) / \alpha_s(\mu_1^2) \geq 1$, or in other words if evolution is performed in the forward direction. The property is not verified for the backward evolution operator as we have already observed.

\subsection*{Acknowledgments}
We thank Jianwei Qiu and Anatoly Radyushkin for stimulating discussions and comments on the manuscript. HD further thanks Valerio Bertone for technical discussions on the APFEL++ evolution code. KO and HD were supported in part  by the U.S.~DOE Grant \mbox{\#DE-FG02-04ER41302}. CJM is supported in part by the U.S.~DOE EC Award \mbox{\#DE-SC0023047}. KO and JK were supported in part by the US Department of Energy (DOE) Contract No.~DE-AC05-06OR23177,
under which Jefferson Science Associates, LLC operates Jefferson Lab. SZ acknowledges support by the French Centre national de la recherche scientifique (CNRS) under an Emergence@INP 2023 project. This work has benefited from the collaboration enabled by the Quark-Gluon Tomography (QGT) Topical Collaboration, U.S.~DOE Award DE-SC0023646.

\bibliography{biblio}
\end{document}